\documentclass[12pt,onecolumn]{article}
\usepackage{color}
\usepackage{amsmath}
\usepackage{amssymb}
\usepackage{amsfonts}
\usepackage{geometry}
\usepackage{setspace}
\usepackage{amsmath,bm}
\usepackage{graphicx}
\usepackage[section]{placeins}% avoid figures cross the section
\usepackage{subfigure}
\usepackage[sort&compress]{natbib}
\usepackage{booktabs}  
\usepackage{makecell}
\usepackage{caption}
\usepackage[colorlinks, citecolor=blue]{hyperref}% link to the superlink
\usepackage{lineno}% line number

\author{Shuwei Zhou$^{1,2}$, Xiaoying Zhuang$^{1,2,*}$}
\title {Phase field modeling of hydraulic fracture propagation in transversely isotropic poroelastic media}
\begin{document}
%\linenumbers % line numbers
% Figure title
\captionsetup[figure]{labelfont={bf},name={Fig.},labelsep=space}
% referecne style
\bibliographystyle{unsrtnat}
\setcitestyle{numbers,square,aysep={},yysep={,}}
% delte the date
\date{}
\maketitle

\spacing {2}
\noindent
1 Department of Geotechnical Engineering, College of Civil Engineering, Tongji University, Shanghai 200092, P.R. China\\
2 Institute of Continuum Mechanics, Leibniz University Hannover, Hannover 30167, Germany\\
$*$ Corresponding author: Xiaoying Zhuang (zhuang@ikm.uni-hannover.de)

%\section*{Highlight}
%	\begin{itemize}
%		\item A phase field model is proposed for hydraulic fracture propagation in transversely isotropic media.
%		\item The proposed method is initially verified by a single-edge-notched square plate subjected to tension and an isotropic porous medium subjected to internal fluid pressure.
%		\item Hydraulic fractures in 2D transversely isotropic media with a notch and two parallel notches are presented.
%		\item Hydraulic fractures in a 3D transversely isotropic medium with a penny-shaped notch are presented.
%	\end{itemize}

\begin{abstract}
\noindent This paper proposes a phase field model (PFM) for describing hydraulic fracture propagation in transversely isotopic media. The coupling between the fluid flow and displacement fields is established according to the classical Biot poroelasticity theory while the phase field model characterizes the fracture behavior. The proposed method uses a transversely isotropic constitutive relationship between stress and strain as well as anisotropy in fracture toughness and permeability. An additional pressure-related term and an anisotropic fracture toughness tensor are added in the energy functional, which is then used to obtain the governing equations of strong form via the variational approach. In addition, the phase field is used to construct indicator functions that transit the fluid property from the intact domain to the fully fractured one. Moreover, the proposed PFM is implemented using the finite element method where a staggered scheme is applied and the displacement and fluid pressure are monolithically solved in a staggered step. Afterwards, two examples are tested to initially verify the proposed PFM: a transversely isotropic single-edge-notched square plate subjected to tension and an isotropic porous medium subjected to internal fluid pressure. Finally, numerical examples of 2D and 3D transversely isotropic media with one or two interior notches subjected to internal fluid pressure are presented to further prove the capability of the proposed PFM in 2D and 3D problems.
\end{abstract}

\noindent Keywords: Phase field model, Hydraulic fracturing, Transverse isotropy, Porous media, Fracture propagation, Staggered scheme
 
%\twocolumn
\section {Introduction}\label{Introduction}

In recent years, the prediction of fracture propagation in porous media has been an attractive and significant research topic in civil, mechanical, environmental, energy, and geological engineering because it is highly important to relevant topics such as hydraulic fracturing (HF), which helps to enhance the exploitation efficient of oil, tight gas, methane, and shale gas from unconventional reservoirs. In HF practice, highly pressurized fluid is injected to a reservoir to form a network of transportation pathways for resource extraction. The reservoirs with low porosity and permeability are therefore fractured, which significantly enhances the overall efficiency of energy exploitation. However, HF is also controversial because in many circumstances it produces unintentionally unfavorable flow channels through which fracturing fluid or gas may leak; this will probably result in the contamination of ground water \citep{osborn2011methane,vidic2013impact}. In addition, the fracturing fluid spills may contaminate surface and reduce air quality \citep{mikelic2013phase}. Therefore, the vast economic benefits and their corresponding risks brought by hydraulic fracturing require more accurate numerical tools to predict complex fracture propagation in porous media.

While most mechanical models of HF treat the rock material as isotropic (for example, \citet{detournay1991plane, detournay2016mechanics, zimmermann2009pressure, schrefler2006adaptive} set the plane strain fractures in isotropic rocks), the reservoir rocks are commonly anisotropic \citep{park2015bonded}, in the reality especially for shale and the evidence for this can been seen in many experimental tests \citep{niandou1997laboratory, tien2006experimental, lin2017experimental}. Anisotropy in rocks mainly results from weak planes, especially such as bedding, foliation, or schistosity in sedimentary and metamorphic rocks. It is believed that these weak planes are the reason for large deformation, strength reduction, and anisotropy of a rock \citep{park2015bonded, lin2017experimental}. Therefore, hydraulic fractures in anisotropic rocks such as shale formation \citep{chen2018anisotropy} are significantly different from those in conventional reservoir formations because the fractures tend to propagate along the weak planes, producing more complex growth patterns.

These observations thus require proper consideration of material anisotropy, which is of great importance and significance for predicting fracture paths accurately in hydraulic fracturing. In this paper, we focus on the transversely isotropic model accounting for the layered structure of shale. It is at present the most relevant and a most widely used model in rock mechanics for anisotropy. For example, the shale formation is considered as transversely isotropic in engineering scale \citep{chen2018anisotropy} where the weak planes are usually believed to have one axis of symmetry \citep{cho2012deformation}. On the transversely isotropy in a porous medium, three aspects are closely related to HF, namely, the anisotropy in elastic modulus, in permeability, and in fracture toughness. Therefore, to predict accurately the hydraulic fractures in the porous medium, the three types of anisotropy, flow field, geomechanics and fracture model should be coupled in the numerical simulation \citep{zeng2019study}.

A suitable fracture model is of the greatest importance in hydraulic fracturing modeling. In the framework of continuum mechanics, the fracture models can be typed into two branches: the discrete and smeared approaches. A discrete approach applies directly or indirectly displacement discontinuity across the crack surface and popular discrete approaches include the extended finite element method (XFEM) \citep{moes2002extended, chen2012extended}, generalized finite element method (GFEM) \citep{fries2010extended}, phantom-node method \citep{chau2012phantom, rabczuk2008new}, and element-erosion method \citep{belytschko1987three, johnson1987eroding}. More specifically, the original discrete approach \citep{ingraffea1985numerical} creates new free stress boundaries for fractures and the mesh topology varies followed by fracture growth, which is similar to the settings in DDA \citep{zheng2018generalized, zheng2018algorithmic, zheng2019modified}. XFEM \citep{moes2002extended} adds extra discontinuous shape functions to represent displacement discontinuity in the cracked element. The element-erosion method \citep{belytschko1987three, johnson1987eroding} setting a zero stress in the fractured elements but this method cannot model fracture branching. 

On the other hand, differently from the discrete approaches, the displacement field is continuous in a smeared approach, which regards the intact and fractured domains as a whole \citep{santillan2017phase}; therefore, the discontinuity-induced issues in a discrete approach are effectively eliminated. Typical smeared approaches are peridynamics, gradient damage models, and phase field models (PFMs). In peridynamics, a solid is composed of material points with a local region named horizon controlling their interaction \citep{ren2017dual, ren2016dual}. In addition, in a gradient model \citep{peerlings1996some} or a phase field model \citep{miehe2010thermodynamically, zhou2019phase2}, the fracture is represented by a localization band with its width proportional to an internal length scale. Compared with the damage gradient models, PFMs are gaining more and more popularity now \citep{miehe2010thermodynamically, miehe2010phase, borden2012phase}; in this method, a continuous auxiliary scalar field (the phase field) is used to diffuse the discrete fracture, and the intact and fully fractured solids are linked through a narrow transition band with its width related to a length scale parameter \citep{zhou2018phase2}. It should be noted that the PFM has a different treatment of localization and length scale compared with the gradient models.

An early phase field model for quasi-static brittle fracture was proposed by \citet{bourdin2008variational} where a variational framework was used and the phase field was first defined to affect the material fracture energy. Afterwards, \citet{miehe2010thermodynamically} presented the thermodynamic consistent phase field framework for brittle fractures; this creative work used a strain decomposition and also a history reference to ensure fracture irreversibility. \citet{miehe2010phase} then proposed a variational framework for rate-independent fractures with robust implementation algorithms. In addition, the PFMs for dynamic fractures are reported in \citep{borden2012phase, hofacker2012continuum, hofacker2013phase}. Recently, \citet{bryant2018mixed, noii2019adaptive} developed the phase field models for anisotropic fractures. All the contributions in PFMs \citep{bourdin2008variational, miehe2010thermodynamically, miehe2010phase, borden2012phase, hofacker2012continuum, hofacker2013phase, bryant2018mixed, noii2019adaptive} have shown the potential of a PFM for simulating hydraulic fracturing due to many advantages: i) all the PF simulations can be performed on a fixed topology without any remeshing technique; ii) PFMs can reproduce well complex fracture patterns such as merging and branching; iii) the fracture path is automatically detected and there is no extra treatment for displacement discontinuity or stress intensity factor as in discrete approaches \citep{fu2013boundary, wu2015simultaneous, fu2018singular, fu2018boundary}; and iv) PFMs can easily simulate fractures in heterogeneous media. Therefore, PFMs are a promising fracture model for hydraulic fracturing and it can easily overcome some essential difficulties in HF such as laborious treatments on complex fracture patterns including merging, curving and branching and heterogeneous materials.

In recent years, many researchers have seeking approaches to couple PFMs to hydraulic fracturing and have made some progress \citep{bourdin2012variational, wheeler2014augmented, mikelic2015quasi, mikelic2015phase,heister2015primal, lee2016pressure,wick2016fluid, yoshioka2016variational, miehe2015minimization, miehe2016phase, ehlers2017phase, santillan2017phase} among which \citet{bourdin2012variational} made an early contribution. The variational method was used for hydraulic fracture modeling by considering the work of fluid pressure along the fracture. \citet{wheeler2014augmented} added an extra poroelastic term to rewrite the energy functional and succeeded in coupling the PFM to poroelastic media. However, the distribution of reservoir and fractured domain was regarded as a moving boundary problem, which requires extra work for numerical implementation. Later, \citet{mikelic2015quasi, mikelic2015phase} proposed a new energy functional to fully couple the elasticity, phase field, and fluid pressure; the Biot equations were enhanced in their approach by using an anisotropic permeability tensor indicating a higher permeability along the fracture. Afterwards, the numerical implementations of PFMs for HF were enhanced using adaptive element schemes by \citet{heister2015primal, lee2016pressure}. \citet{wick2016fluid, yoshioka2016variational} then directly implemented PFMs in reservoir simulators. In addition, \citet{miehe2015minimization, miehe2016phase} presented robust phase field modeling theory for fractures in poroelastic media, and in their framework the phase field was driven by an effective stress in the solid medium with a specified threshold; moreover, a Darcy flow was also assumed in the medium with an anisotropic permeability tensor. Recently, \citet{ehlers2017phase} coupled the phase-field model with the theory of porous media to characterize dynamic hydraulic fracturing. \citet{santillan2017phase} proposed an immersed-fracture formulation for an impermeable porous media. \citet{zhou2018phase2} presented a novel phase field modeling approach for fluid-driven fractures based on the COMSOL platform. However, to the authors' best knowledge, none of the above-mentioned contributions on PFM of hydraulic fractures focus on transversely isotropic porous media and the fracture evolution is isotropic but not anisotropic.

To this end, this paper proposes a new phase field model for hydraulic fracture propagation in transversely isotopic media. The coupling between the fluid flow field and geomechanics is established based on the classical Biot poroelasticity theory while the phase field model is used to reflect the fracture behavior. It should be noted that due to the transversely isotropy, the transversely isotropic constitutive relationship between stress and strain is applied with consideration of the anisotropy in fracture toughness and permeability. Firstly, we add an additional pressure-related term and an anisotropic fracture toughness tensor in the energy functional, which is then used to achieve the governing equations of strong form through the variational method. Secondly, the phase field is used to construct indicator functions that transit the fluid property from the intact domain to the fully fractured one. The phase field model is implemented by using the finite element method where a staggered scheme is applied. In addition, two examples are used to initially verify the proposed PFM: a transversely isotropic single-edge-notched square plate subjected to tension and an isotropic porous medium subjected to internal fluid pressure. Finally, three representative numerical examples are presented to further prove the capability of the proposed PFM in 2D and 3D problems, namely, internal fluid-driven fracture propagation in a 2D transversely isotropic medium with an interior notch, a 2D medium with two parallel interior notches, and a 3D medium with a penny-shaped notch.

This paper is organized as follows. We give the mathematical models for hydraulic fracture propagation in a transversely isotropic medium in Section \ref{Phase field model for transversely isotropic poroelastic media} followed by Section \ref{Numerical algorithm} which shows the numerical algorithm on the phase field modeling. In Section \ref{Validation of the PFM}, we initially verify the proposed PFM via two cases: a transversely isotropic single-edge-notched plate subjected to tension and a 2D isotropic medium subjected to an internal fluid pressure. Afterwards, we present some 2D and 3D examples of transversely isotropic media subjected to internal fluid pressure in Section \ref{Hydraulic fractures in the transversely isotropic medium}. Finally, we end with concluding remarks regarding our work in Section \ref{Conclusions}. 

\section {Phase field model for transversely isotropic poroelastic media}\label{Phase field model for transversely isotropic poroelastic media}
\subsection {Energy functional}

\begin{figure}[htbp]
	\centering
	\subfigure[Porous domain]{\includegraphics[height = 5cm]{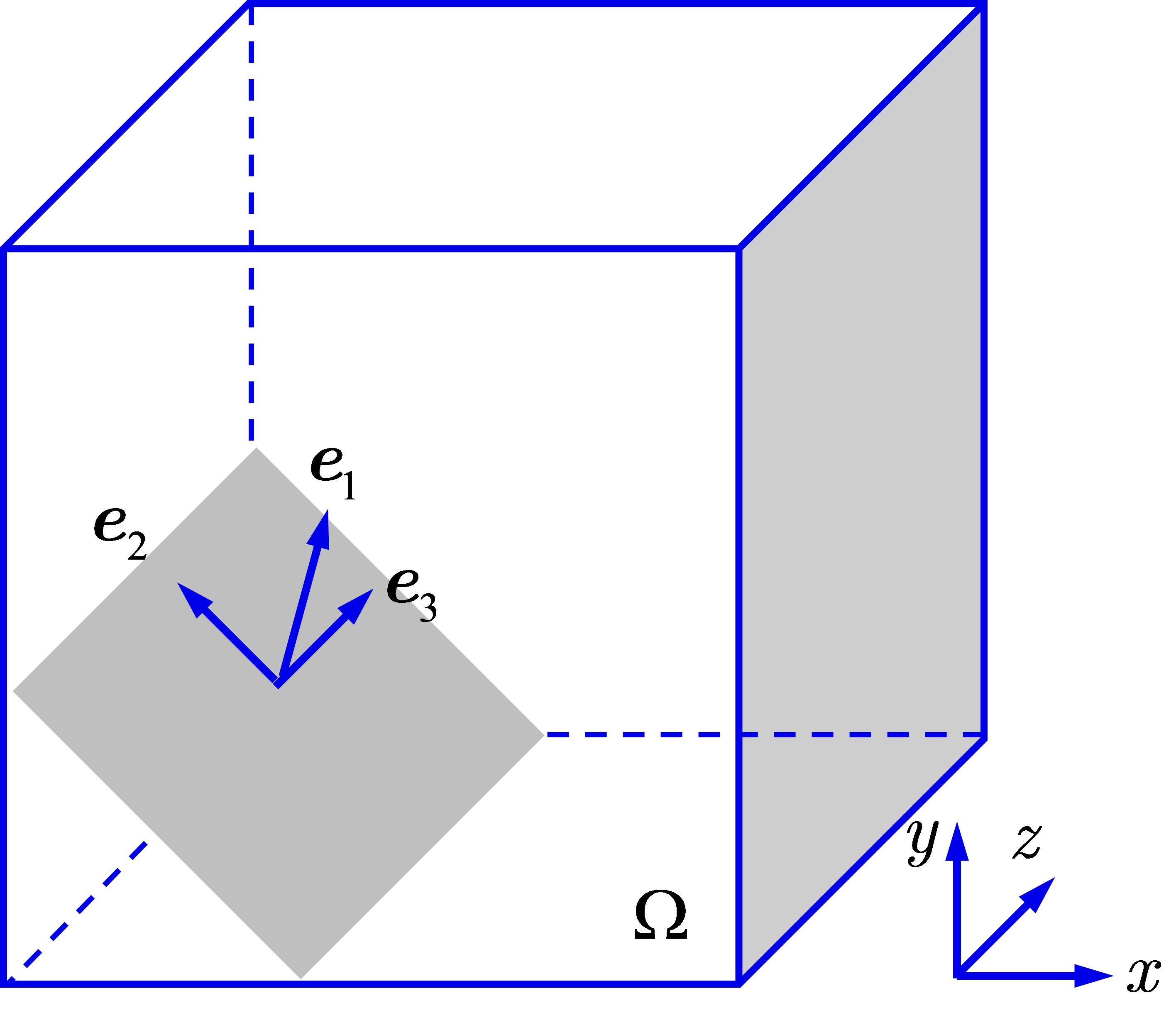}}
	\subfigure[2D profile]{\includegraphics[height = 5cm]{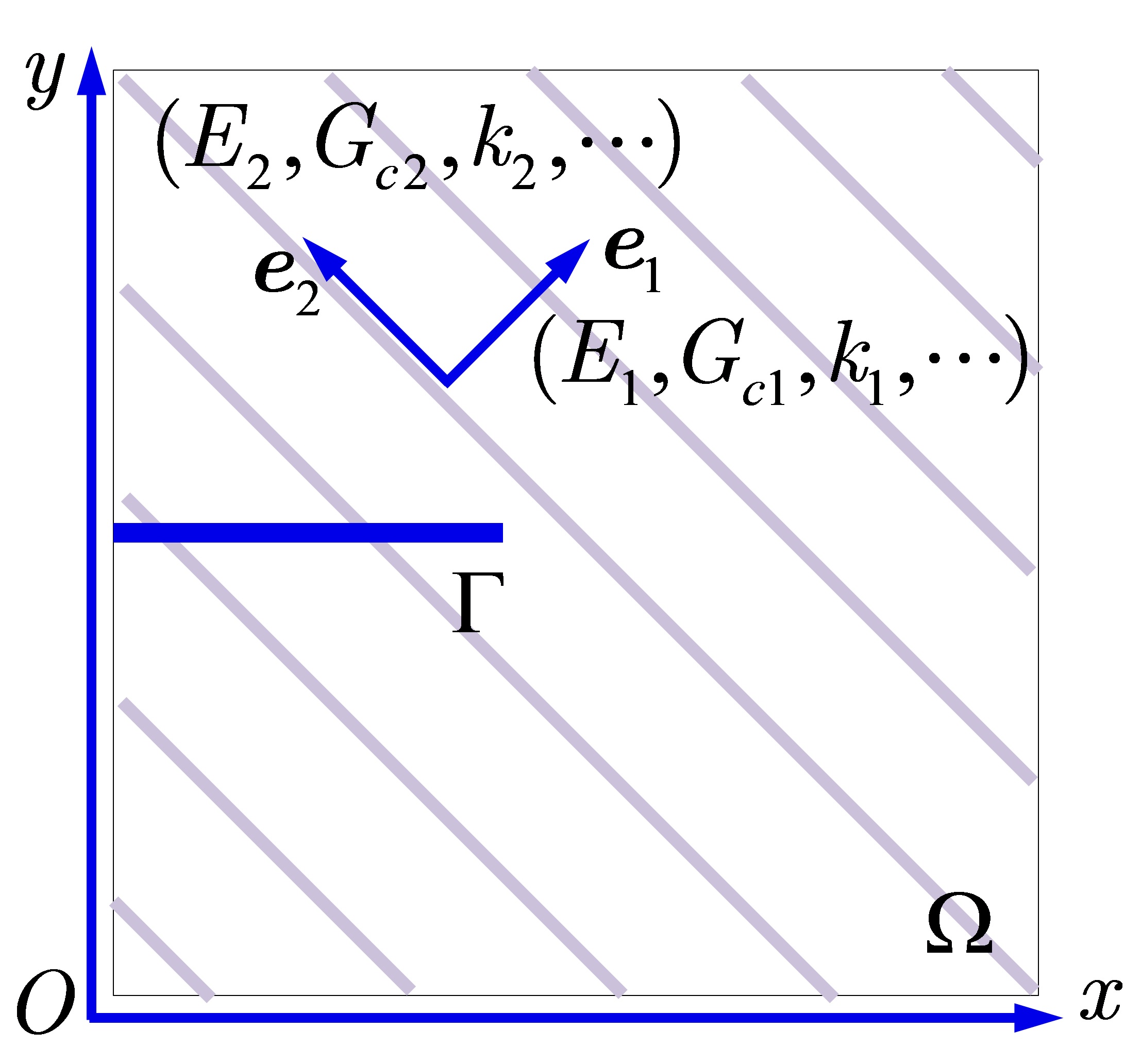}}
	\subfigure[Realistic layered rock \citep{shang2018numerical}]{\includegraphics[width = 3cm, height=5cm]{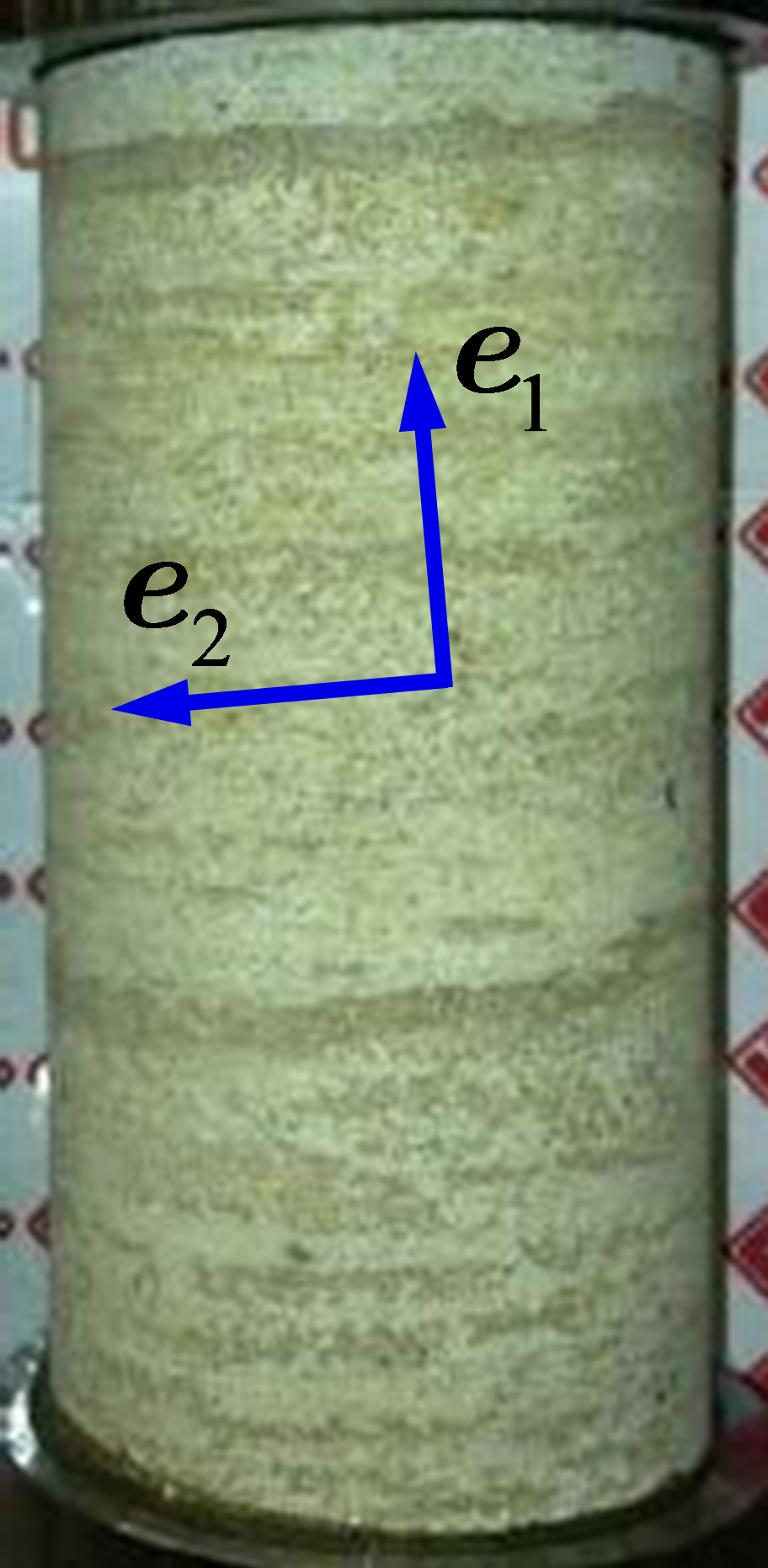}}
	\caption{Transversely isotropic porous medium}
	\label{Sharp fracture in a transversely isotropic porous medium}
	\end{figure}

Let us consider a cracked transversely isotropic permeable porous solid $\Omega$ in Fig. \ref{Sharp fracture in a transversely isotropic porous medium} where the domain boundary is denoted as $\partial \Omega$ with its outward unit normal vector described as $\bm n$. For the displacement field, $\partial \Omega$ includes the time-dependent Dirichlet boundary $\partial \Omega_u$ and Neumann boundary $\partial \Omega_t$; $\partial \Omega_t \cap \partial \Omega_u = \emptyset$ and $\overline{\partial \Omega_t \cup \partial \Omega_u} = \partial \Omega$. A specified displacement is applied on $\partial \Omega_u$ and a traction $\bm {t}$ on $\partial \Omega_t$. In addition, in Fig. \ref{Sharp fracture in a transversely isotropic porous medium}b a sharp fracture $\Gamma$ emerges in the domain and Fig. \ref{Sharp fracture in a transversely isotropic porous medium}c represents a realistic layered rock \citep{shang2018numerical}. 

The medium $\Omega$ is assumed to have an axis of rotational symmetry along the $\bm e_1$ direction; therefore, the material property in any direction in a plane perpendicular to this rotation axis is identical. We then define another two orthogonal vectors $\bm e_2$ and $\bm e_3$ to define the transverse plane. It should be noted that in this work, the direction $\bm e_3$ is set to coincide with the $z-$axis of the global coordinate system for simplicity. The relationship between the material directions and global coordinate system can be also seen in Fig \ref{Sharp fracture in a transversely isotropic porous medium}a and b. After this, we construct the space of the displacement $\bm u (\bm x)$ and its variation $\delta \bm u$ before presenting the energy functional governing the phase field modeling:
	\begin{equation}
	\bm u (\bm x) \in U_u \mathrel{\mathop:}= \left \{ \bm u | \bm u(\bm x) = \bm {u^*} \hspace {0.25 cm}\forall \bm x \in \partial \Omega _u \right \}
	\end{equation}
	\begin{equation}
	\delta \bm u (\bm x) \in V_u \mathrel{\mathop:}= \left \{ \delta \bm u |\delta \bm u(\bm x) = \bm 0 \hspace {0.25 cm}\forall \bm x \in \partial \Omega _u \right \}
	\end{equation} 

\noindent where $\bm x$ describes the position.

We establish the energy functional $L$ within the framework of a variational method extended from the Griffith's theory \citep{francfort1998revisiting}; this type of energy functional is commonly composed of the elastic energy $\Psi_{\varepsilon}(\bm \varepsilon)$, fracture energy $\Psi_f$ and external work $W_{ext}$, especially for a material without the fluid phase\citep{zhou2018phase, zhou2018phase3}. However, for the porous medium $\Omega$, the influence of fluid pressure $p$ must be involved in the energy functional $L$ \citep{mikelic2015quasi, mikelic2015phase, lee2016pressure, zhou2018phase2}: 
	\begin{equation}
	\Psi(\bm u,p,\Gamma) = \underbrace{\int_{\Omega}\psi_{\varepsilon}(\bm \varepsilon) \mathrm{d}{\Omega}}_{\Psi_{\varepsilon}}\underbrace{-\int_{\Omega}\alpha p \cdot (\nabla \cdot \bm u) \mathrm{d}{\Omega}}_{\text{pressure-related term}}+\underbrace{\int_{\Gamma}G_{ce} \mathrm{d}\Gamma}_{\Psi_f}\underbrace{-\int_{\Omega} \bm b\cdot{\bm u}\mathrm{d}{\Omega} - \int_{\partial\Omega_{t}} \bm {t^*}\cdot{\bm u}\mathrm{d}S}_{W_{ext}}
	\label{functional2}
	\end{equation}

\noindent where $\psi_{\varepsilon}(\bm \varepsilon)$ is the elastic energy density; $\alpha\in [\epsilon_p,1]$ is the Biot coefficient with $\epsilon_p$ being the porosity \citep{biot1962mechanics}; $G_{ce}$ is a comprehensive critical energy release rate for the transversely isotropic medium; $\bm b$ and $\bm t^*$ are the body force and traction on the Neumann boundary on $\mathrm{d}S$ for the displacement field, respectively. In addition, $\bm \varepsilon$ is the linear strain tensor defined as
	\begin{equation}
	\varepsilon_{ij}=\frac 1 2 \left(\frac{\partial u_i}{\partial x_j}+\frac{\partial u_j}{\partial x_i}\right)
	\end{equation}
	
\subsection{Phase field description}

\begin{figure}[htbp]
	\centering
	\includegraphics[height = 5cm]{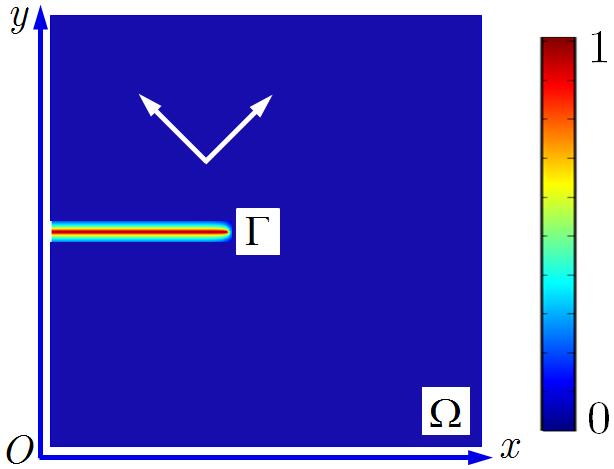}
	\caption{Phase field representation of the sharp fracture}
	\label{Phase field representation of the sharp fracture}
	\end{figure}

It can be seen from Eq. \eqref{functional2} that the total energy functional involves sharp fracture surface $\Gamma$, which makes the minimization of this functional rather difficult to apply. Therefore, a phase field $\phi(\bm x,t)\in[0,1]$ is used to diffuse the sharp fracture in the porous medium and facilitate numerical implementation; see also Fig. \ref{Phase field representation of the sharp fracture} for the phase field representation with a finite width. Specifically speaking, in this work $\phi=0$ and $\phi=1$ represent an intact and a fully broken state, respectively.

\begin{figure}[htbp]
	%\centering	
	\subfigure[sharp fracture topology]{\includegraphics[height = 5cm]{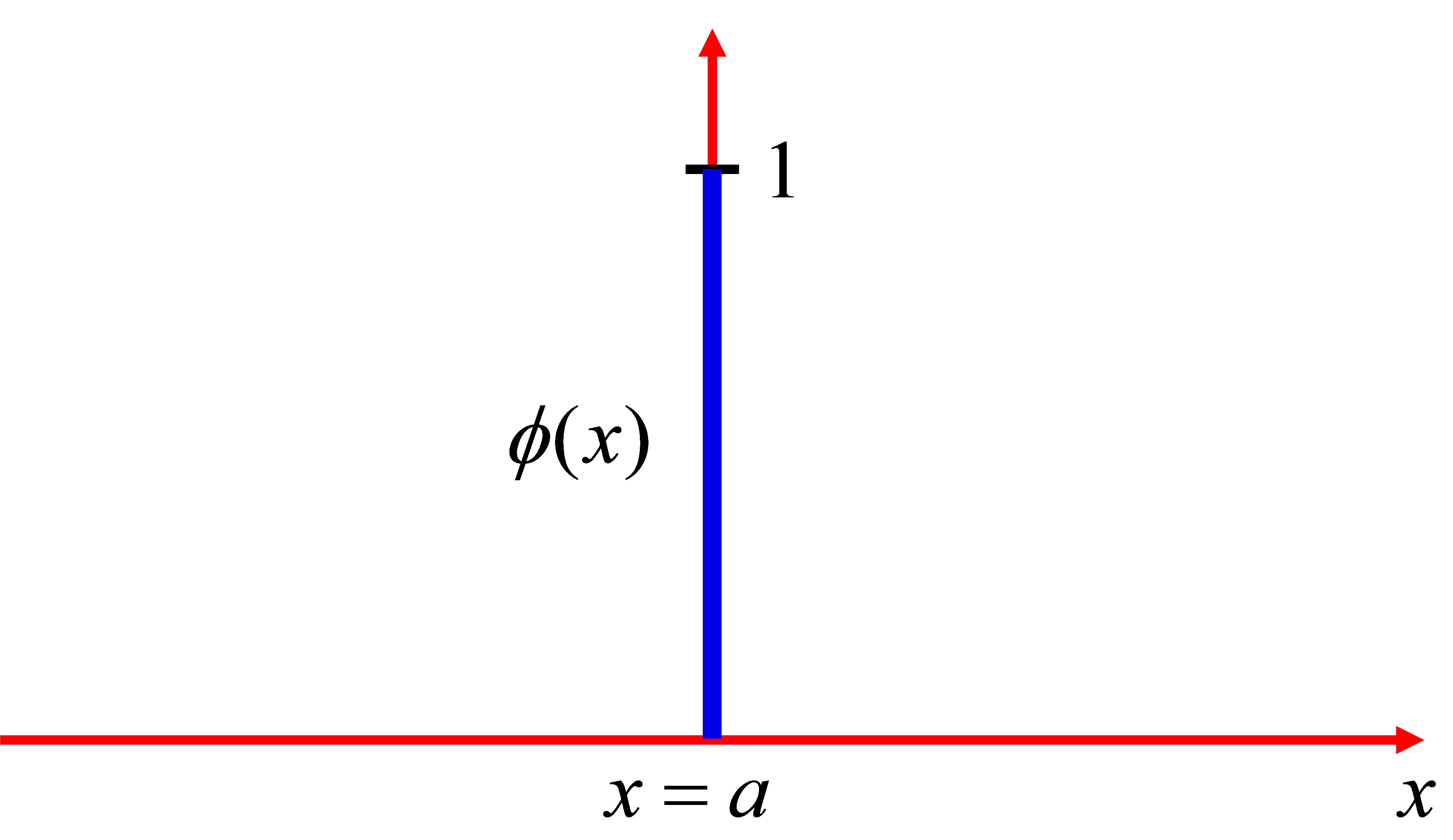}}
	\subfigure[smeared fracture topology]{\includegraphics[height = 5cm]{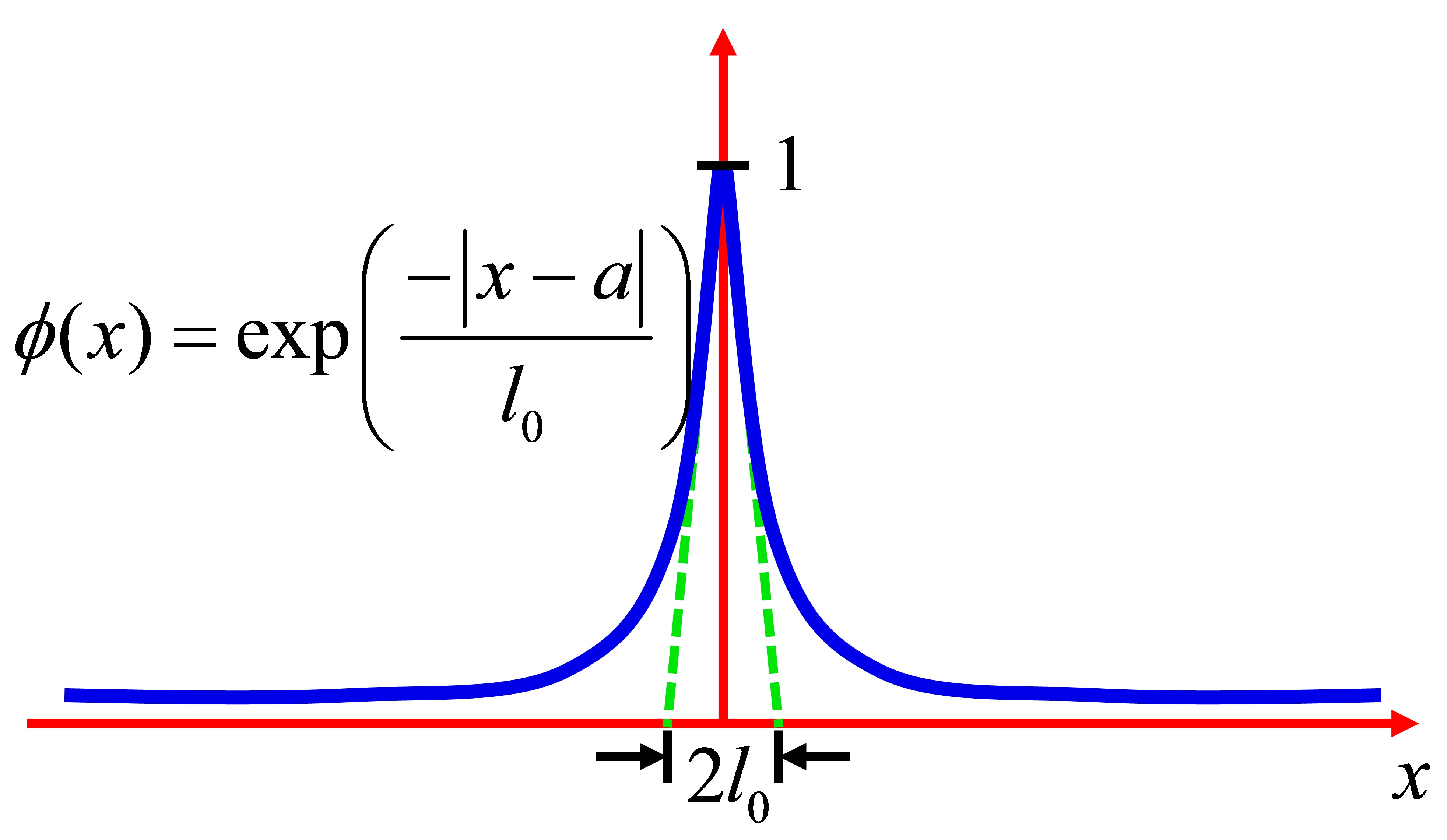}} \\
	\caption{Fracture topology in a 1D bar}
	\label{sharp fracture and smeared fracture topology}
\end{figure}

For a fracture in a 1D bar, its sharp and smeared topology are illustrated in Fig. \ref{sharp fracture and smeared fracture topology} where the diffused fracture is represented by an inverse exponential function $\phi(x)$ \citep{miehe2010phase}:
	\begin{equation}
	\phi(x) = \mathrm{exp}\left(-\frac{|x-a|}{l_0}\right)
	\end{equation}

\noindent where $x=a$ is the fracture location and $l_0$ denotes the intrinsic length scale parameter. The latter controls the width of the diffused fracture while many research \citep{borden2012phase, zhou2018phase, zhou2018phase2, zhou2018phase3, zhou2018adaptive} show that a solid has a larger fracture width and lower nominal tensile strength as the length scale $l_0$ increases. It should be also noted that the sharp fracture surface can be recovered when $l_0$ tends to zero, as is indicated by the $\Gamma$-convergence. In addition, the pore size of the porous medium $\Omega$ is assumed to be much smaller than the length scale parameter $l_0$ in this study.

For 2D and 3D problems in an isotropic medium, the crack surface density per unit volume is composed of the phase field and its gradient \citep{miehe2010phase}:
	\begin{equation}
	\gamma(\phi,\bigtriangledown\phi)=\frac{\phi^2}{2l_0}+\frac{l_0}2\frac{\partial\phi}{\partial x_i}\frac{\partial\phi}{\partial x_i}
	\label{phase field approximation}
	\end{equation}

The resulting fracture energy in the isotropic medium then can be expressed as \citep{miehe2010phase}  
\begin{equation}
\Psi_f = \int_{\Gamma}G_c \mathrm{d}\Gamma \approx \int_{\Omega}G_c\left(\frac{\phi^2}{2l_0}+\frac{l_0}2|\nabla\phi|^2 \right) \mathrm{d}\Omega
\label{psi_f}
\end{equation}

\noindent where $G_c$ is the isotropic energy release rate.

However, for the transversely isotropic medium, the crack surface density $\gamma(\phi,\bigtriangledown\phi)$ is modified according to \citet{bleyer2018phase} and the fracture energy $\Psi_f$ is expressed as
\begin{equation}
\Psi_f =  \int_{\Omega}G_{ce} \left(\frac{\phi^2}{2l_0}+\frac{l_0}2 \nabla\phi \cdot \bm B \cdot \nabla\phi \right) \mathrm{d}\Omega
\label{new psi_f}
\end{equation}

\noindent where $\bm B=\lambda_1 \bm e_1 \otimes \bm e_1+\lambda_2(\bm e_2 \otimes \bm e_2+\bm e_3 \otimes \bm e_3)$; $G_{ce}=\mathrm{min}(G_{c2},G_{c1})$ with $G_{c1}$ and $G_{c2}$ being the critical energy release rate along the rotational axis $\bm e_1$ and in the transverse isotropic plane defined by $\bm e_2$ and $\bm e_3$, respectively. In addition, $\lambda_1=G_{c1}/G_{ce}$, and $\lambda_2=G_{c2}/G_{ce}$. It should be noted that comparing with the energy release rate $G_{c1}$ used in \citet{bleyer2018phase}, Eq. \eqref{new psi_f} adopts a smaller fracture resistance, which is more consistent with the fact that a fracture initiates more easily if a solid has a lower energy release rate \citep{zhou2018phase, zhou2019phase}.

In a phase field method, the fracture evolution is prompted by the elastic energy where a degradation function $g(\phi)$ is inserted. Therefore, the elastic energy density $\psi_\varepsilon$ for the transversely isotropic medium is expressed as
\begin{equation}
\psi_\varepsilon =  \frac 1 2 g(\phi) \bm\varepsilon : \mathbb{C}_0 : \bm\varepsilon = g(\phi)\psi_0
\label{new psi_e}
\end{equation}

\noindent where $\mathbb{C}_0$ is the undegraded elasticity tensor. Although there are many forms for the degradation function, we use a quadratic form of $g(\phi)=(1-k)(1-\phi)^2+k$ with $k=10^{-9}$ being a rather small parameter to prevent numerical singularity when $\phi=0$. Therefore, the total energy functional \eqref{functional2} is rewritten as
\begin{multline}
\Psi=\int_{\Omega}\left[(1-k)(1-\phi)^2+k\right]\psi_0 d{\Omega}-\int_{\Omega}\alpha p \cdot (\nabla \cdot \bm u) d{\Omega}+\int_{\Omega}G_{ce}\left[\frac{\phi^2}{2l_0}+\frac{l_0}2\frac{\partial\phi}{\partial x_i}B_{ij}\frac{\partial\phi}{\partial x_j}\right]d{\Omega}-\\ \int_{\Omega} b_iu_id{\Omega}-\int_{\partial\Omega_{t}} t^*_iu_idS
\label{final functional}
\end{multline}

\subsection{Deformation of the transversely isotropic medium}

We employ the variational method to obtain the governing equations on medium deformation and thus fracture initiation and growth is considered as a process to minimize the total energy functional $\Psi$. Assuming $\delta 
\bm u$ and $\delta \phi$ are slight variations in the displacement $\bm u$ and phase field $\phi$, we naturally achieve the first variation of the energy functional $\Psi$ as
	\begin{multline}
	\delta \Psi=\underbrace{\int_{\partial \Omega_t}\left[(\sigma_{ij}-\alpha p\delta_{ij})n_j-t_i^* \right] \delta u_i d{S}}_{\textcircled{1}}
-\underbrace{\int_{\Omega}\left[(\sigma_{ij}-\alpha p \delta_{ij})_{,j}+b_i\right]\delta  u_i d{\Omega}}_{\textcircled{2}}
-\\ \underbrace{\int_{\Omega} \left[ 2(\phi-1)(1-k)\psi_{\varepsilon}^+ + \frac {G_{ce} \phi}{l_0}-G_{ce} l_0\frac{\partial}{\partial x_i}(B_{ij}\frac{\partial \phi}{\partial x_j}) \right]\delta\phi d{\Omega}}_{\textcircled{3}} + \underbrace{\int_{\partial\Omega}\left(B_{ij} \frac{\partial\phi}{\partial x_j}n_i\right)\delta\phi dS}_{\textcircled{4}}
	\label{first variation of the functional}
	\end{multline}

\noindent where the effective stress tensor $\bm \sigma(\bm\varepsilon)$ is given by:
	\begin{equation}
	\sigma_{ij}=\left [(1-k)(1-\phi)^2+k \right]\frac {\partial{\psi_0}}{\partial {\varepsilon_{ij}}}
	\end{equation}
	\begin{equation}
	\bm \sigma=\left [(1-k)(1-\phi)^2+k \right] \mathbb{C}_0: \bm\varepsilon
	\end{equation}

\noindent with $\bm I$ being the identity tensor.

The Cauchy stress tensor $\bm\sigma^{por}$ is then given by
	\begin{equation}
	\bm \sigma^{por}(\bm\varepsilon)=\bm \sigma(\bm\varepsilon)-\alpha p \bm I,\hspace{0.5cm} \mathrm{in} \hspace{0.1cm} \Omega
	\end{equation}

In the variational method, the first variation $\delta \Psi=0$ must hold for any admissible $\delta \bm u$, therefore Eq. \eqref{first variation of the functional}$\textcircled{2}$ gives the governing equation of strong form for the displacement filed as
	\begin{equation}
	\frac {\partial {\sigma_{ij}^{por}}}{\partial x_j}+b_i=0
	\end{equation}

Furthermore, Eq. \eqref{first variation of the functional}$\textcircled{1}$ yields the Neumann condition for the displacement field as
	\begin{equation}
	\sigma_{ij}^{por}n_j=t^*_i, \hspace{1cm} \mathrm{on}\hspace{0.5cm} \partial\Omega_{t}
	\label{boundary condition of the displacement field}
	\end{equation}

For the transversely isotropic medium, we use the Voigt matrix notation; therefore, the constitutive relation between the stress and strain is expressed first in the local material coordinate system as \citep{zeng2019study}
\begin{equation}
\left [(1-k)(1-\phi)^2+k \right] \left\{\begin{array}{c}
	\varepsilon_{11}^e\\\varepsilon_{22}^e\\\varepsilon_{33}^e\\\varepsilon_{23}^e\\\varepsilon_{13}^e\\\varepsilon_{12}^e
	\end{array}\right\}=\left[\begin{array}{cccccc}
	\frac 1 {E_1} & -\frac {\nu_{21}} {E_2} & -\frac {\nu_{21}} {E_2} & 0 &0 &0\\
	  & \frac {1} {E_2} & -\frac {\nu_{23}} {E_2} & 0 &0 &0\\
	  &  & \frac 1 {E_2} & 0 &0 &0\\
	  &  &  & \frac{1}{G_{23}} &0 &0\\
	  &  &  &  &\frac{1}{G_{12}} &0\\
	  & \mathrm{Symmetric} &  &  &  &\frac{1}{G_{12}}
	\end{array}\right]
	\left\{\begin{array}{c}
	\sigma_{11}^e\\\sigma_{22}^e\\\sigma_{33}^e\\\sigma_{23}^e\\\sigma_{13}^e\\\sigma_{12}^e
	\end{array}\right\}
	\label{Constitutive model 1}
\end{equation}

\noindent where $\bm\sigma^e=\left\{\sigma_{11}^e,\sigma_{22}^e,\sigma_{33}^e,\sigma_{23}^e,\sigma_{13}^e,\sigma_{12}^e\right\}^{\rm{T}}$ and $\bm\varepsilon^e=\left\{\varepsilon_{11}^e,\varepsilon_{22}^e,\varepsilon_{33}^e,\varepsilon_{23}^e,\varepsilon_{13}^e,\varepsilon^e_{12}\right\}^{\rm{T}}$ are the stress and strain in the local coordinate system; $E_i$ denotes the elastic modulus along three principal material directions ($\bm e_1$, $\bm e_2$, $\bm e_3$); $G_{ij}$ is the shear modulus in the orthogonal plane defined by $\bm e_i$ and $\bm e_j$; $\nu_{ij}=\nu_{ji} E_i/E_j $ is the Poisson’s ratio in the orthogonal plane, also denoting resulting strain in $\bm e_j$ direction due to loading in $\bm e_i$ direction.

The elasticity matrix $\bm S^e$ in the local material coordinate system can be then derived from Eq. \eqref{Constitutive model 1}:
\begin{equation}
\bm S^e=\left [(1-k)(1-\phi)^2+k \right] \left[\begin{array}{cccccc}
Q_{11} & Q_{12} & Q_{12} & 0 & 0 & 0\\
Q_{12} & Q_{22} & Q_{23} & 0 & 0 & 0\\
Q_{12} & Q_{23} & Q_{22} & 0 & 0 & 0\\
0 & 0 & 0& Q_{44} & 0 & 0 \\
0 & 0 & 0& 0 & Q_{66} & 0 \\
0 & 0 & 0& 0 & 0& Q_{66}\\
\end{array}\right]
\end{equation}

\noindent where

\begin{equation}
Q_{11}=\frac{1-\nu_{23}\nu_{32}}{E_2 E_3 \Lambda}
\end{equation}
\begin{equation}
Q_{12}=\frac{\nu_{12}+\nu_{32}\nu_{13}}{E_1 E_3 \Lambda}
\end{equation}
\begin{equation}
Q_{22}=\frac{1-\nu_{31}\nu_{13}}{E_1 E_3 \Lambda}
\end{equation}
\begin{equation}
Q_{23}=\frac{\nu_{23}+\nu_{21}\nu_{13}}{E_1 E_2 \Lambda}
\end{equation}
\begin{equation}
Q_{44}=G_{23}=\frac{Q_{22}-Q_{23}}{2}
\end{equation}
\begin{equation}
Q_{66}=G_{12}
\end{equation}

\noindent with
\begin{equation}
\Lambda = \frac{1-\nu_{12}\nu_{21}-\nu_{23}\nu_{32}-\nu_{13}\nu_{31}-2\nu_{12}\nu_{23}\nu_{31}}{E_1 E_2 E_3} 
\end{equation}

In these equations, $E_3$=$E_2$ while the different Poisson's ratios can be calculated through their definitions. However, in the global coordinate system $(x,y,z)$, the elasticity matrix $\bm S$ must be determined from $\bm S^e$ by using a coordinate frame transformation. Thus, the following global elasticity matrix $\bm S$ is used:
\begin{equation}
\bm S = \bm T \bm S^e \bm T^{\rm T}
\end{equation}

\noindent where $\bm T$ is the transformation matrix defined by 
\begin{equation}
\bm T = \left[\begin{array}{cccccc}
l_1^2 & l_2^2 & l_3^2& 2l_2l_3 & 2l_3l_1 & 2l_1l_2\\
m_1^2 & m_2^2 & m_3^2& 2m_2m_3 & 2m_3m_1 & 2m_1m_2\\
n_1^2 & n_2^2 & n_3^2& 2n_2n_3 & 2n_3n_1 & 2n_1n_2\\
m_1n_1 & m_2n_2 & m_3n_3& m_2n_3 +m_3n_2 & m_3n_1+m_1n_3& m_1n_2+m_2n_1\\
n_1l_1 & n_2l_2 & n_3l_3& n_2l_3 +n_3l_2 & n_3l_1+n_1l_3& n_1l_2+n_2l_1\\
l_1m_1 & l_2m_2 & l_3m_3& l_2m_3 +l_3m_2 & l_3m_1+l_1m_3& l_1m_2+l_2m_1
\end{array}
\right]
\end{equation}

\noindent in which $l_i=\mathrm{cos}(\bm e_i,x)$, $m_i=\mathrm{cos} (\bm e_i,y)$, and $n_i=\mathrm{cos}(\bm e_i,z)$ ($i=1,2,3$).

\subsection{Fracture evolution}

In the phase field modeling, extra fracture criterion is not required and the growth of a fracture is only governed by an evolution equation. Therefore, due to the fact that Eq. \eqref{first variation of the functional} must always hold for all possible $\delta \phi$, Eq. \eqref{first variation of the functional}$\textcircled{3}$ produces the initial form for the fracture evolution:
	\begin{equation}	  
	 \left[\frac{2l_0(1-k)\psi_0}{G_{ce}}+1\right]\phi-l_0^2\frac{\partial}{\partial x_i}\left(B_{ij}\frac{\partial \phi}{\partial {x_j}}\right)=\frac{2l_0(1-k)\psi_0}{G_{ce}}
	\label{governing equations 2}
	\end{equation}

However, Eq. \eqref{governing equations 2} cannot distinguish different fracture mechanisms and some unrealistic fracture patterns may be achieved. Therefore, many researchers have attempted to decompose the elastic energy into different parts to reflect those experiment-supported fracture modes \citep{miehe2010phase, miehe2010thermodynamically, amor2009regularized, bleyer2018phase}. For example, \citet{miehe2010phase, miehe2010thermodynamically} decomposed the elastic energy into compressive and positive parts; \citet{amor2009regularized} used the positive volumetric strain and deviatoric strain induced elastic energy to drive the fracture growth. In this paper, we follow the decomposition of \citet{bleyer2018phase} and set the positive elastic energy as
\begin{equation}
\psi^+=\frac 1 2 (\bm\varepsilon_+^e)^{\rm T} \bm S^e \bm\varepsilon_+^e
\end{equation}

\noindent where 
\begin{equation}
\bm\varepsilon_+^e=\left\{
\langle \varepsilon_{11}^e \rangle_+\quad 
\langle \varepsilon_{22}^e \rangle_+\quad
\langle \varepsilon_{33}^e \rangle_+\quad
\varepsilon_{23}^e \quad \varepsilon_{13}^e \quad \varepsilon_{12}^e\right\}^{\rm T}
\end{equation}

\noindent and $\langle \star \rangle_+=\left(\star+|\star|\right)/2$ calculates the positive part of $\star$.

The phase field modeling also requires the irreversibility condition $\Gamma(\bm x,s)\in\Gamma(\bm x,t)(s<t)$; therefore, we establish a history energy reference field $H(\bm x,t)$ to record the maximum positive elastic energy density in the time interval $[0,t]$:
	\begin{equation}
	H(\bm x,t) = \max \limits_{s\in[0,t]}\psi^+(\bm x,s)
	\end{equation}

By using the history field $H$, it can be ensured that the phase field increases monotonically during compression or unloading. Furthermore, substituting $\psi^+$  by  $H(\bm x,t)$  into Eq. \eqref{governing equations 2}, the strong form of evolution equation is now rewritten as
\begin{equation}	  
\left[\frac{2l_0(1-k)H}{G_{ce}}+1\right]\phi-l_0^2\frac{\partial}{\partial x_i}\left(B_{ij}\frac{\partial \phi}{\partial {x_j}}\right)=\frac{2l_0(1-k)H}{G_{ce}}
\end{equation}

In addition, Eq. \eqref{first variation of the functional} $\textcircled{4}$ yields the Neumann condition for the  phase field,
	\begin{equation}
	B_{ij} \frac{\partial\phi}{\partial x_j}n_i=0
	\end{equation}

\subsection{Fluid flow in the transversely isotropic medium}

In this study, the fluid in the transversely isotropic medium is assumed to be compressible and viscous, and then for the fluid flow the calculation domain is decomposed into three parts: the unbroken domain (reservoir domain) $\Omega_r(t)$, fracture domain $\Omega_f(t)$ and transition domain $\Omega_t(t)$ \citep{zhou2018phase2}. Two phase field values $c_1$ and $c_2$ are used to define these subdomains as illustrated in Table \ref{Subdomain definition}.

\begin{table}[htbp]
	\caption{Subdomain definition}
	\label{Subdomain definition}
	\centering
	\begin{tabular}{ll}
		\toprule[1pt]
		Subdomain&Phase field\\
		Reservoir domain & $\phi\le c_1$\\
		Transition domain & $c_1<\phi<c_2$\\
		Fracture domain & $\phi\ge c_2$\\
		\bottomrule[1pt] 
	\end{tabular}
\end{table}

In the transition domain, the hydraulic parameters are assumed to be linearly interpolated from those in the reservoir and fracture domains. Therefore, these two indicator functions $\chi_r$ and $\chi_f$ \citep{lee2016pressure} are established:
	\begin{equation}
	\chi_r(\cdot,\phi)=\left\{
		\begin{aligned}
		&1,\hspace{2 cm}&\phi\le c_1\\ &\frac{c_2-\phi}{c_2-c_1} &c_1<\phi<c_2
\\&0,&\phi\ge c_2
		\end{aligned}\right.
	\label{function1}
	\end{equation} 
	\begin{equation}
	\chi_f(\cdot,\phi)=\left\{
		\begin{aligned}
		&0,\hspace{2 cm}&\phi\le c_1\\ &\frac{\phi-c_1}{c_2-c_1} &c_1<\phi<c_2
\\&1,&\phi\ge c_2
		\end{aligned}\right.
	\label{function2}
	\end{equation}

We follow our previous study \citep{zhou2018phase2} and use the Darcy's law to characterize the fluid flow in the whole calculation domain for the transversely isotropic medium: 
	\begin{equation}
	\rho S \frac{\partial p}{\partial t}+\nabla\cdot(\rho\bm v)=q_m-\rho\alpha\chi_r\frac{\partial \varepsilon_{vol}}{\partial t}
	\label{mass conservation of the whole domain}
	\end{equation}

\noindent where $\rho$, $S$, $\bm v$, $\varepsilon_{vol}=\nabla\cdot\bm u$, and $q_m$ represent the fluid density, storage coefficient, flow velocity, volumetric strain, and fluid source term, respectively. $\rho=\rho_r\chi_r+\rho_f\chi_f$ with $\rho_r$ and $\rho_f$ the fluid densities in the reservoir and fracture domains. Similarly, $\alpha=\alpha_r\chi_r+\alpha_f\chi_f$. The Biot coefficient $\alpha=1$ is set for the fracture domain; therefore, $\alpha=\alpha_{r}\chi_r+\chi_f$ with $\alpha_{r}$ being the Biot coefficient of the reservoir domain.

In addition, the storage coefficient $S$ is established in terms of the fluid property and Biot coefficient  \citep{zhou2018phase2}:
\begin{equation}
S=\varepsilon_pc+\frac{(\alpha-\varepsilon_p)(1-\alpha)}{K_{Vr}}
\end{equation}

\noindent where $\varepsilon_p$, $c$, and $K_{Vr}$ are the porosity, fluid compressibility, and bulk modulus of the calculation domain, respectively. Given that $c_r$ and $c_f$ are the fluid compressibility in the reservoir and fracture domains, we have $c=c_r\chi_r+c_f\chi_f$. In addition, because the fracture domain is fully damaged, we set $\varepsilon_p=1$ for this domain and then $\varepsilon_p=\varepsilon_{pr}\chi_r+\chi_f$ with $\varepsilon_{pr}$ the porosity of the reservoir domain.

For the transverse isotropic porous medium, the Darcy's velocity $\bm v$ is defined by
	\begin{equation}
	\bm v=-\frac{\bm K}{\mu}(\nabla p+\rho\bm g)
	\label{velocity of the whole domain}
	\end{equation} 

\noindent where $\bm g$ and $\mu$ denote the gravity and fluid viscosity. $\mu=\mu_{r}\chi_r+\mu_f \chi_f$ in which $\mu_{r}$ and $\mu_{f}$ are the fluid viscosity in the reservoir and fracture domains, respectively. In addition, $\bm K$ is the effective permeability tensor. Given that the principal direction $\bm e_3$ coincides with the $z-$axis, if the rotation angle between the principal direction $\bm e_1$ and $x-$axis is set as $\beta$, the permeability $\bm K$ can be expressed in a matrix form as \citep{zeng2019study}
\begin{equation}
\bm K = \left(\begin{array}{ccc}
k_{xx} & k_{xy} & k_{xz}\\
k_{xy} & k_{yy} & k_{yz}\\
k_{xz} & k_{yz} & k_{zz}\\
\end{array}\right)=\left(\begin{array}{ccc}
k_1\cos ^2\beta+k_2\sin^2\beta&(k_1-k_2)\sin\beta\cos\beta&0\\
(k_1-k_2)\sin\beta\cos\beta&k_1\sin ^2\beta+k_2\cos^2\beta&0\\
0&0&k_2
\end{array}\right)
\end{equation}

\noindent where $k_1=k_{r1}\chi_r+k_{f1}\chi_f$ with $k_{r1}$ and $k_{f1}$ being the permeability along the material direction $\bm e_1$ for the reservoir and fracture domains; $k_2=k_{r2}\chi_r+k_{f2}\chi_f$ with $k_{r2}$ and $k_{f2}$ being the permeability in the transverse plane for the reservoir and fracture domains.

Finally, the governing equation for fluid flow in the transversely isotropic medium $\Omega$ is expressed in terms of the fluid pressure $p$:
	\begin{equation}
	\rho S \frac{\partial p}{\partial t}-\nabla\cdot \frac{\rho \bm K}{\mu}(\nabla p+\rho\bm g)=q_m-\rho\alpha\chi_r\frac{\partial \varepsilon_{vol}}{\partial t}
	\label{governing equation of the whole domain}
	\end{equation}

It should be noted that determination of the coefficients in the PFM for hydraulic fracture propagation in the transversely isotropic porous medium will be tackled in future research. Although the permeability $k_{1f}$ and $k_{2f}$ in the fracture domain can be coupled to the volumetric strain \citep{Zhuang2017} due to the damage-based nature of the PFM, we still use fixed values for $k_{1f}$ and $k_{2f}$ for simplicity and eliminating the influence of extra coupling terms on calculation convergence. This treatment is also proved to produce favorable fracture patterns \citep{zhou2018phase2}.

Particularly, when $\phi\le c_1$, Eq. \eqref{governing equation of the whole domain} yields the mass conservation in the reservoir domain as	
	\begin{equation}
	\rho_r S_r \frac{\partial p}{\partial t}+\nabla\cdot(\rho_r\bm v_r)=q_r-\rho_r\alpha_r\frac{\partial \varepsilon_{vol}}{\partial t}
	\label{governing equation of the reservoir domain}
	\end{equation}
	  
\noindent and then the storage coefficient $S$ can degenerate into a classical storage model proposed by \citet{biot1962mechanics}: 
 	\begin{equation}
	S_r=\varepsilon_{pr}c_r+\frac{(\alpha_r-\varepsilon_{pr})(1-\alpha_r)}{K_{Vr}}
	\label{S_r}
	\end{equation}

On the other hand, when $\phi\ge c_2$, the volumetric strain vanishes in the governing equation, and Eq. \eqref{governing equation of the whole domain} yields the mass conservation in the fracture domain as
	\begin{equation}
	\rho_f S_f \frac{\partial p}{\partial t}+\nabla\cdot(\rho_f\bm v_f)=q_f
	\label{governing equation of the fracture domain}
	\end{equation} 
\noindent where $S_f$ and $q_f$ represent the storage coefficient and source term of the fracture domain.

In addition to the mass conservation equation, describing the fluid flow also requires the Dirichlet condition on $\partial\Omega_D$ and Neumann condition on $\partial\Omega_N$ with $\partial\Omega_D\cap \partial\Omega_N =\emptyset$ \citep{zhou2018phase2}:
\begin{equation}
p=p_D \hspace{2cm}\mathrm{on}\quad\partial\Omega_D
\end{equation}
\begin{equation}
-\bm n \cdot \rho\bm v=M_N  \hspace{2cm}\mathrm{on}\quad\partial\Omega_N
\end{equation}

\noindent where $p_D$ and $M_N$ are the prescribed pressure and mass flux.

\section {Numerical algorithm}\label {Numerical algorithm}

In this section, we describe the numerical algorithm on implementation of the phase field modeling of fracture propagation in a transversely isotropic medium. We use the finite element method to discretize the spatial domain and apply a finite difference method for the time domain discretization. Finally, we use a staggered scheme to solve the coupling equations of multi-field problem and for pursuing a higher convergence rate.

\subsection{Finite element discretization}
We first derive the weak form of all the governing equations as
	\begin{equation}
	\int_{\Omega}\left[-(\bm\sigma-\alpha p\bm I):\delta \bm {\varepsilon}\right] \mathrm{d}\Omega +\int_{\Omega}\bm b \cdot \delta \bm u \mathrm{d}\Omega +\int_{\Omega_{t}}\bm {t^*} \cdot \delta \bm u \mathrm{d}S=0
	\label{weak form 1}
	\end{equation}
	\begin{equation}
	\int_{\Omega}-2(1-k)H(1-\phi)\delta\phi\mathrm{d}\Omega+\int_{\Omega}G_{ce}\left(l_0\bm B\nabla\phi\cdot\nabla\delta\phi+\frac{1}{l_0}\phi\delta\phi\right)\mathrm{d}\Omega=0
	\label{weak form 2}
	\end{equation}
	\begin{equation}
	\int_{\Omega} \rho S \frac{\partial p}{\partial t}\delta p\mathrm{d}\Omega-\int_{\Omega} \rho \bm v \cdot\nabla\delta p \mathrm{d}\Omega=\int_{\partial\Omega}M_n \delta p \mathrm{d}S+\int_{\Omega}\left(q_m-\rho\alpha\chi_r\frac{\partial \varepsilon_{vol}}{\partial t}\right) \delta p \mathrm{d}\Omega
	\label{weak form 3}
	\end{equation}  

For the three fields $\bm u$, $\phi$, and $p$, we define the nodal values as $\bm u_i$, $\phi_i$, and $p_i$ ($i=1,2,\cdots,n$ with $n$ the node number), and under standard vector-matrix notation system, the field discretization is expressed as
	\begin{equation}
	\bm u = \bm N_u \bm d,\hspace{0.5cm} \phi = \bm N_{\phi} \hat{\bm\phi}, \hspace{0.5cm} p = \bm N_{p} \hat{\bm p}
	\end{equation}
\noindent where $\bm d$, $\hat{\bm\phi}$, and $\hat{\bm p}$ represent the vectors composed of node values $\bm u_i$, $\phi_i$ and $p_i$. In addition, $\bm N_u$, $\bm N_{\phi}$, and  $\bm N_{p}$ are the corresponding shape function matrices for the displacement, phase field, and fluid pressure defined by
	\begin{equation}		
			\bm N_u = \left[ \begin{array}{ccccccc}
			N_{1}&0&0&\dots&N_{n}&0&0\\
			0&N_{1}&0&\dots&0&N_{n}&0\\
			0&0&N_{1}&\dots&0&0&N_{n}\\
			\end{array}\right], \hspace{0.5cm}
			\bm N_\phi = \bm N_p=\left[ \begin{array}{cccc}
			N_{1}&N_{2}&\dots&N_{n}
			\end{array}\right]
	\end{equation}

\noindent where $N_i$ is the shape function of node $i$. 

The test functions have the same discretization:
\begin{equation}
	\delta \bm u = \bm N_u \delta \bm d,\hspace{0.5cm} \delta \phi = \bm N_{\phi} \delta \hat{\bm\phi},\hspace{0.5cm} \delta p = \bm N_{p} \delta \hat{\bm p}
	\end{equation}

\noindent where $\delta \bm d$, $\delta \hat{\bm\phi}$ and $\delta \hat{\bm p}$ are the vectors composing of node values of the test functions.

We then define the gradients of the trial and test functions as 
	\begin{equation}
	\bm \varepsilon =  \bm B_u \bm d,\hspace{0.5cm} \nabla\phi = \bm B_\phi \hat{\bm\phi}, \hspace{0.5cm} \nabla p = \bm B_p \hat{\bm p}
	\end{equation}
	\begin{equation}
	 \delta \bm \varepsilon =  \bm B_u \delta \bm d,\hspace{0.5cm} \nabla\delta\phi = \bm B_\phi \delta \hat{\bm\phi},\hspace{0.5cm} \nabla \delta p = \bm B_p \delta \hat{\bm  p}
	\end{equation}

\noindent where $\bm B_u$, $\bm B_\phi$ and  $\bm B_p$ are the derivatives of the shape functions:
	\begin{equation}
	\bm B_u=\left[
		\begin{array}{ccccccc}
		N_{1,x}&0&0&\dots&N_{n,x}&0&0\\
		0&N_{1,y}&0&\dots&0&N_{n,y}&0\\
		0&0&N_{1,z}&\dots&0&0&N_{n,z}\\
		0&N_{1,z}&N_{1,y}&\dots&0&N_{n,z}&N_{n,y}\\
		N_{1,z}&0&N_{1,x}&\dots&N_{n,z}&0&N_{n,x}\\
		N_{1,y}&N_{1,x}&0&\dots&N_{n,y}&N_{n,x}&0\\
		\end{array}\right],\hspace{0.2cm}
		\bm B_\phi = \bm B_p=\left[
		\begin{array}{cccc}
		N_{1,x}&N_{2,x}&\dots&N_{n,x}\\
		N_{1,y}&N_{2,y}&\dots&N_{n,y}\\
		N_{1,z}&N_{2,z}&\dots&N_{n,z}\\
		\end{array}\right]
		\label{BiBu}
	\end{equation}

It should be noted that Eqs. \eqref{weak form 1} to \eqref{weak form 3} must hold for any admissible test functions, therefore the discrete equations of the three fields are derived as follows,
	\begin{equation}
	\bm R^u = \bm K_u \bm d - \bm f_u^{ext} = \bm 0
	\label{Discrete equation 1}
	\end{equation}
	\begin{equation}
	\bm R^\phi = \bm K_\phi \hat{\bm \phi} - \bm f_\phi^{ext} = \bm 0
	\label{Discrete equation 2}
	\end{equation}
	\begin{equation}
	\bm R^p = \bm S_p \frac{\mathrm{d} {\hat{\bm p}}}{\mathrm {d} t} + \bm H_p \hat{\bm p} - \bm q_p^{ext} = \bm 0
	\label{Discrete equation 3}
	\end{equation}

\noindent where $\bm R^u$, $\bm R^\phi$, and $\bm R^p$ are the residuals of the three fields. In addition, $\bm K_u$ and $\bm K_\phi$ are the stiffness matrices of the displacement and phase fields; $\bm S_p$ is the compressibility matrix, and $\bm H_p$ is the permeability matrix. Specifically, the matrices of $\bm K_u$, $\bm K_\phi$, $\bm S_p$, and $\bm H_p$ are defined by
	\begin{equation}
	\bm K_u = \int_{\Omega}[\bm B_u]^{\mathrm T}\bm S \bm B_u \mathrm{d}\Omega
	\end{equation}
	\begin{equation}
	\bm K_\phi = \int_{\Omega}\left\{\bm B_\phi^{\mathrm{T}} G_{ce} l_0 \bm B\bm B_\phi +\bm N_\phi^{\mathrm{T}} \left [ \frac{G_{ce}}{l_0} + 2(1-k)H \right ] \bm N_\phi \right \} \mathrm{d}\Omega
	\end{equation}
	\begin{equation}
	\bm S_p = \int_{\Omega} \bm N_p^{\mathrm T} \rho S \bm N_p \mathrm{d}\Omega
	\end{equation}
	\begin{equation}
	\bm H_p = \int_{\Omega}[\bm B_p]^{\mathrm T}\frac{\rho \bm K}{\mu} \bm B_p \mathrm{d}\Omega
	\end{equation}

Moreover, $\bm f_u^{ext}$ and $\bm f_\phi^{ext}$ are the external force vectors of the displacement and phase fields, and $\bm q_p^{ext}$ is the external flux vector of the flow field; these matrices are defined by
	\begin{equation}
	\bm f_u^{ext}=  \int_{\Omega} \bm N_u^{\mathrm T}\bm b \mathrm{d}\Omega+ \int_{\Omega_{t}}  \bm N_u^{\mathrm T}\bm {t^*} \mathrm{d}S+ \int_{\Omega}[\bm B_{u}]^{\mathrm T}\alpha p \bm I \mathrm{d}\Omega	
	\end{equation}
	\begin{equation}
	\bm f_\phi^{ext}=  \int_{\Omega}2(1-k)H\bm N_\phi^{\mathrm{T}} \mathrm{d}\Omega	
	\end{equation}
	\begin{equation}
	\bm q_p^{ext} =  \int_{\Omega} \bm N_p^{\mathrm T} \left( q_m-\rho\alpha\chi_r\frac{\partial \varepsilon_{vol}}{\partial t} \right) \mathrm{d}\Omega+ \int_{\partial\Omega_{N}}\bm N_p^{\mathrm T} M_N \mathrm{d}S	
	\end{equation}

\subsection{Staggered scheme}\label{Staggered scheme}

We implement the aforementioned numerical algorithm in the commercial software, COMSOL Multiphysics due to its high capability in dealing with multiple field problems. The powerful suitability and practicability of this software for phase field modeling have also been verified in \citet{zhou2018phase, zhou2018phase2, zhou2018phase3, zhou2018adaptive, zhou2018propagation}, where the readers can acquire more detail on the COMSOL implementations. In these numerical schemes, time domain discretization is coupled to an implicit Generalized-$\alpha$ method \citep{borden2012phase}, which ensures unconditional numerical stability for time integration. On the other hand, a staggered scheme \citep{miehe2010phase} is used to solve the coupled discrete equations of the three fields (Eqs. \eqref{Discrete equation 1}, \eqref{Discrete equation 2}, and \eqref{Discrete equation 3}). 

Three staggered steps, which correspond to the fields $(\bm u,p)$, $H$, and $\phi$, are set in the staggered scheme. Note that the displacement $\bm u$ and pressure $p$ are monolithically solved in a staggered step under fixed $H$ and $\phi$ for improving the convergence efficiency. Therefore, the staggered scheme solves all the three staggered steps sequentially and independently in a fixed time interval. These implementation detail are also depicted in Table \ref{Solution flow-chart for the fracture propagation in the transversely isotropic medium} where a solution flow-chart in the staggered scheme is shown for modeling hydraulic fracture propagation in the transversely isotropic medium. Due to the strong nonlinearity in the displacement equations, the Newton-Raphson iterations are required with a maximum number of 150 set in the simulations. Furthermore, the stabilization function of COMSOL is switched on and the Anderson acceleration technique \citep{zhou2018phase2} is applied to enrich the numerical stability. The COMSOL codes can be downloaded in this website: ``https://sourceforge.net/projects/phasefieldmodelingcomsol/".

	\begin{table}[htbp]
	\caption{Solution flow-chart for fracture propagation modeling in the transversely isotropic medium}
	\label{Solution flow-chart for the fracture propagation in the transversely isotropic medium}
	\centering	
	\begin{tabular}{l}
	\toprule[1pt]
	%\hline
	\textbf{Time step}\\
	\hspace{1cm} $t=t_0,t_1,\cdots,t_{n-1},t_{n},\cdots$\\
	\textbf{Initiation}\\
	\hspace{1cm} $\bm u_0$, $\phi_0$, $H_0$, $p_0$ (for $t=t_0$)\\
	\textbf{After obtaining} $\bm u_{n-1}$, $\phi_{n-1}$, $H_{n-1}$, $p_{n-1}$ (for $t=t_{n-1}$)\\
	\textbf{Go to}\\
	\hspace{1cm} $t=t_{n}$\\
	\hspace{1cm}\textbf{Initiation}\\
			\hspace{2cm} Construct the initial guess $(\bm u, p)_{n}^{j=0}$, $H_{n}^{j=0}$, and $\phi_{n}^{j=0}$ by using $(\bm u, p)_{n-1}$, $H_{n-1}$, and $\phi_{n-1}$\\
	\hspace{1cm}\textbf{For} every successive iteration step $j$ \textbf{do}\\
			\hspace{2cm} Step 1. \\
			\hspace{3cm} Input $(\bm u, p)_{n}^{j-1}$ and $\phi_{n}^{j-1}$\\
			\hspace{3cm} Solve Eqs. \eqref{Discrete equation 1} and \eqref{Discrete equation 3}\\
			\hspace{3cm} Output $(\bm u, p)_{n}^{j}$\\
			\hspace{2cm} Step 2. \\
			\hspace{3cm} Input $(\bm u, p)_{n}^{j}$ and $H_{n}^{j-1}$\\
			\hspace{3cm} Output $H_{n}^{j}$\\
			\hspace{2cm} Step 3. \\
			\hspace{3cm} Input $H_{n}^{j}$ and $\phi_{n}^{j-1}$\\
			\hspace{3cm} Solve Eq. \eqref{Discrete equation 2}\\
			\hspace{3cm} Output $\phi_{n}^{j}$\\
			\hspace{2cm} Step 4. \\
			\hspace{3cm} Evaluate the global relative error of $(\bm u, p, H, \phi)$\\
			\hspace{3cm} \textbf{If} ERROR$\le \varepsilon_t=1\times10^{-4}$\\
			\hspace{4cm} $(\bm u, p, H, \phi)_n$ = $(\bm u, p, H, \phi)_n^j$, go to next time step $t=t_{n+1}$\\
			\hspace{3cm} \textbf{If} ERROR$> \varepsilon_t=1\times10^{-4}$\\
			\hspace{4cm}$j=j+1$, go to Step 1, continue the iteration\\
	\bottomrule[1pt]
	\end{tabular}
	\end{table}

\section{Validation of the PFM}\label{Validation of the PFM}

The following section presents two numerical examples for initially verifying our phase field model for transversely isotropic medium. The first example is a single-edge-notched transversely isotropic square plate under tension; however, the fluid pressure field is not considered in this example. The second example is an isotropic elastic porous medium subjected to internal fluid pressure.

\subsection{Single-edge-notched plate under tension}

\begin{figure}[htbp]
	\centering
	\includegraphics[width = 7.5cm]{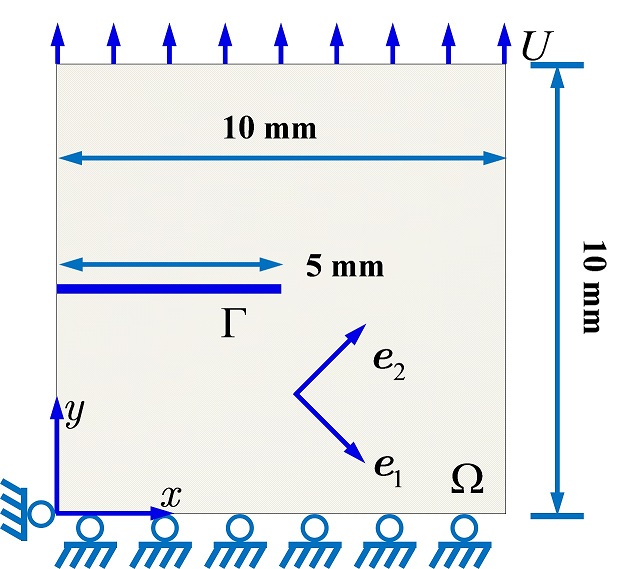}
	\caption{Geometry and boundary conditions for the single-edge-notched square plate}
	\label{Geometry and boundary conditions for the single-edge-notched square plate}
\end{figure}

We first investigate a pre-notched square plate subjected to static tension loading. The geometry and boundary condition of this example is shown in Fig. \ref{Geometry and boundary conditions for the single-edge-notched square plate}. The pre-existing notch has a length of 5 mm, and this example has been also investigated by \citet{teichtmeister2017phase}. The displacement on the bottom edge of the plate is fixed in the normal direction while the left bottom corner is fully constrained. In addition, a vertical displacement $U$ is applied on the upper edge of the plate. The material parameters for calculation are listed in Table \ref{Parameters for calculation1} where the length scale $l_0=0.075$ mm and the stability parameter $k=1\times10^{-9}$. Note that we use the parameters in Table \ref{Parameters for calculation1} to reflect material anisotropy and match the elastic free energy used in \citet{teichtmeister2017phase}.

\begin{table}[htbp]
	\small
	\caption{Parameters for the single-edge-notched square plate}
	\label{Parameters for calculation1}
	\centering
	\begin{tabular}{llllllll}%{p{10pt}p{50pt}p{10pt}p{50pt}p{10pt}p{50pt}p{20pt}p{100pt}}
		\hline
		$Q_{11}$ & 140 GPa & $Q_{22}$ & 139 GPa& $Q_{12},Q_{23}$& 77.8 GPa& $Q_{44}$& 30.6 GPa\\
		$Q_{66}$ & 30.6 GPa & $G_{c2}$ & 20 N/m& $G_{c1}$ & 40, 120, 1020 N/m& &\\
		\hline
		\normalsize
	\end{tabular}
\end{table}

The square is discretized by four node quadrilateral elements with their maximum size around $h=0.0375$ mm. In addition, we apply a displacement increment $\Delta u = 1\times 10^{-5}$ mm for each time step. The inclination angle between $\bm e_1$ and $x-$axis $\beta$ is set as $45^\circ$ in this example and three $G_{c1}$ (40, 120, 1020 N/m) are investigated while $G_{c2}=20$ N/m is fixed. 

\begin{figure}[htbp]
	\centering
	\subfigure[$\lambda_c=1$, $U=5.4\times10^{-3}$ mm]{\includegraphics[height = 5.5cm]{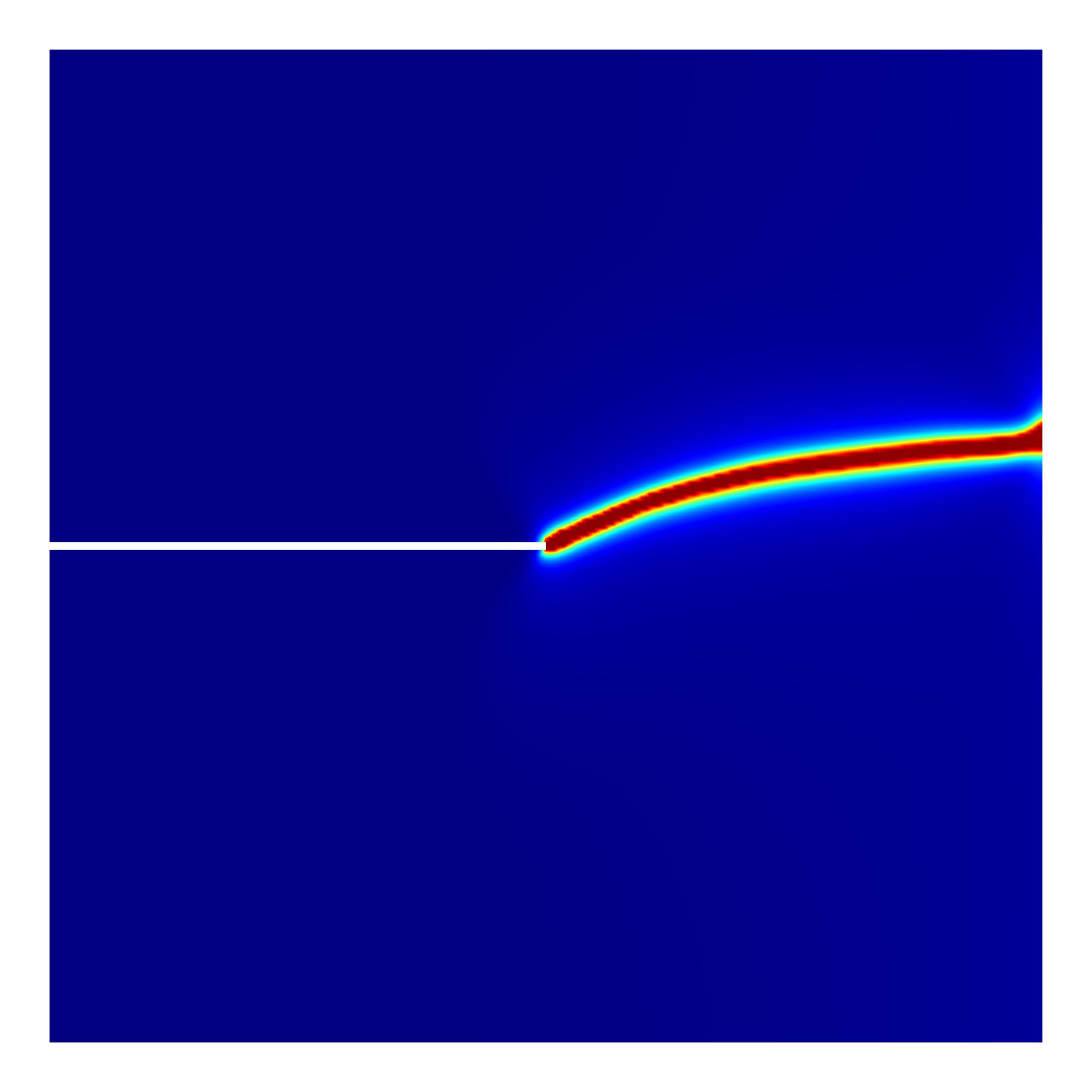}}
	\subfigure[$\lambda_c=5$, $U=9.4\times10^{-3}$ mm]{\includegraphics[height = 5.5cm]{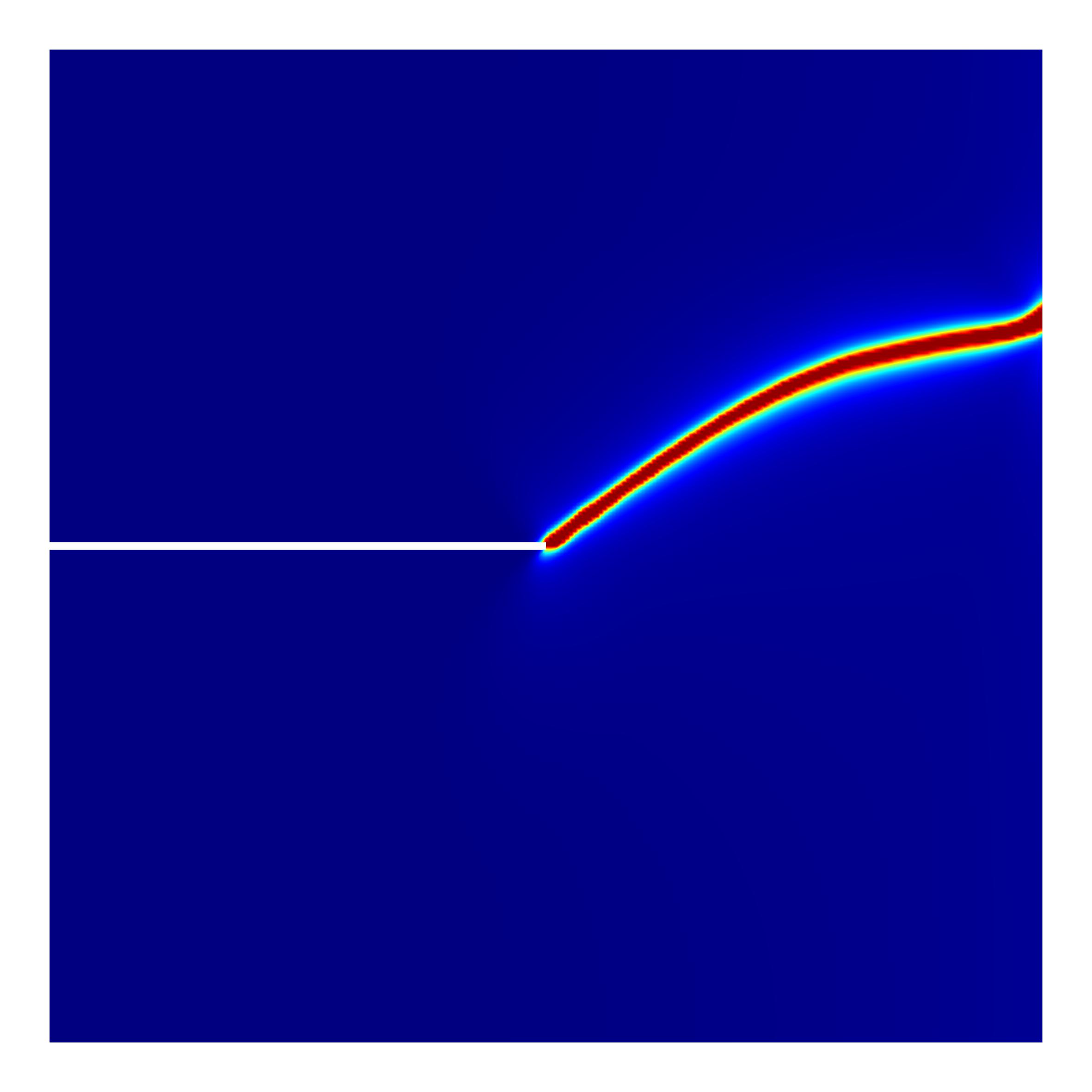}}
	\subfigure[$\lambda_c=50$, $U=2.56\times10^{-2}$ mm]{\includegraphics[height = 5.5cm]{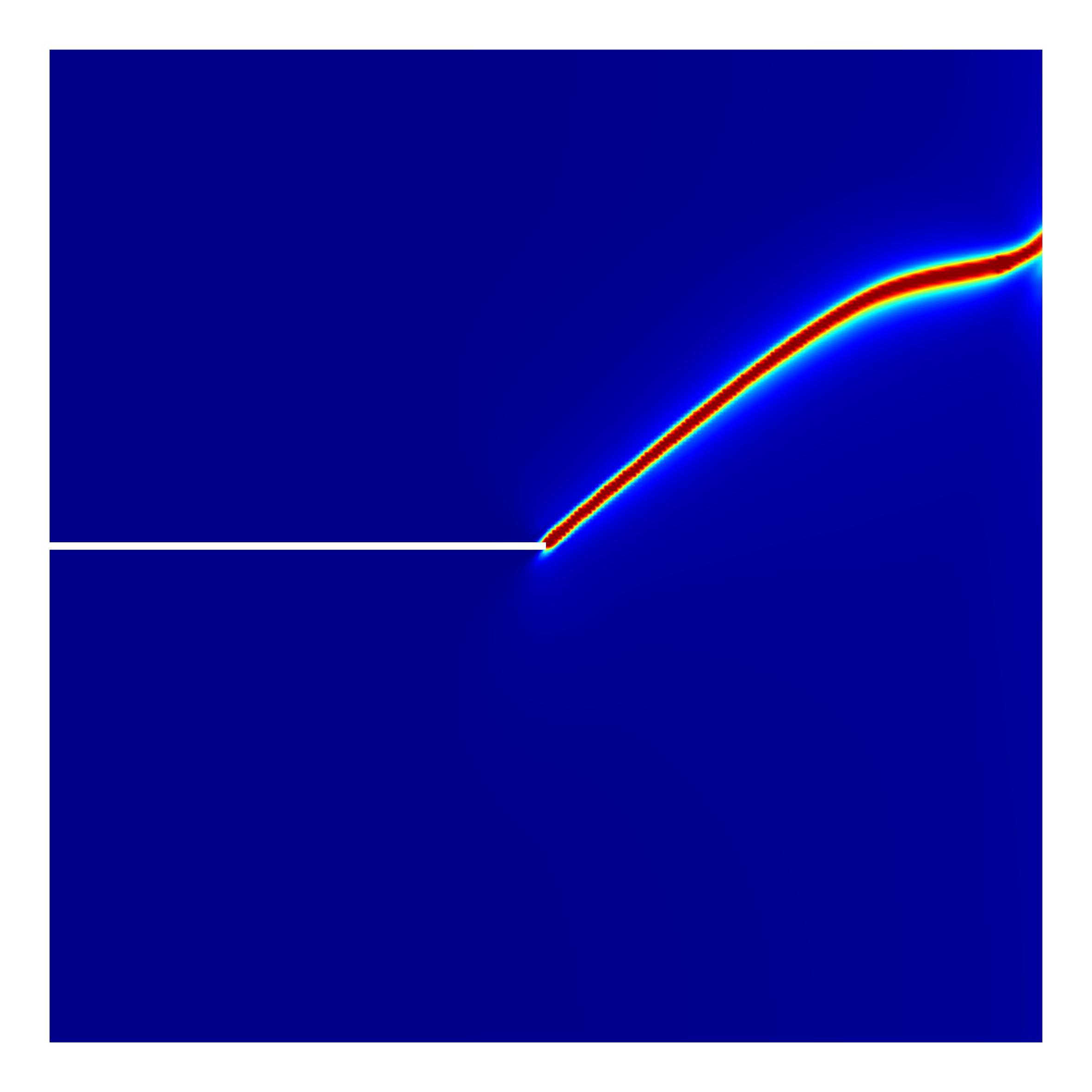}}\\
	\caption{Fracture pattern in the single-edge-notched plate subjected to tension}
	\label{Fracture pattern in the single-edge-notched plate subjected to tension}
	\end{figure}

\begin{figure}[htbp]
	\centering
	\includegraphics[height = 6cm]{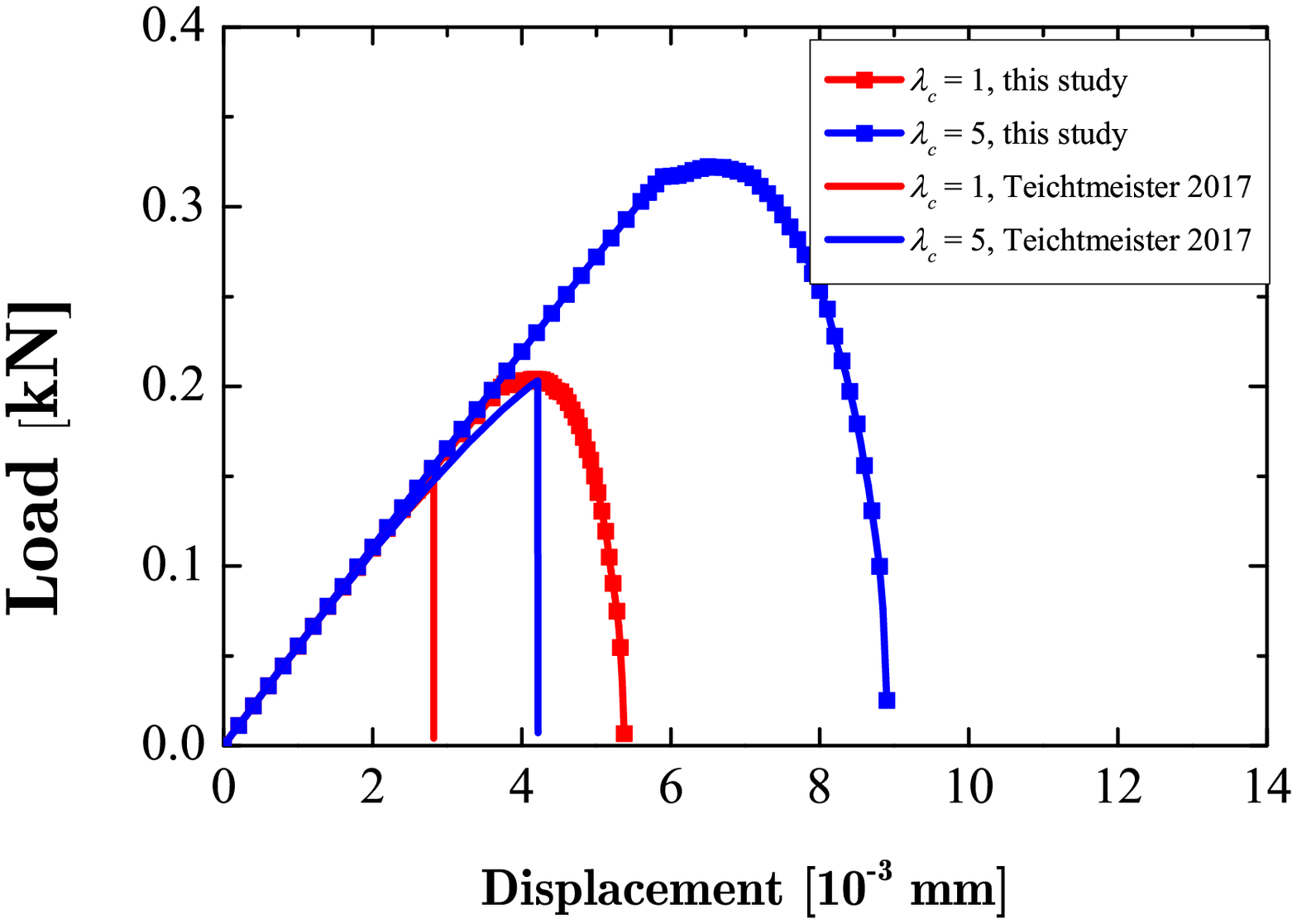}
	\caption{Load-displacement curve of the single-edge-notched plate subjected to tension}
	\label{Load-displacement curve of the single-edge-notched plate subjected to tension}
\end{figure}

Figure \ref{Fracture pattern in the single-edge-notched plate subjected to tension} presents the final fracture patterns for different $\lambda_c=G_{c1}/G_{c2}-1$, which corresponds to the parameter $\alpha$ defined in \citet{teichtmeister2017phase}. As observed, the fractures propagate at an inclination angle with the horizontal direction due to the transversely isotropic mechanical properties. In addition, the inclination angles of these observed fractures increase as the ratio $\lambda_c$ rises. The simulated fracture paths are in good agreement with those predicted by \citet{teichtmeister2017phase}. Table \ref{Fracture angles for different lambda_c} also compares the fracture angles obtained by our phase field model and the theoretical and numerical predictions of \citet{teichtmeister2017phase}. It can be seen from this table that our model reproduces well the fracture angles compared with the previous study of \citet{teichtmeister2017phase}. Moreover, Fig. \ref{Load-displacement curve of the single-edge-notched plate subjected to tension} shows the load-displacement (LD) curves of the notched plate when $\lambda_c=1$ and $\lambda_c=5$; the load is calculated on the upper boundary of the plate while the thickness of the plate is set as 1 mm. It can be seen from Fig. \ref{Load-displacement curve of the single-edge-notched plate subjected to tension} that the peak load increases with the increase in the ratio $\lambda_c$. In addition, the LD curves obtained by the model of \citet{teichtmeister2017phase} are also depicted in Fig. \ref{Load-displacement curve of the single-edge-notched plate subjected to tension} for comparison. As observed, the models of \citet{teichtmeister2017phase} and this study share the same elastic stage in the LD curve. However, they are different in the peak load, post-peak stage, and final displacement. The main reason for the discrepancy is that the two methods apply different energy decomposition strategies and driving forces of phase field.

\begin{table}[htbp]
	\small
	\caption{Fracture angles for different $\lambda_c$}
	\label{Fracture angles for different lambda_c}
	\centering
	\begin{tabular}{llll}%{p{10pt}p{50pt}p{10pt}p{50pt}p{10pt}p{50pt}p{20pt}p{100pt}}
		\toprule
		$\lambda_c$ &This study & Theoretical prediction \citep{teichtmeister2017phase} & Numerical model \citep{teichtmeister2017phase}\\
		\midrule
		1&$20^\circ$ & $18.4^\circ$ & $14^\circ$\\
		5&$36^\circ$ & $35.5^\circ$ & $33^\circ$\\
		50&$42^\circ$ & $43.8^\circ$ & $43^\circ$\\
		\bottomrule
		\normalsize
	\end{tabular}
\end{table}

\subsection{Isotropic specimen subjected to internal fluid pressure}

\begin{figure}[htbp]
	\centering
	\includegraphics[width = 8cm]{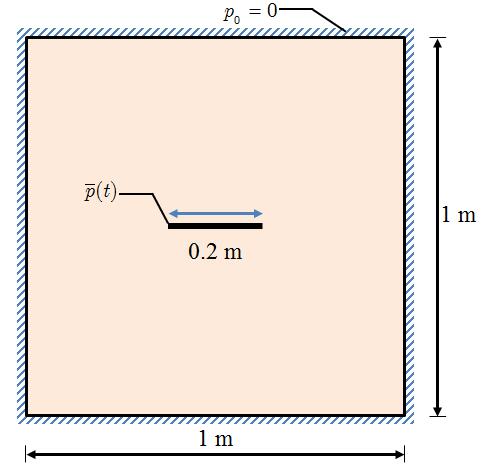}
	\caption{Geometry and boundary conditions for an isotropic specimen subjected to internal fluid pressure}
	\label{Geometry and boundary conditions for an isotropic specimen subjected to internal fluid pressure}
	\end{figure}

In this example, we investigate the fluid-driven fracture propagation in an isotropic medium due to an increasing fluid pressure on a pre-existing notch. The geometry and boundary condition of the specimen are shown in Fig. \ref{Geometry and boundary conditions for an isotropic specimen subjected to internal fluid pressure} while all the outer boundaries are fixed for the displacement field. In addition, all the parameters for calculation are listed in Table \ref{Parameters for an isotropic specimen subjected to internal fluid pressure}; the rotation angle between $\bm e_1$ and $x-$axis is set as $0^\circ$. Moreover, the fluid pressure on the notch surface is $\bar{p}=5\times10^{4}$ $\mathrm{Pa/s}\cdot t$. It should be noted that the parameters used in the proposed method for transversely isotropic medium are adjusted to satisfy the isotropic constitutive relation in this example.

We discretize all the fields with uniform Q4 elements and the maximum size is set as $h=5$ mm while the time step $\Delta t = 0.02$s is adopted. Subsequently, an analytical solution is used to verify the numerical results. Assuming a single crack having a length of $2l_c$ in $y=0$ plane, under the plane strain assumption the analytical solution \citep{sneddon1969crack} gives the pressure induced displacement as 
\begin{equation}
u^+(0,x)=\frac{2pl_c}{E_p}\left(1-\frac{x^2}{l_c^2}\right)^{1/2}
\label{u+}
\end{equation}

\noindent where $E_p=E/(1-\nu^2)$ is the plane strain Young's modulus with $E$ and $\nu$ being the respective Young's modulus and Poisson's ratio.

\begin{figure}[htbp]
	\centering
	\includegraphics[width = 8cm]{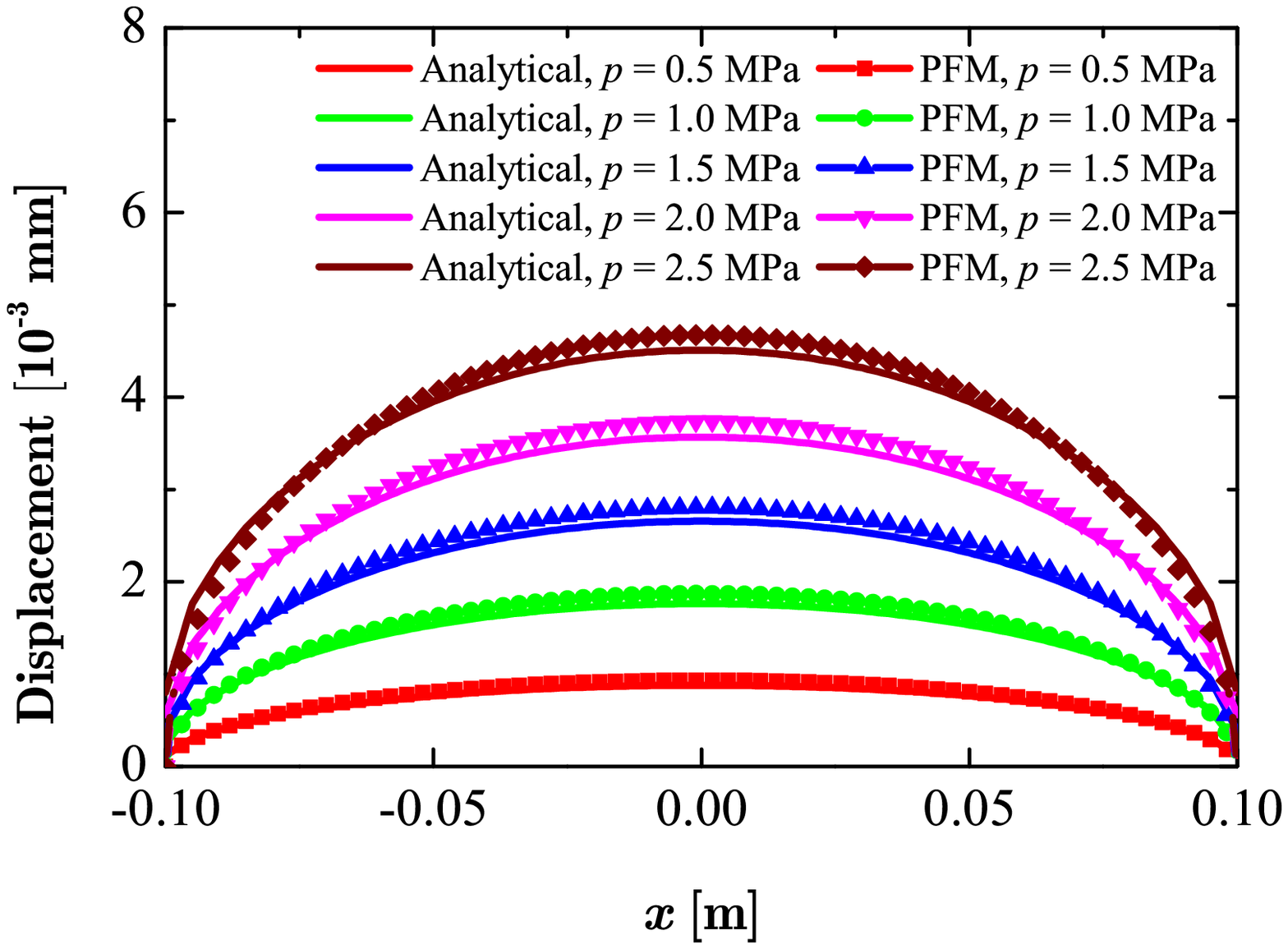}
	\caption{Comparison of the displacement along the notch}
	\label{Comparison of the displacement along the notch}
\end{figure}

\begin{figure}[htbp]
	\centering
	\subfigure[$t=58.5$ s]{\includegraphics[width = 5.5cm]{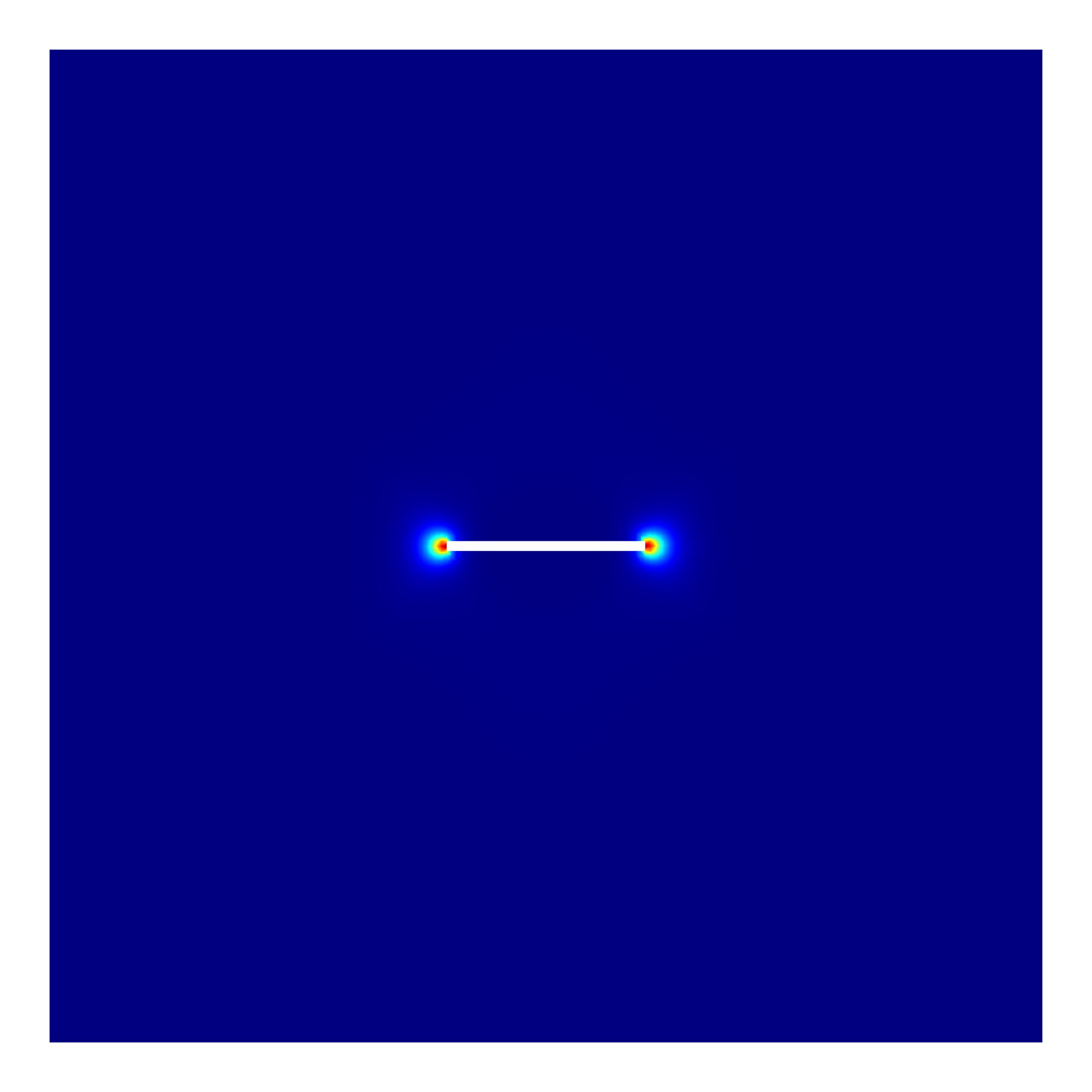}}
	\subfigure[$t=60$ s]{\includegraphics[width = 5.5cm]{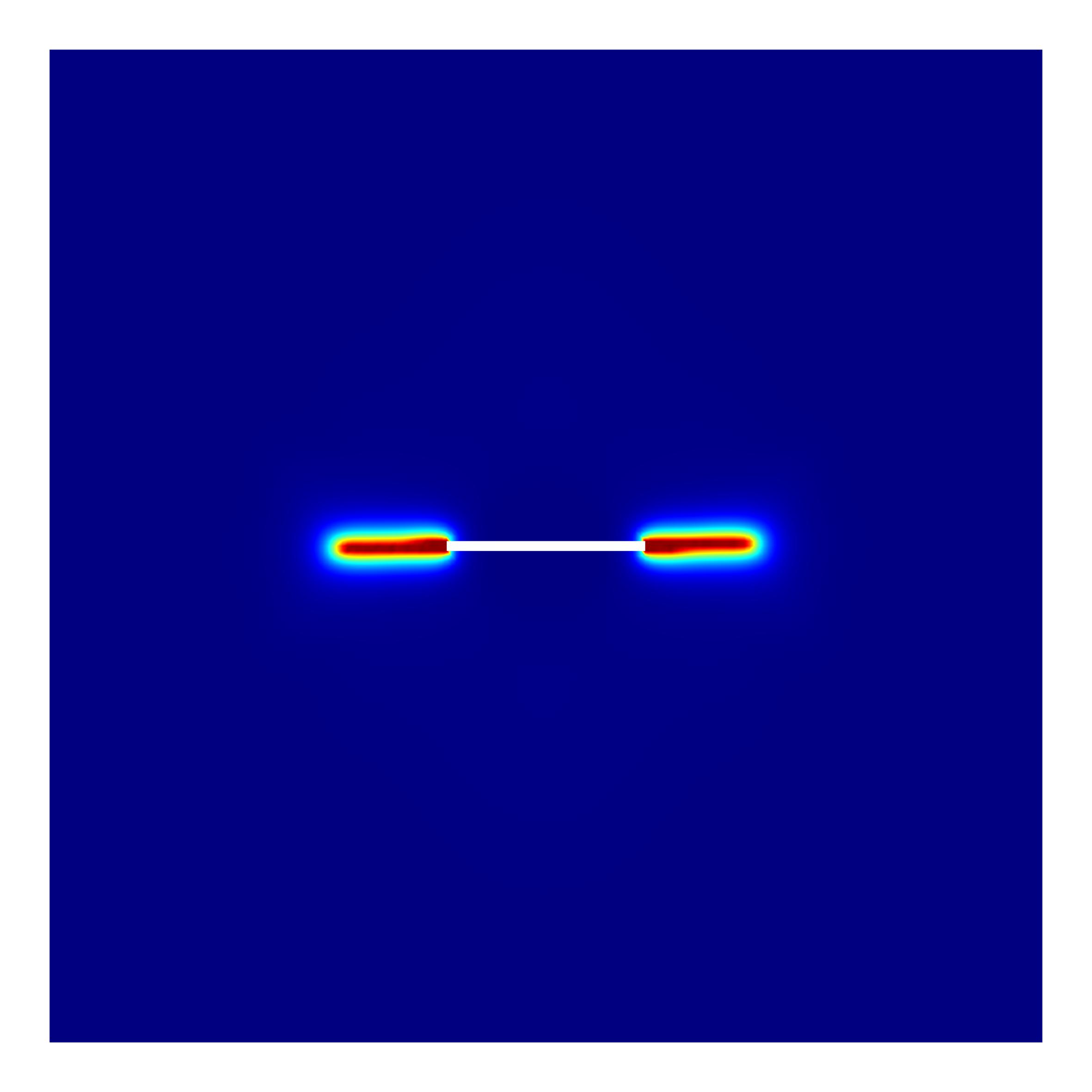}}
	\subfigure[$t=63.3$ s]{\includegraphics[width = 5.5cm]{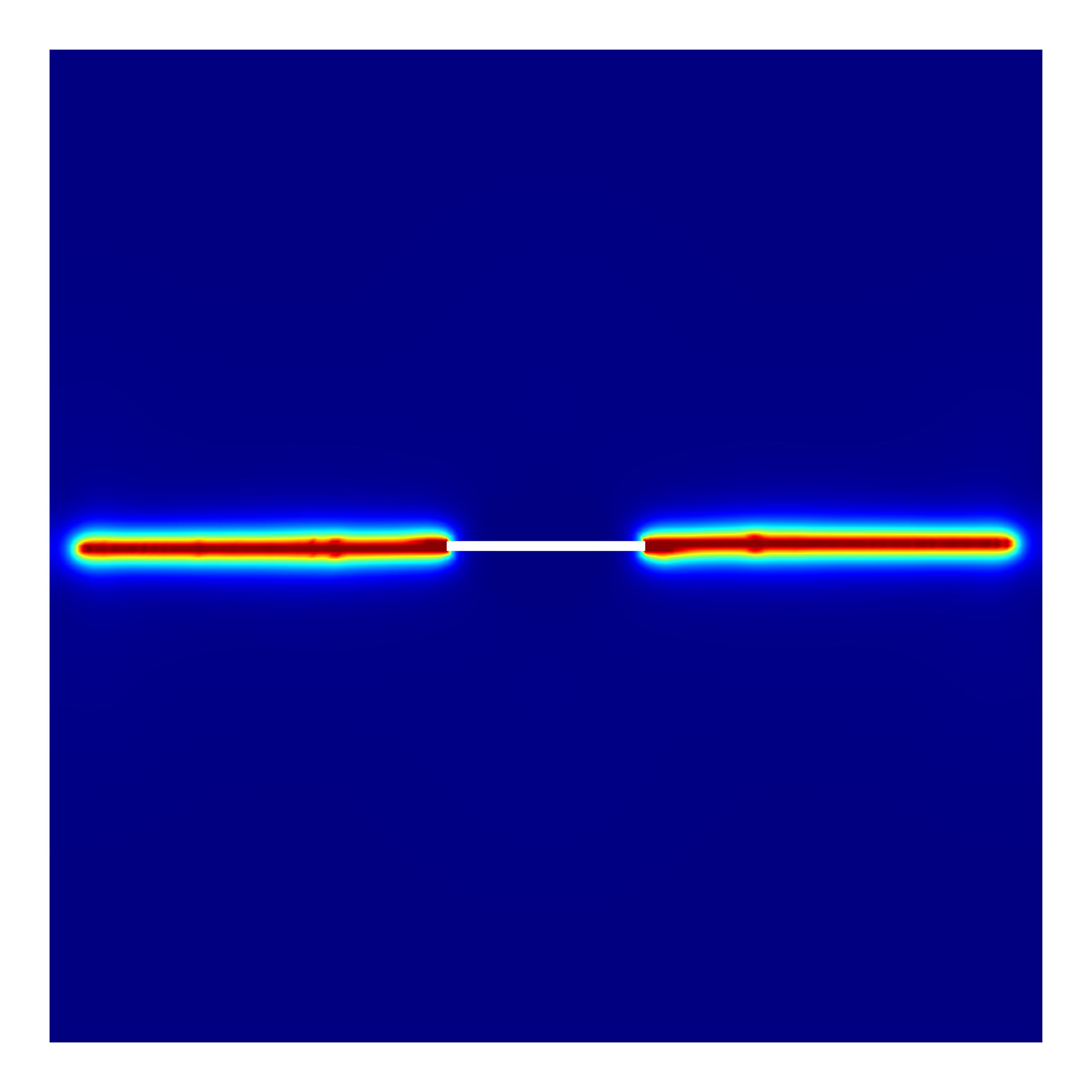}}
	\caption{Fracture initiation and propagation in the isotropic medium}
	\label{Fracture initiation and propagation in the isotropic medium}
	\end{figure}

Figure \ref{Comparison of the displacement along the notch} compares the displacement along the notch obtained by our phase field model and the analytical solution of \citet{sneddon1969crack}. It is observed that our numerical results are consistent with the analytical solution. The minor difference mainly results from the mesh size and that a small notch width of $l_0$ is set in this example. In addition, Fig. \ref{Fracture initiation and propagation in the isotropic medium} shows the hydraulic fracture initiation and propagation in the isotropic porous medium. As observed, the fracture initiates at $t=58.5$ s and propagates along the horizontal direction at $t=60$ s and 63.3 s. This fracture pattern is also consistent with the previous numerical simulations of \citet{ehlers2017phase, zhou2018phase2}. In summary, the successful applications of the proposed phase field model in the transversely isotropic plate subjected to tension and the isotropic porous medium subjected to internal fluid pressure initially prove the practicability and feasibility of our PFM.

	\begin{table}[htbp]
	\small
	\caption{Parameters for an isotropic specimen subjected to internal fluid pressure}
	\label{Parameters for an isotropic specimen subjected to internal fluid pressure}
	\centering
	\begin{tabular}{llllllll}%{p{10pt}p{50pt}p{10pt}p{50pt}p{10pt}p{50pt}p{20pt}p{100pt}}
	\hline
	$E_1$ & 100 GPa & $E_2$ & 100 GPa & $G_{12}$ & 40 Gpa & $\nu_{12}$ & 0.25\\
	$\nu_{23}$ & 0.25 & $l_0$ & $1\times10^{-2}$ m & $\alpha_r$ &0.002 & $\rho_{r}$ & $1.0\times10^{3}$ kg/m$^3$ \\
	$k$ &$1\times10^{-9}$ & $c_1$ & 0.4 & $c_2$ & 1.0 & $\varepsilon_{pr}$ & 0.002 \\
	$\rho_{f}$& $1.0\times10^{3}$ kg/m$^3$ & $c_r$ & $1\times10^{-8}$ 1/Pa & $c_f$ & $1\times10^{-8}$ 1/Pa & $\mu_r$ & $1\times10^{-3}$ Pa$\cdot$s \\
	$\mu_f$ & $1\times10^{-3}$ Pa$\cdot$s & $q_r$ & 0 & $q_f$ & 0 & $k_{r1}$ & $1\times10^{-15}$ m$^2$ \\
	$k_{r2}$ & $1\times10^{-15}$ m$^2$ &  $k_{f1}$ & $1\times10^{-6}$ m$^2$ &  $k_{f2}$ & $1\times10^{-6}$ m$^2$& $G_{c1}$ & 10 N/m \\ 
	$G_{c2}$ & 10 N/m &&&&&&\\
	\hline
	\normalsize
	\end{tabular}
	\end{table}

\section{Hydraulic fractures in the transversely isotropic medium}\label{Hydraulic fractures in the transversely isotropic medium}

In this section, we present three numerical problems for hydraulic fracture propagation in the transversely isotropic porous medium. These numerical examples include a 2D medium with an interior notch subjected to fluid pressure, a 2D medium with two parallel interior notches, and a 3D medium with a penny-shaped notch subjected to internal fluid pressure. These problems are used to further validate the practicability and feasibility of our proposed phase field model while they present some basic characteristics that the proposed formulation yields in the multiple filed modeling.

\subsection{2D medium with an interior notch}

Let us first consider a 2D transversely isotropic medium with an interior notch, which is subjected to an increasing fluid pressure of $\bar{p}=5\times10^{4}$ $\mathrm{Pa/s}\cdot t$. The geometry and boundary condition of this problem is the same as those in Fig. \ref{Geometry and boundary conditions for an isotropic specimen subjected to internal fluid pressure} while a rotation angle $\beta$ between the material direction $\bm e_1$ and $x-$axis is set. In addition, the parameters for calculation are identical to those in Table \ref{Parameters for an isotropic specimen subjected to internal fluid pressure} with the critical energy release rate in the transverse plane being changed to 50 N/m. Note that the focus of this study is to present a phase field framework  that can be easily implemented and applied to transversely isotropic porous media. Therefore, any possible permeability model can be applied in future research. For example, the dependence of the permeability in the fractured domain on the strain or stress can be easily applied within COMSOL. However, the fracture opening cannot be directly extracted in COMSOL because of the smooth representation of the sharp fracture in a PFM. In this case, the permeability models in \citet{miehe2015minimization, lee2017iterative, heider2018modeling}, which adopt the fracture opening as an input in the modeling, are difficult to apply in our phase field modeling. Therefore, for simplicity, an unchanged permeability is adopted in our presented examples, which also shows favorable results.

The length and width of the notch are 0.2 m and 0.01 m, respectively. Q4 elements are used to discretize the domain and the element size is set as $h=5$ mm. In addition, the time step $\Delta t =0.02$ s is adopted for the phase field modeling. Subsequently, the influence of material direction, modulus anisotropy, and permeability anisotropy are analyzed and discussed in the following subsections.
 
\subsubsection{Influence of material direction}

\begin{figure}[htbp]
	\centering
	\subfigure[$t=122.3$ s, $\beta=0^\circ$]{\includegraphics[width = 4.2cm]{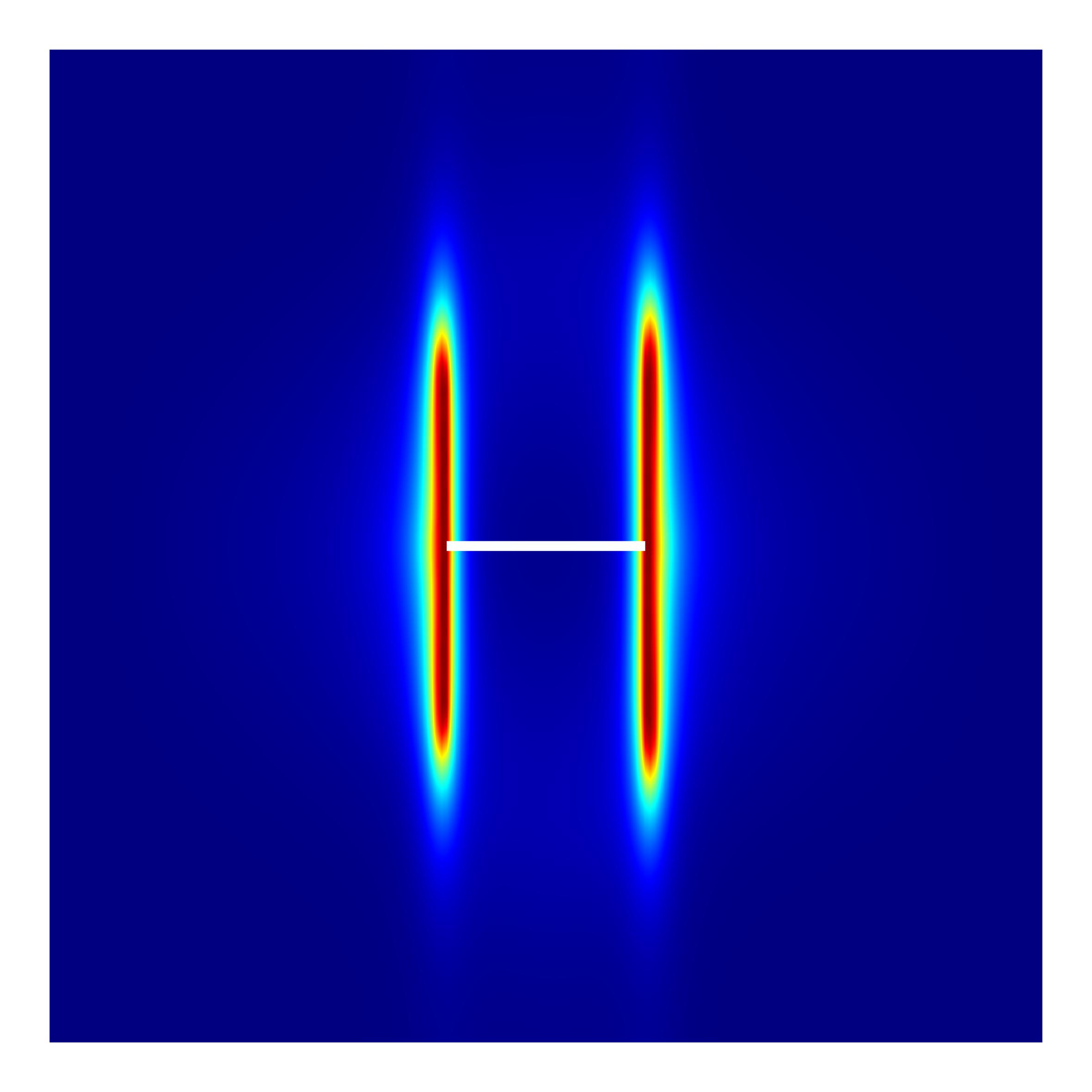}}
	\subfigure[$t=124.6$ s, $\beta=-15^\circ$]{\includegraphics[width = 4.2cm]{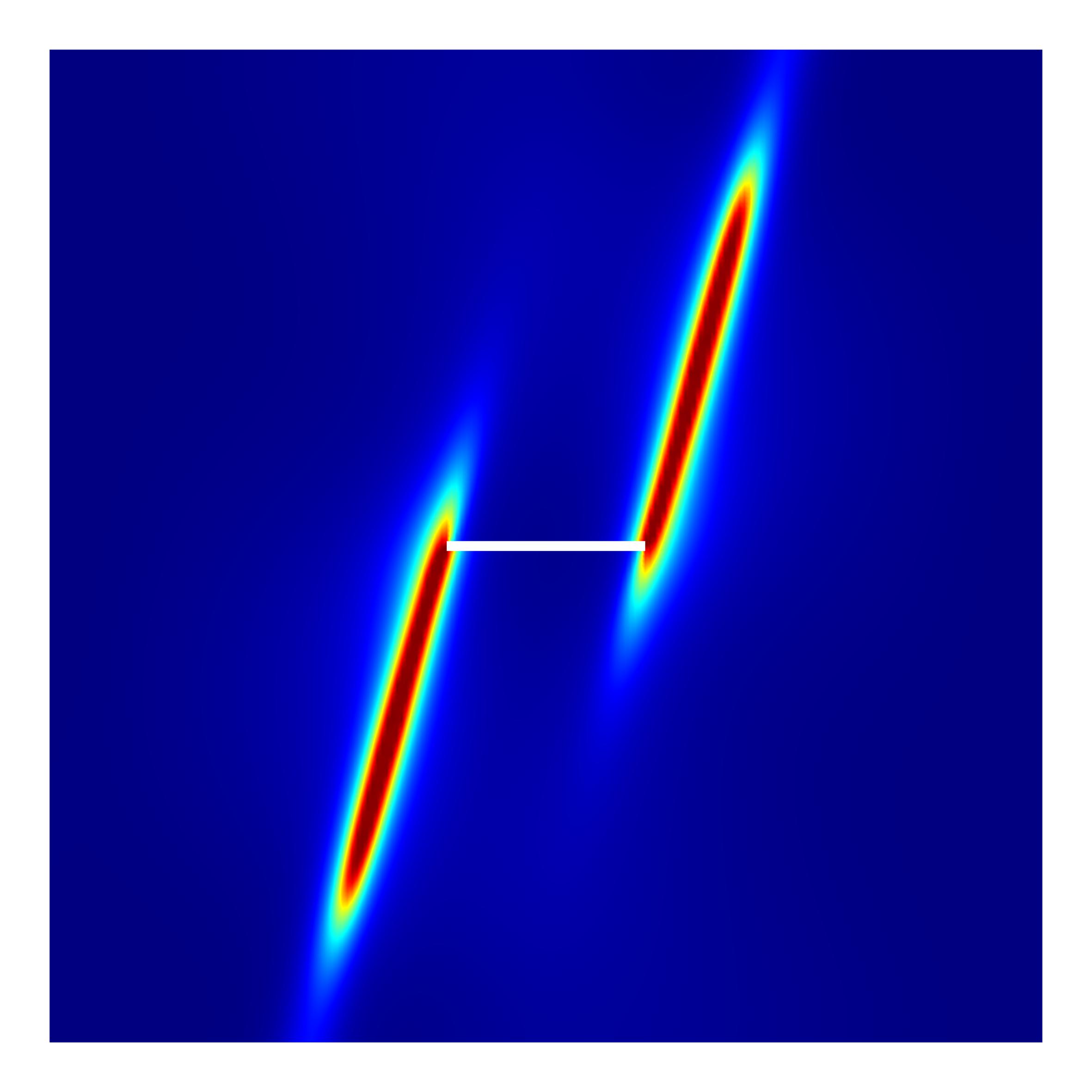}}
	\subfigure[$t=127.37$ s, $\beta=-30^\circ$]{\includegraphics[width = 4.2cm]{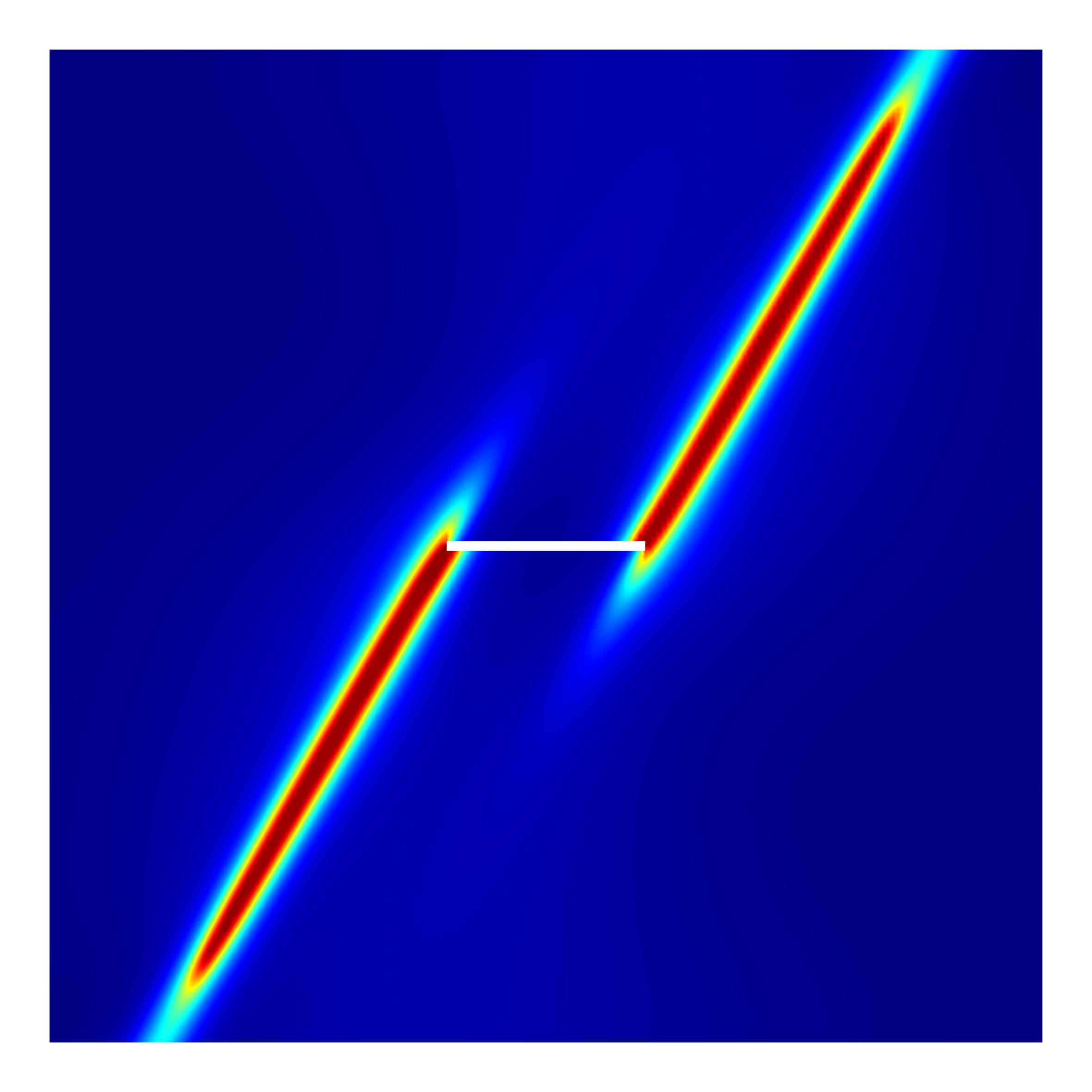}}
	\subfigure[$t=121.79$ s, $\beta=-45^\circ$]{\includegraphics[width = 4.2cm]{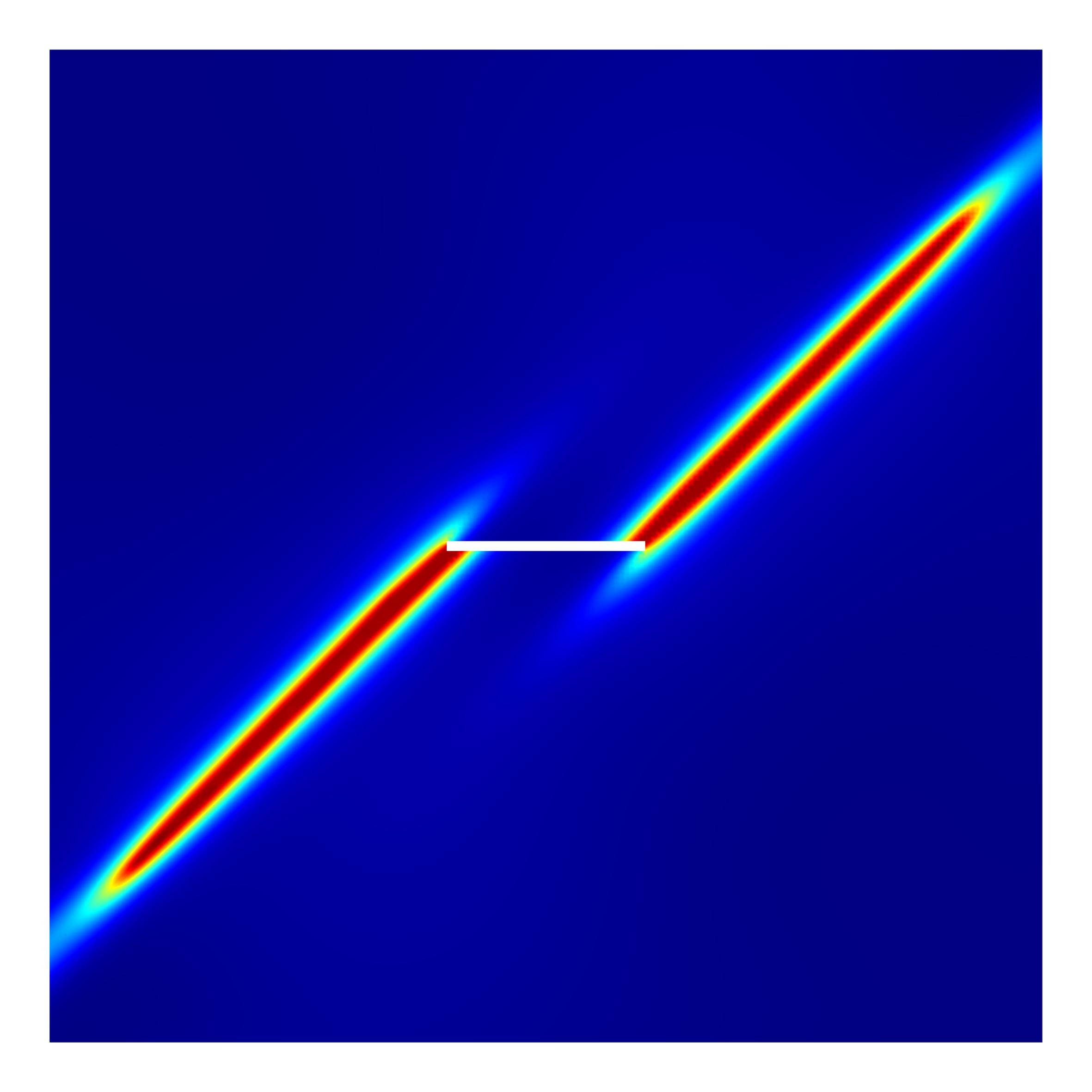}}\\
	\subfigure[$t=112.3$ s, $\beta=-60^\circ$]{\includegraphics[width = 4.2cm]{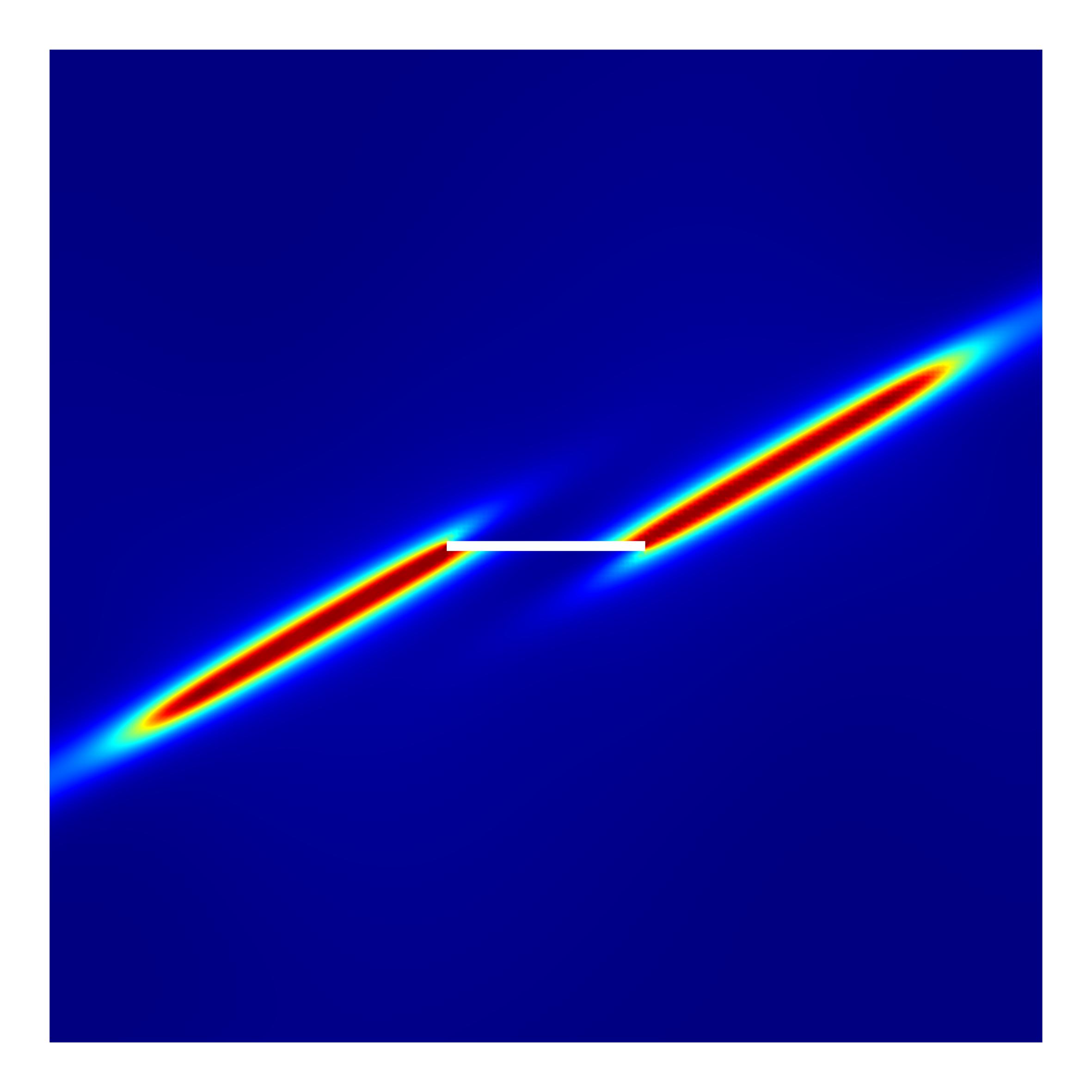}}
	\subfigure[$t=105.9$ s, $\beta=-75^\circ$]{\includegraphics[width = 4.2cm]{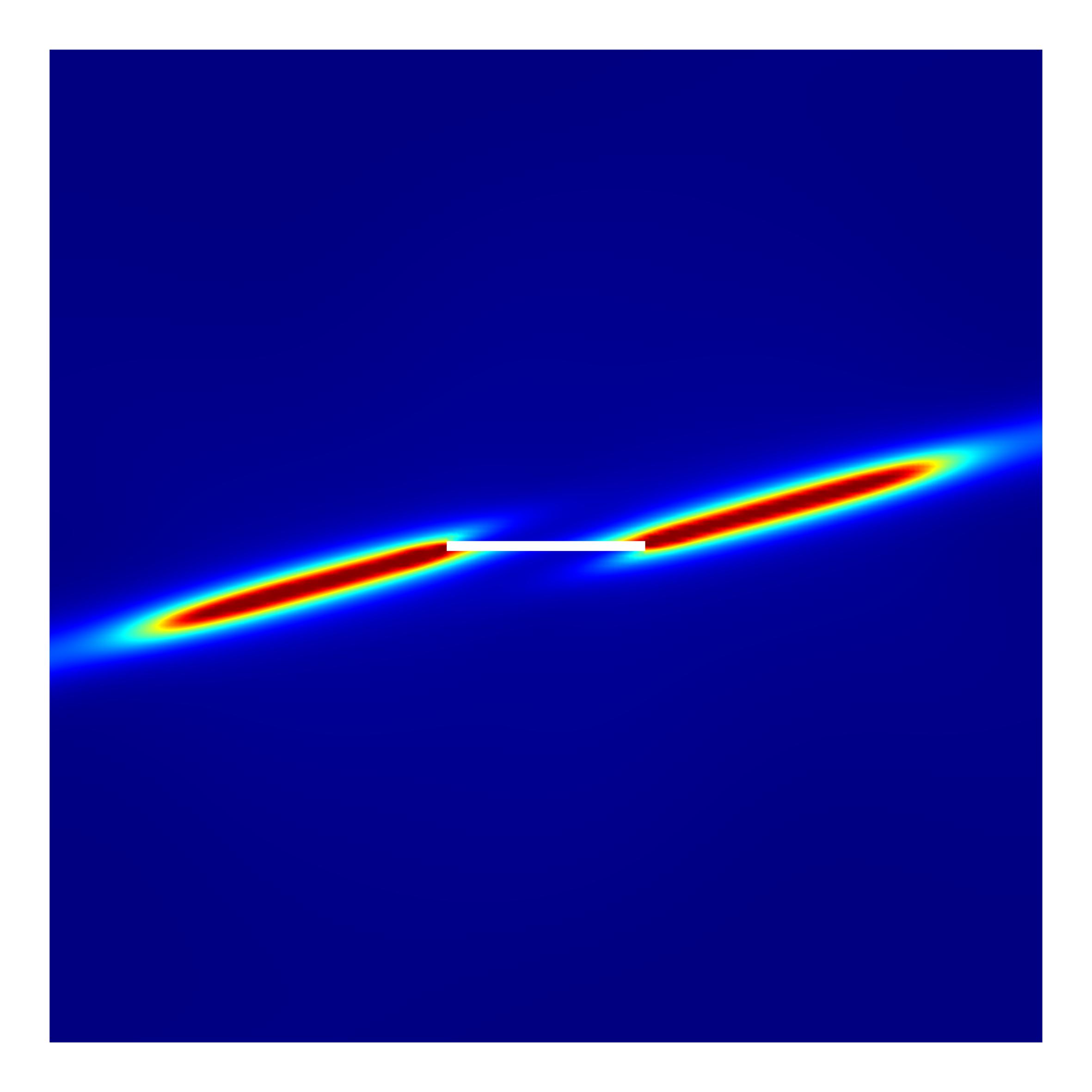}}
	\subfigure[$t=103.95$ s, $\beta=-90^\circ$]{\includegraphics[width = 4.2cm]{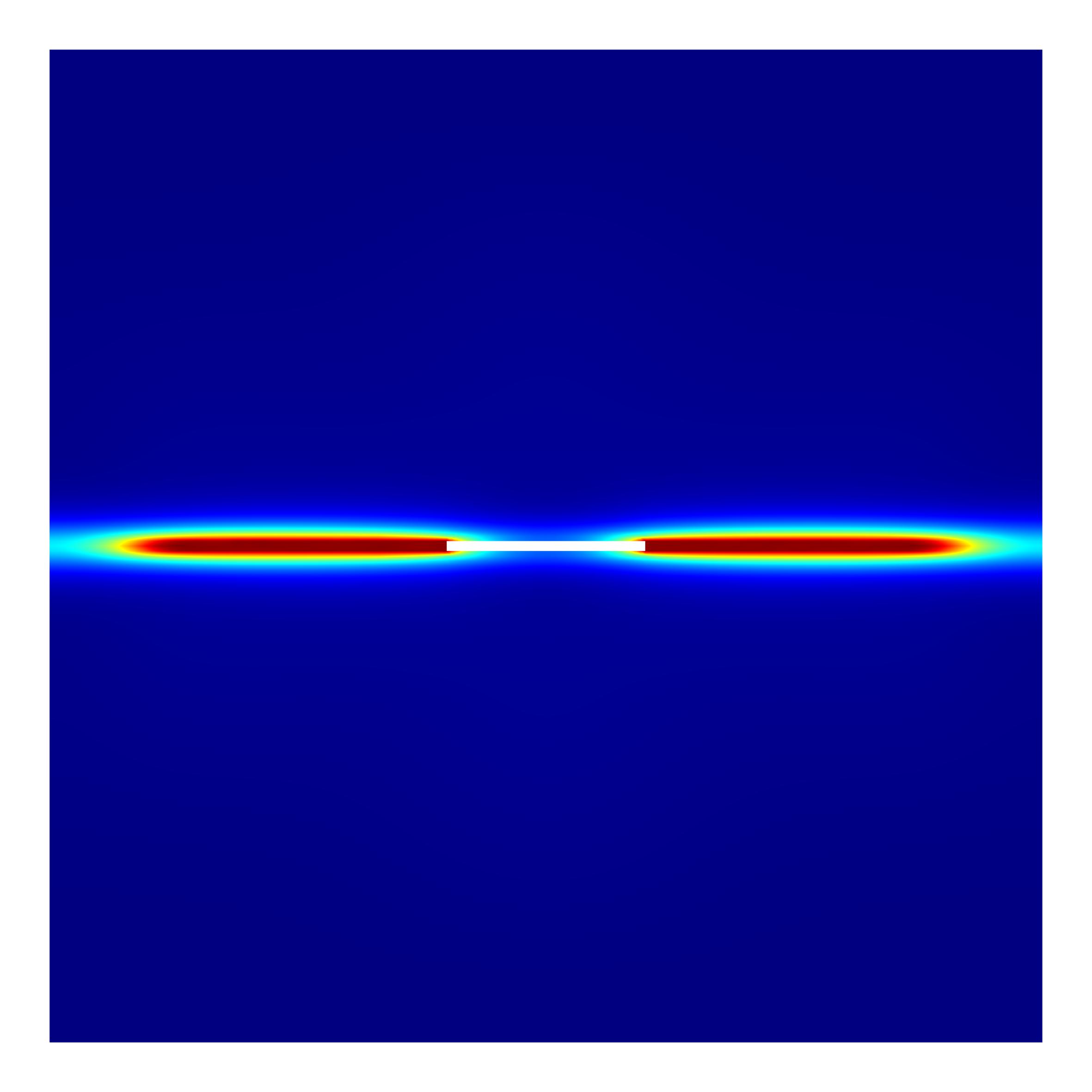}}
	\caption{Fracture evolution of the transversely isotropic porous medium with an interior notch under different $\beta$}
	\label{Fracture evolution of the transversely isotropic porous medium with an interior notch under different beta}
\end{figure}

A total of seven rotation angles are tested while the other parameters are unchanged. That is, $\beta=0^\circ$, $-15^\circ$, $-30^\circ$, $-45^\circ$, $-60^\circ$, $-75^\circ$, and $-90^\circ$. Figure \ref{Fracture evolution of the transversely isotropic porous medium with an interior notch under different beta} shows the fracture patterns of the transversely isotropic porous medium under different rotation angles $\beta$. The fractures are observed to be fully affected by the rotation angle $\beta$ and they propagate along the direction of $\bm e_2$, which has the maximum critical energy release rate $G_{c2}$. Specifically, the fractures are parallel to the pre-existing notch when $\beta=-90^\circ$. If $\beta$ varies from $-15^\circ$ to $-75^\circ$, the fractures propagate at an inclination angle with the horizontal direction. However, for $\beta=0^\circ$, the fractures are along the vertical direction ($y-$axis) and bifurcate at the tips of the pre-existing notch due to the symmetry in this problem. 

\begin{figure}[htbp]
	\centering
	\subfigure[$t=122.3$ s, $\beta=0^\circ$]{\includegraphics[width = 4.2cm]{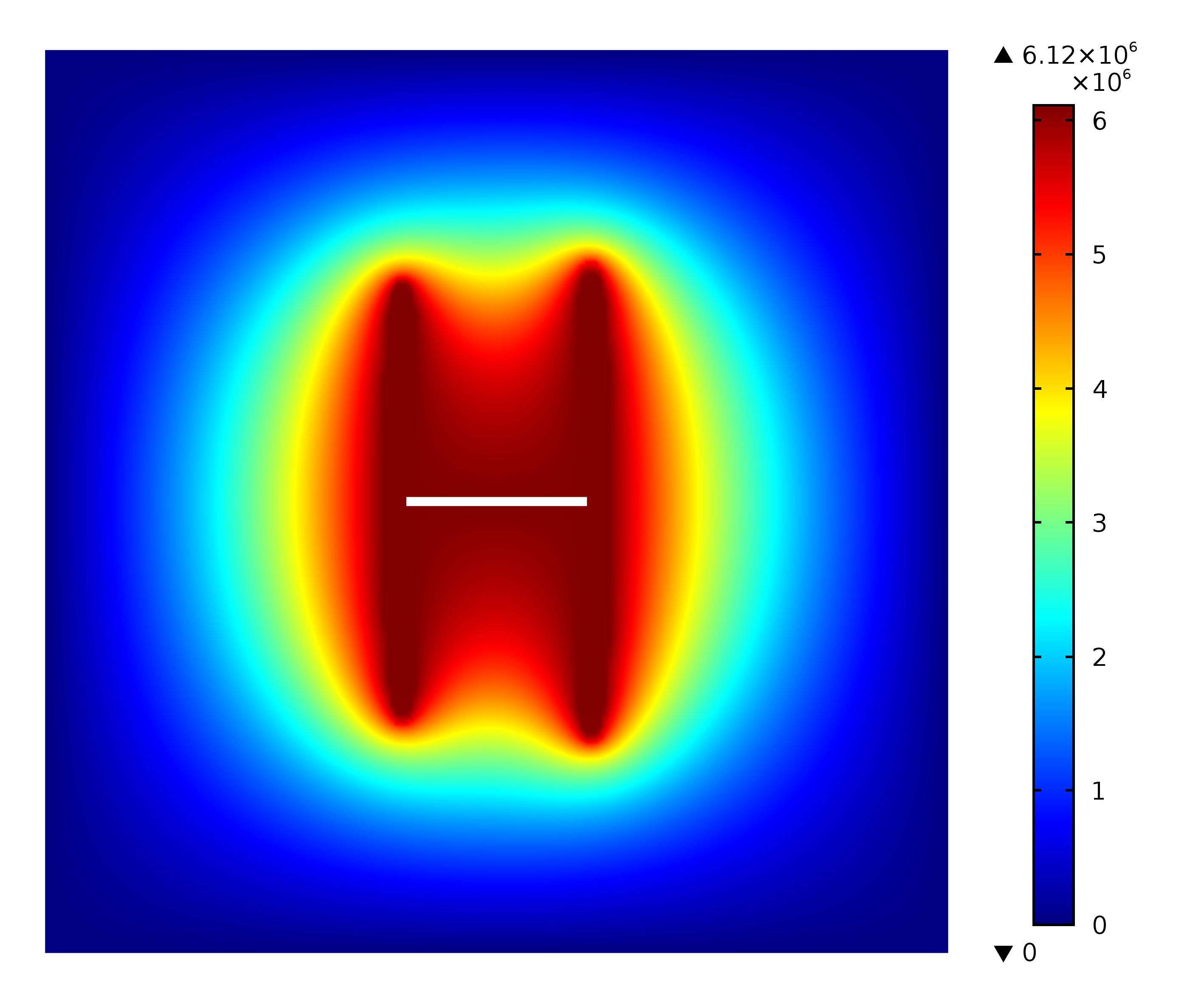}}
	\subfigure[$t=124.6$ s, $\beta=-15^\circ$]{\includegraphics[width = 4.2cm]{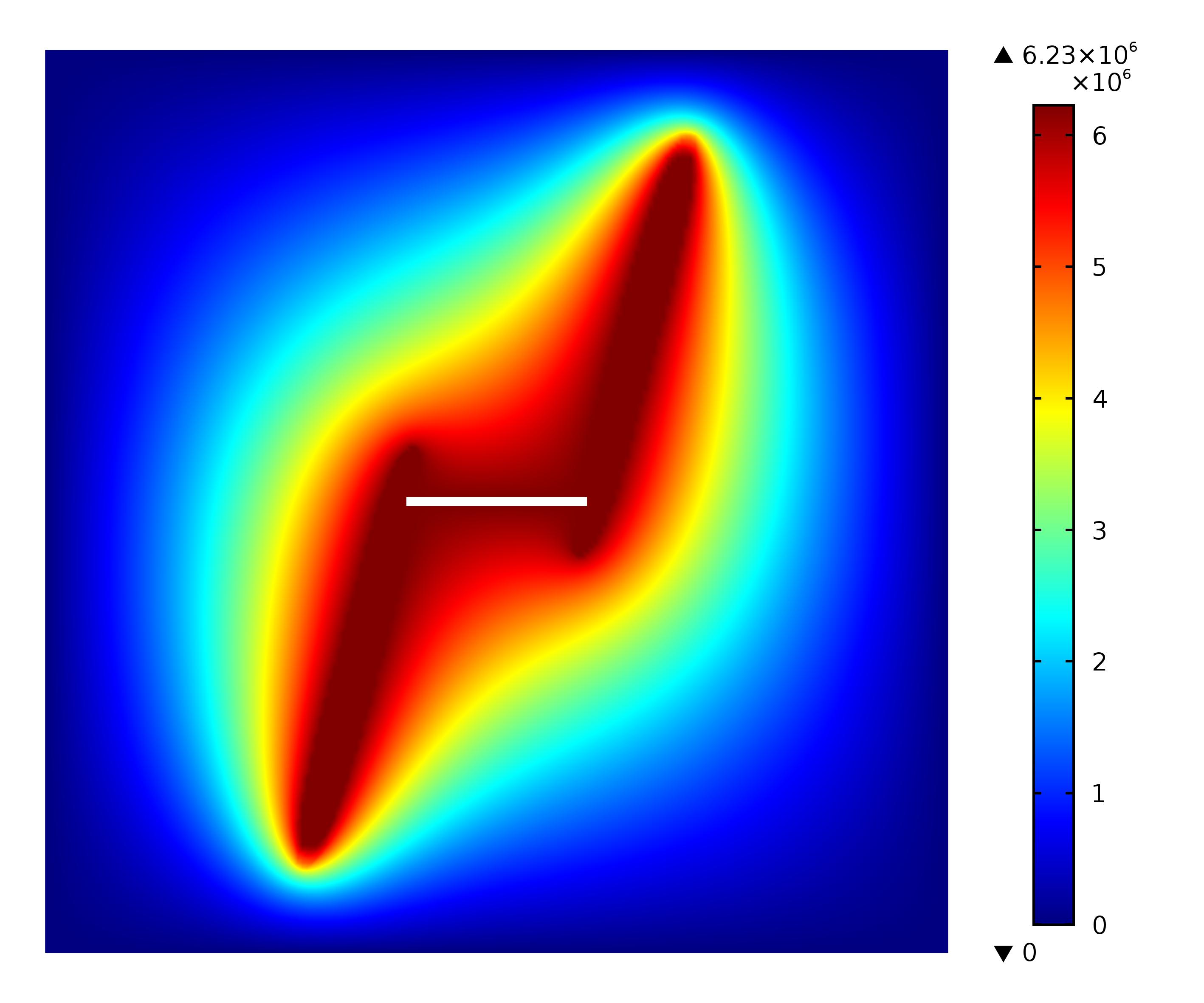}}
	\subfigure[$t=127.37$ s, $\beta=-30^\circ$]{\includegraphics[width = 4.2cm]{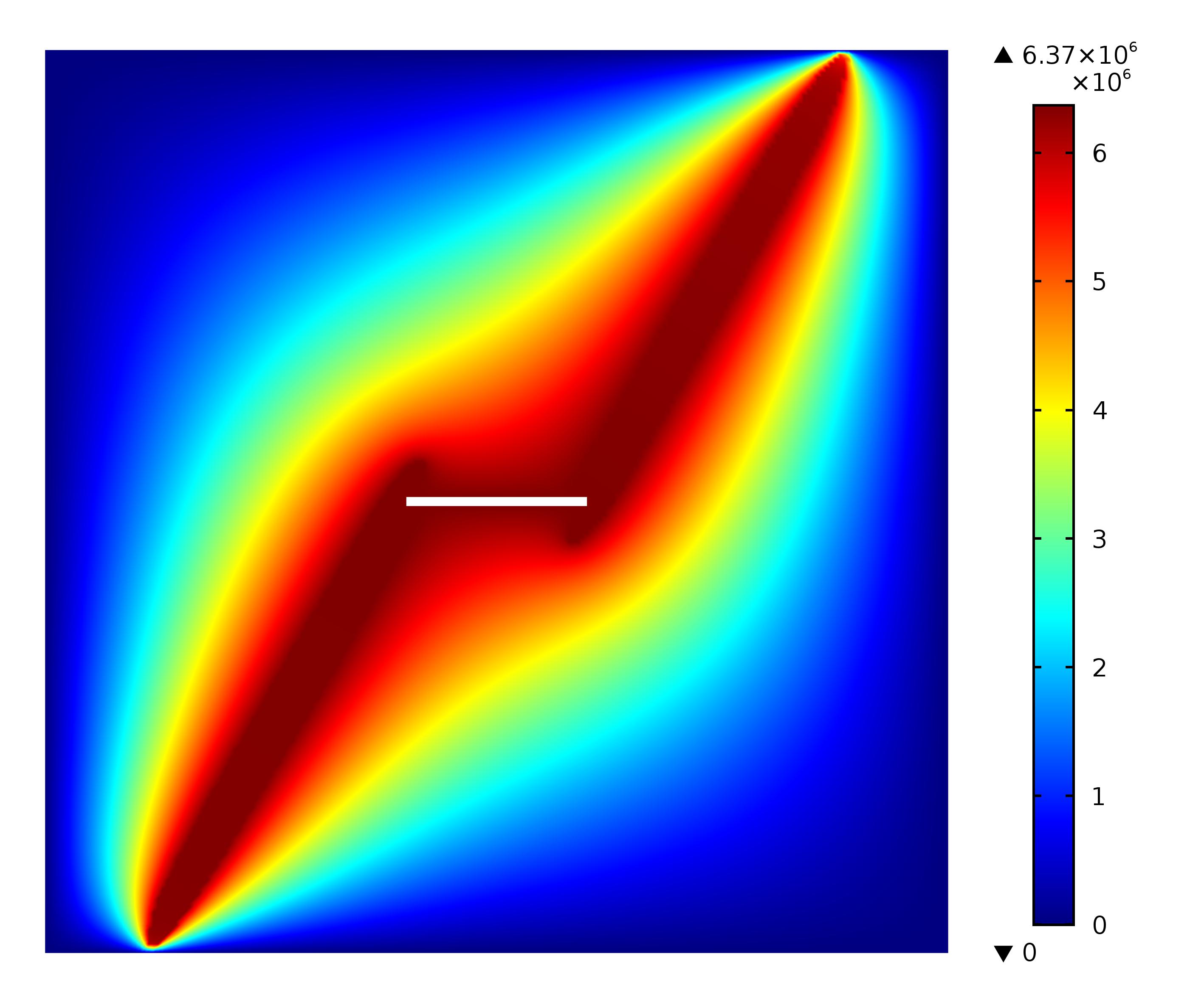}}
	\subfigure[$t=121.79$ s, $\beta=-45^\circ$]{\includegraphics[width = 4.2cm]{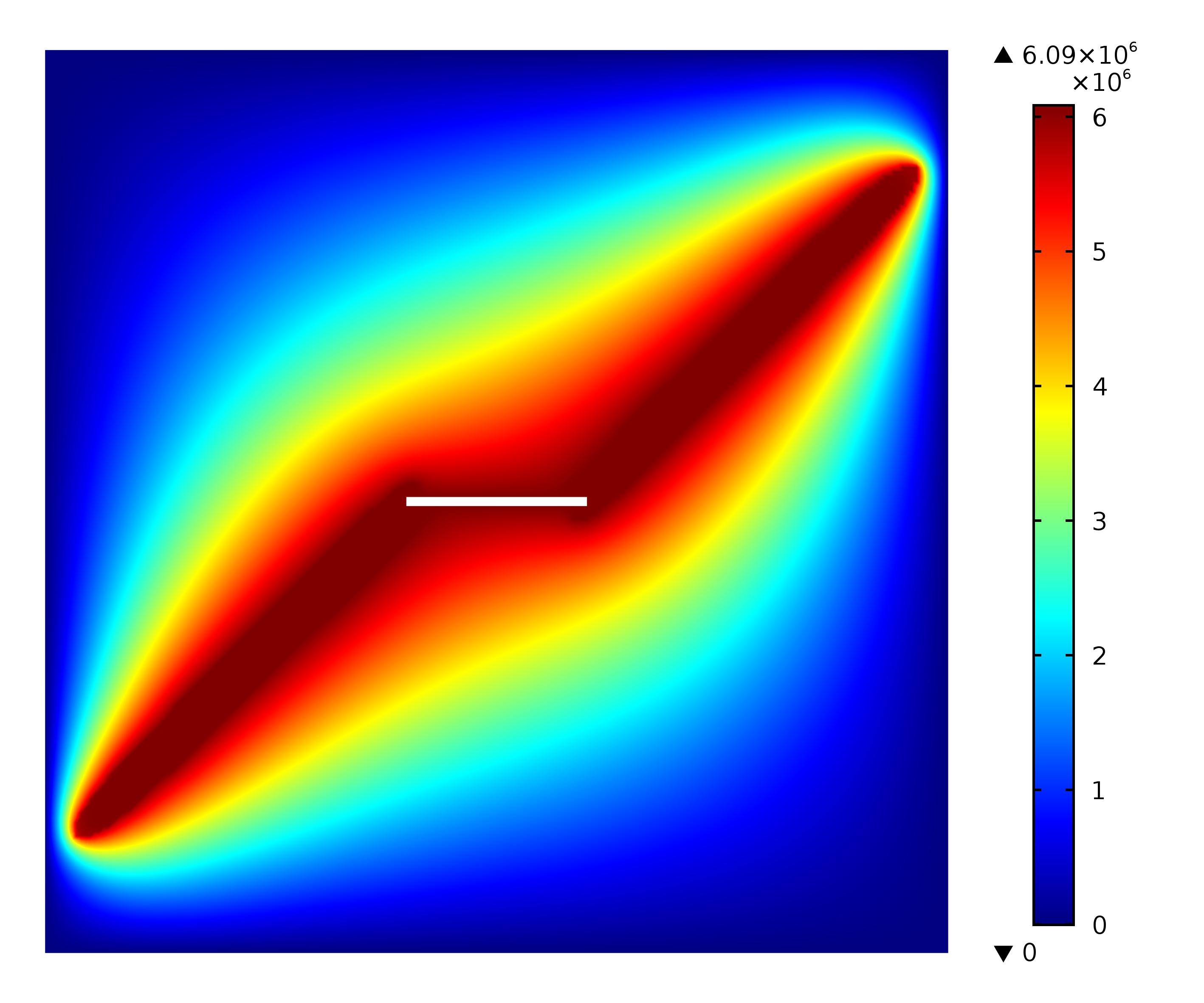}}\\
	\subfigure[$t=112.3$ s, $\beta=-60^\circ$]{\includegraphics[width = 4.2cm]{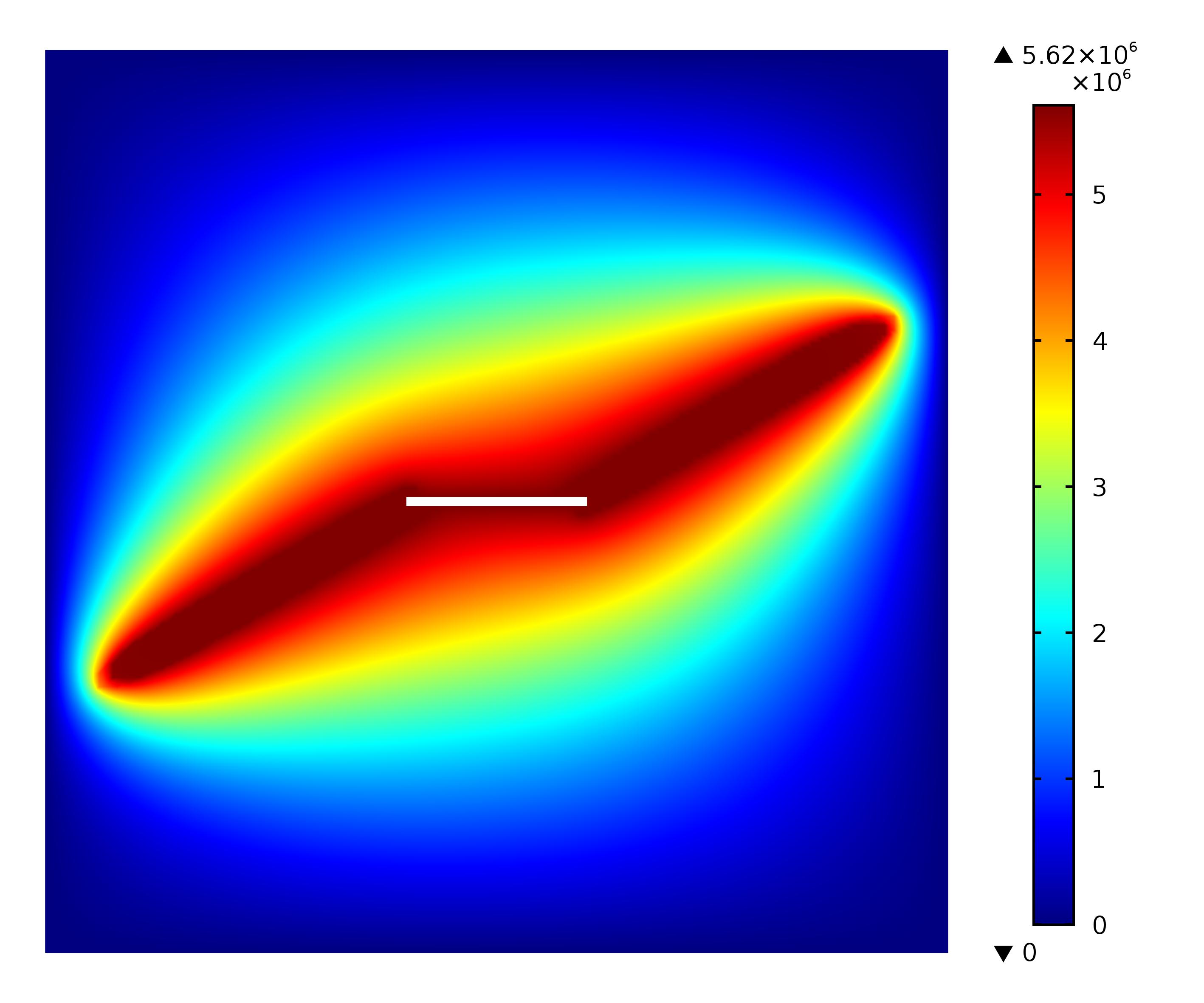}}
	\subfigure[$t=105.9$ s, $\beta=-75^\circ$]{\includegraphics[width = 4.2cm]{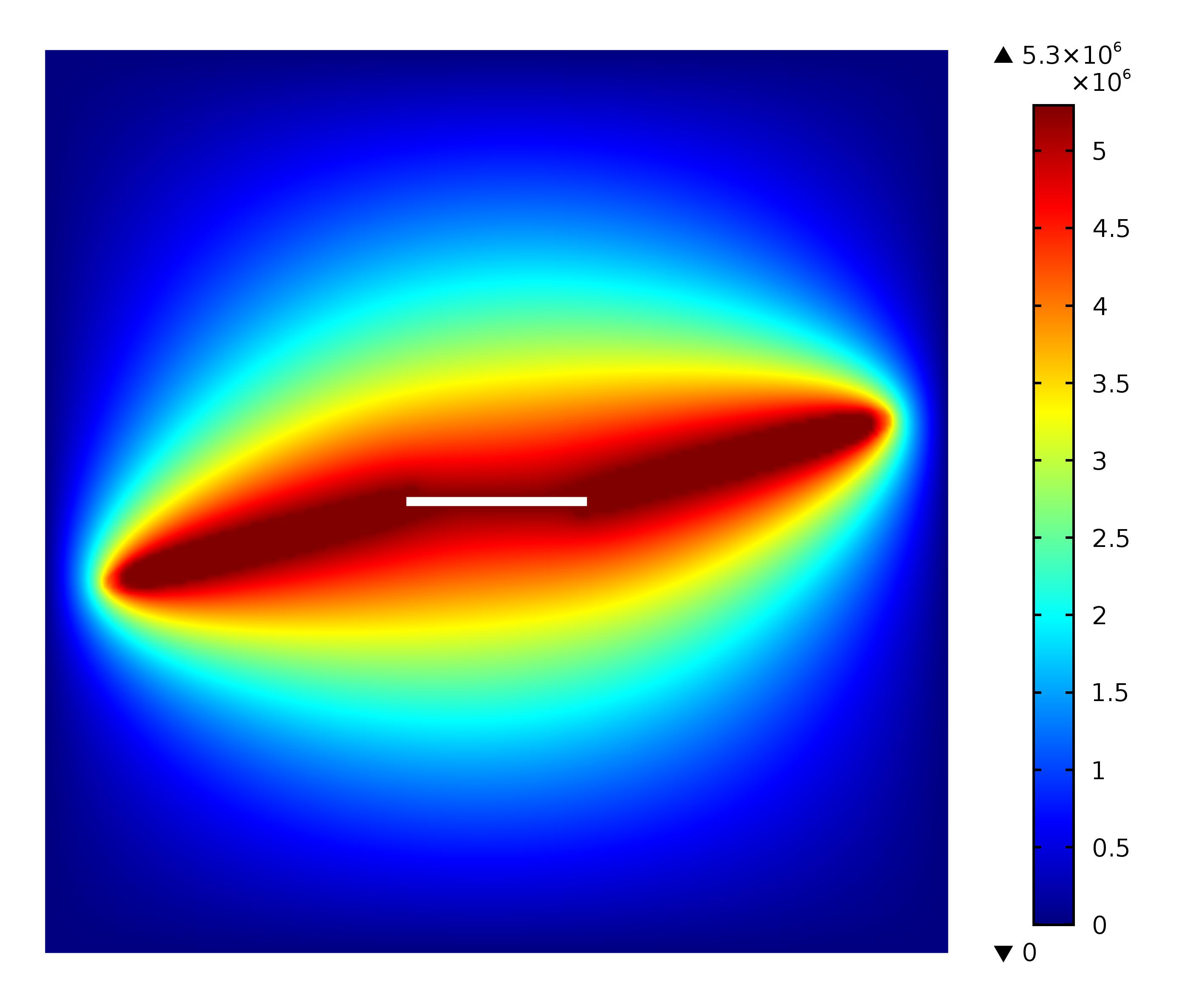}}
	\subfigure[$t=103.95$ s, $\beta=-90^\circ$]{\includegraphics[width = 4.2cm]{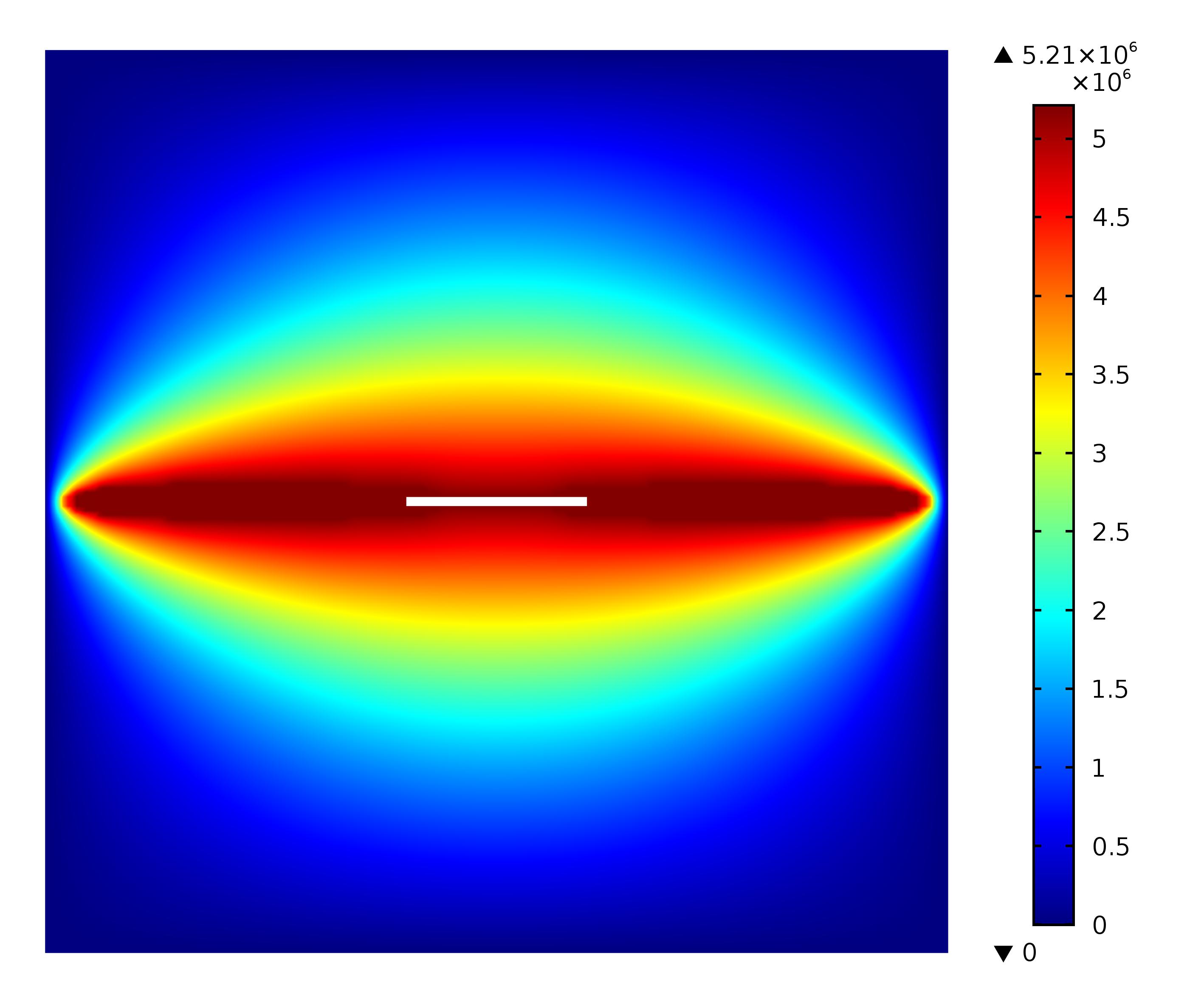}}
	\caption{Fluid pressure field of the transversely isotropic porous medium with an interior notch under different $\beta$ (unit: Pa)}
	\label{Fluid pressure field of the transversely isotropic porous medium with an interior notch under different beta}
\end{figure}

\begin{figure}[htbp]
	\centering
	\includegraphics[width = 10cm]{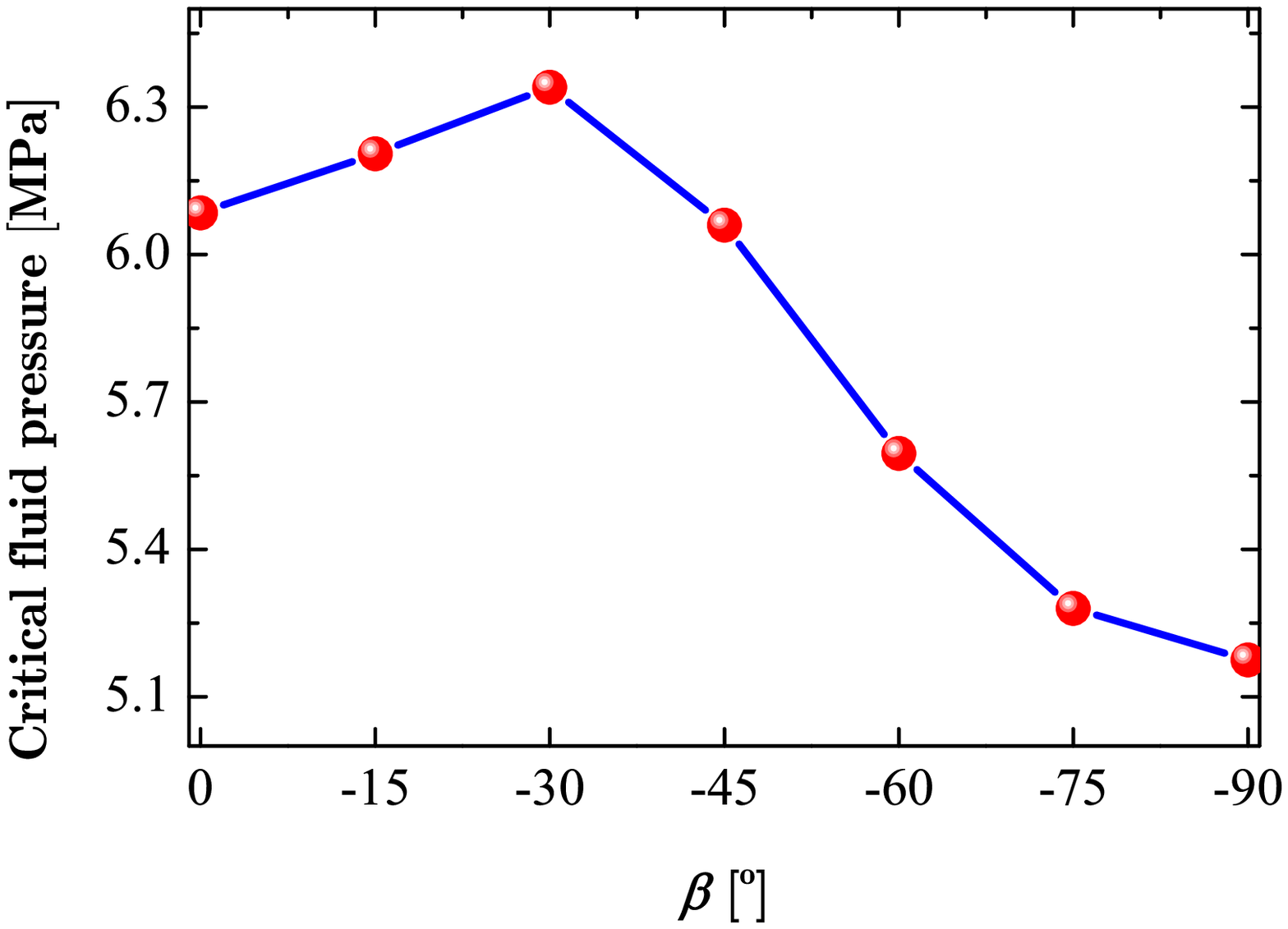}
	\caption{Critical fluid pressure for fracture initiation under different rotation angles $\beta$}
	\label{Critical fluid pressure for fracture initiation under different rotation angle beta}
\end{figure}

Figure \ref{Fluid pressure field of the transversely isotropic porous medium with an interior notch under different beta} shows the fluid pressure field for the transversely isotropic medium with an interior notch under different rotation angles $\beta$. It is observed that the fluid pressure field has a consistent shape with the phase field as shown in Fig. \ref{Fracture evolution of the transversely isotropic porous medium with an interior notch under different beta}; the fluid pressure has a maximum value in the fracture domain. In addition, Fig. \ref{Critical fluid pressure for fracture initiation under different rotation angle beta} indicates the influence of rotation angle $\beta$ on the critical fluid pressure for fracture initiation, which means the internal fluid pressure when the phase field reaches 1. It can be seen from this figure that the critical fluid pressure is sensitive to the rotation angle. The critical pressure is thus related to the material direction; on the other hand, its maximum value occurs for $\beta=-30^\circ$ but not $\beta=-45^\circ$ due to the multiple effects of the parameters used in the phase field modeling. More specifically, the critical pressure increases when $\beta$ varies from $0^\circ$ to $-30^\circ$ while it decreases as the rotation angle increases after $\beta=-30^\circ$.

\subsubsection{Influence of modulus anisotropy}

\begin{figure}[htbp]
	\centering
	\subfigure[$t=141$ s, $E_2/E_1=0.5$]{\includegraphics[width = 4.2cm]{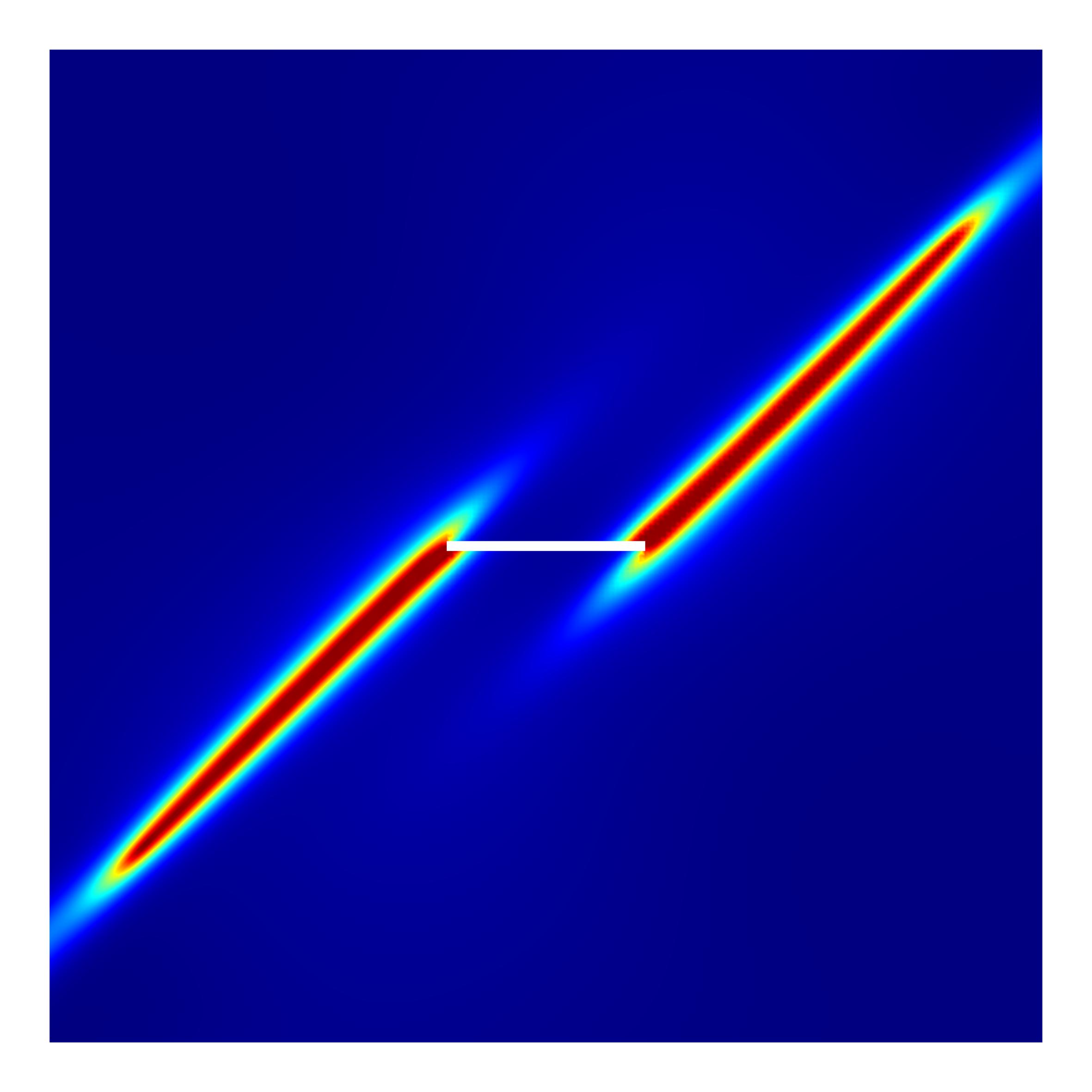}}
	\subfigure[$t=121.79$ s, $E_2/E_1=1$]{\includegraphics[width = 4.2cm]{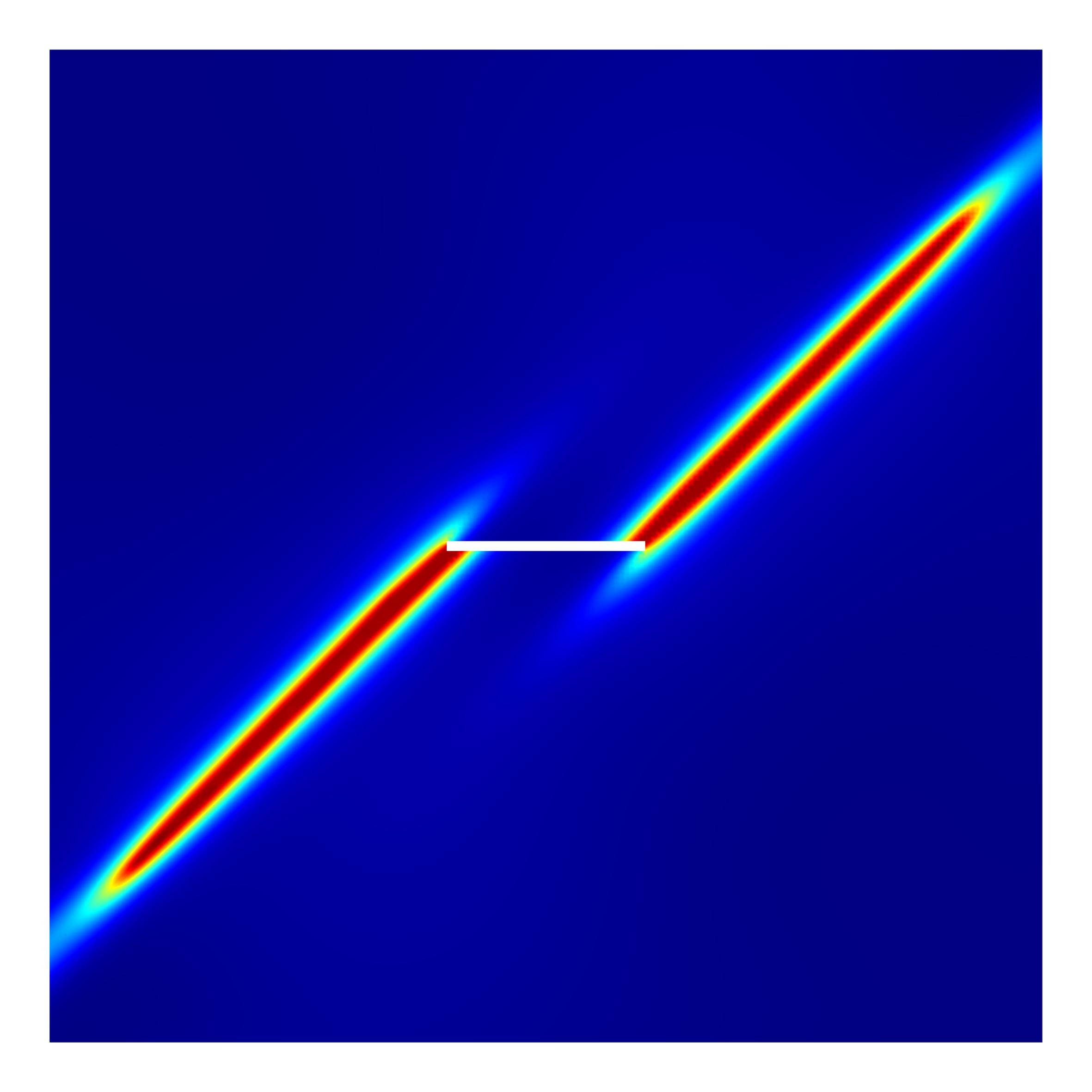}}
	\subfigure[$t=101.3$ s, $E_2/E_1=2$]{\includegraphics[width = 4.2cm]{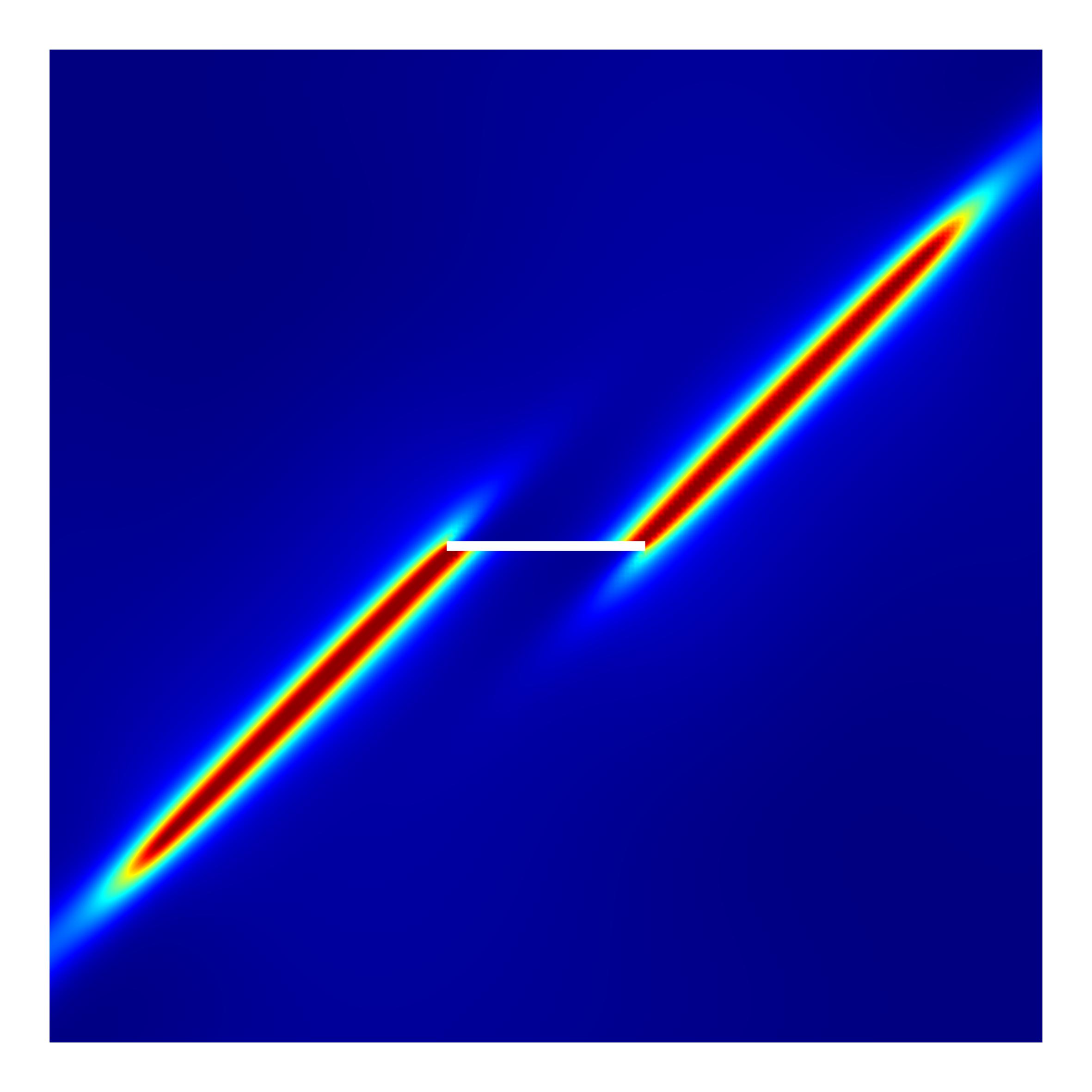}}
	\subfigure[$t=68.6$ s, $E_2/E_1=5$]{\includegraphics[width = 4.2cm]{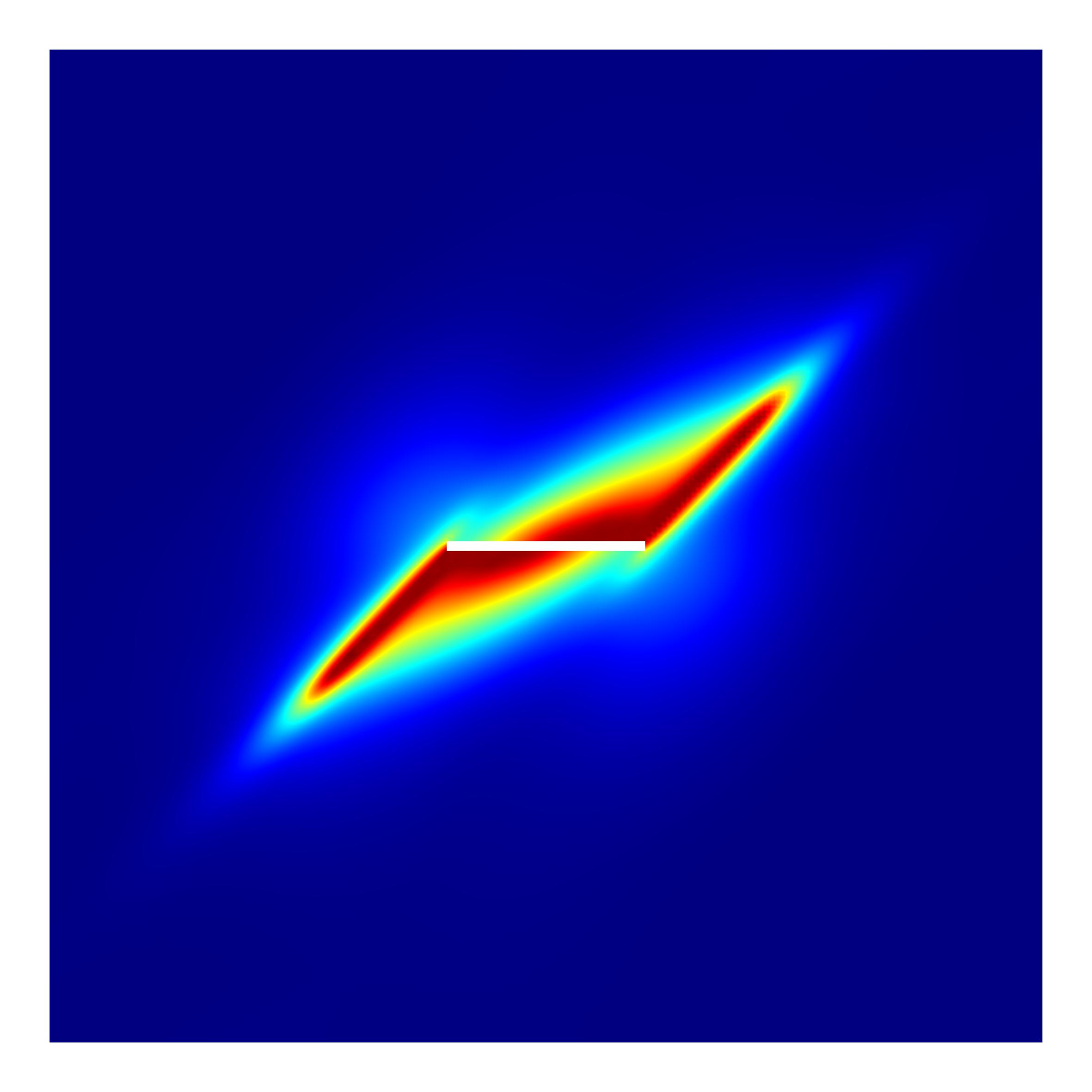}}
	\caption{Fracture evolution of the transversely isotropic porous medium with an interior notch under different $E_2/E_1$}
	\label{Fracture evolution of the transversely isotropic porous medium with an interior notch under different E_2E_1}
\end{figure}

We now test the influence of the difference between $E_1$ and $E_2$ on the fracture patterns and fluid pressure field. The modulus $E_1$ is then changed to 20, 50, 100, and 200 GPa while $E_2=100$ GPa is fixed and the other parameters are identical to those in Table \ref{Parameters for an isotropic specimen subjected to internal fluid pressure}. Figure \ref{Fracture evolution of the transversely isotropic porous medium with an interior notch under different E_2E_1} represents the fracture patterns of the transversely isotropic medium with an interior notch under different ratios of $E_2$ to $E_1$. It is observed from this figure that the fracture pattern is slightly affected by $E_2/E_1$ and most of them are similar to Fig. \ref{Fracture evolution of the transversely isotropic porous medium with an interior notch under different beta}d. However, for $E_2/E_1=5$, more fractured regions are found around the pre-existing notch due to the drastic decrease in the modulus $E_1$ along the material direction $\bm e_1$.

\begin{figure}[htbp]
	\centering
	\subfigure[$t=141$ s, $E_2/E_1=0.5$]{\includegraphics[width = 4.2cm]{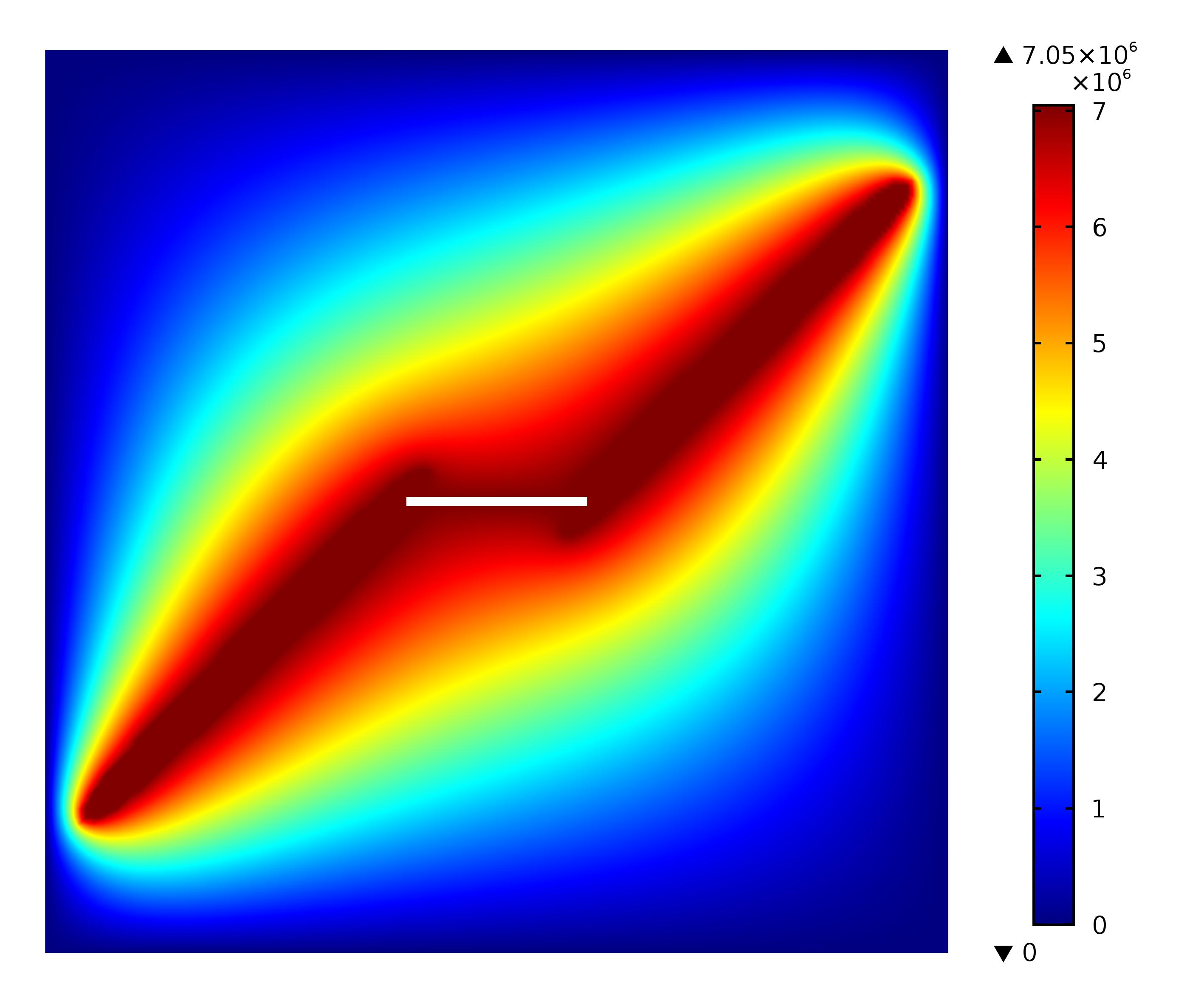}}
	\subfigure[$t=121.79$ s, $E_2/E_1=1$]{\includegraphics[width = 4.2cm]{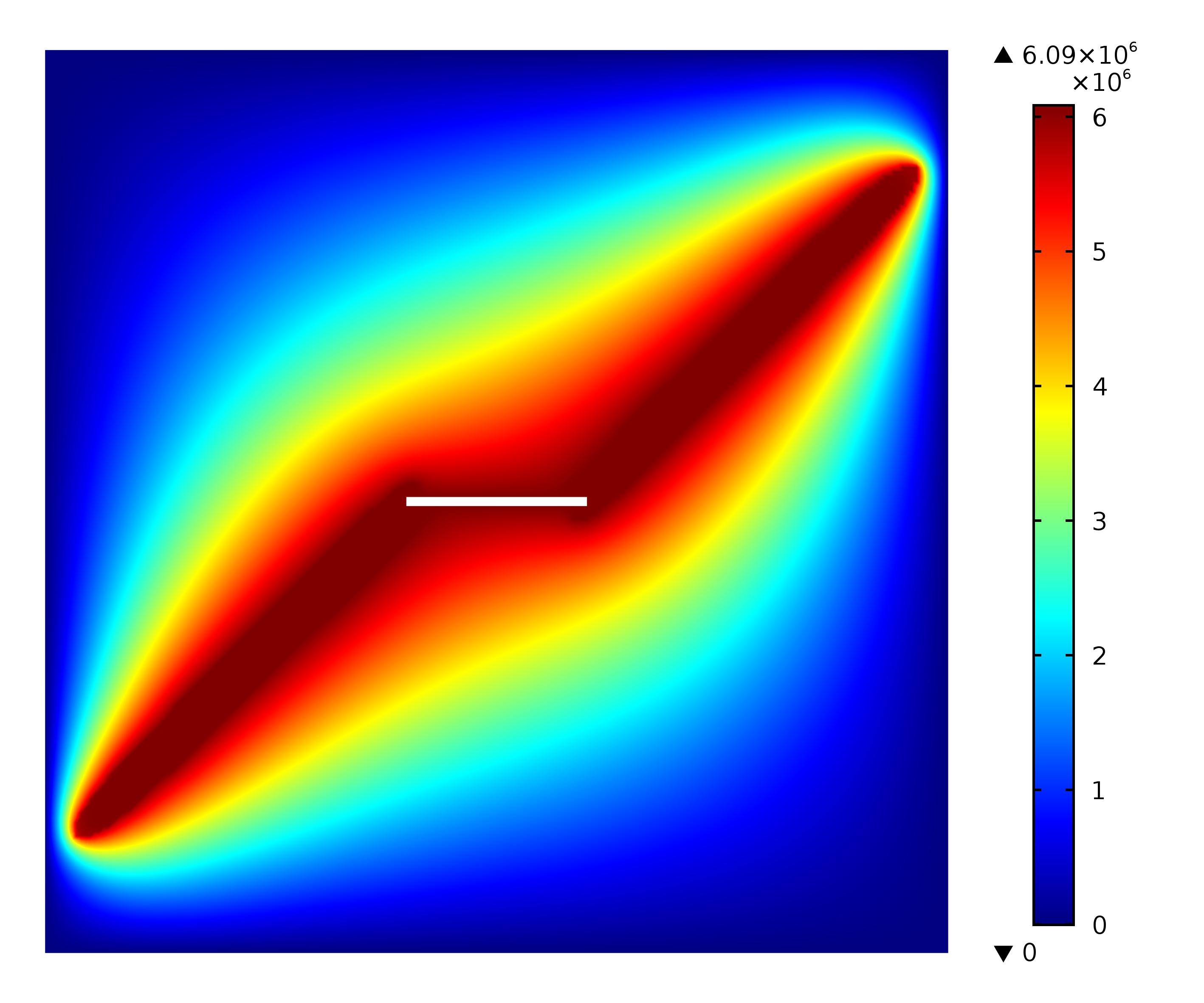}}
	\subfigure[$t=101.3$ s, $E_2/E_1=2$]{\includegraphics[width = 4.2cm]{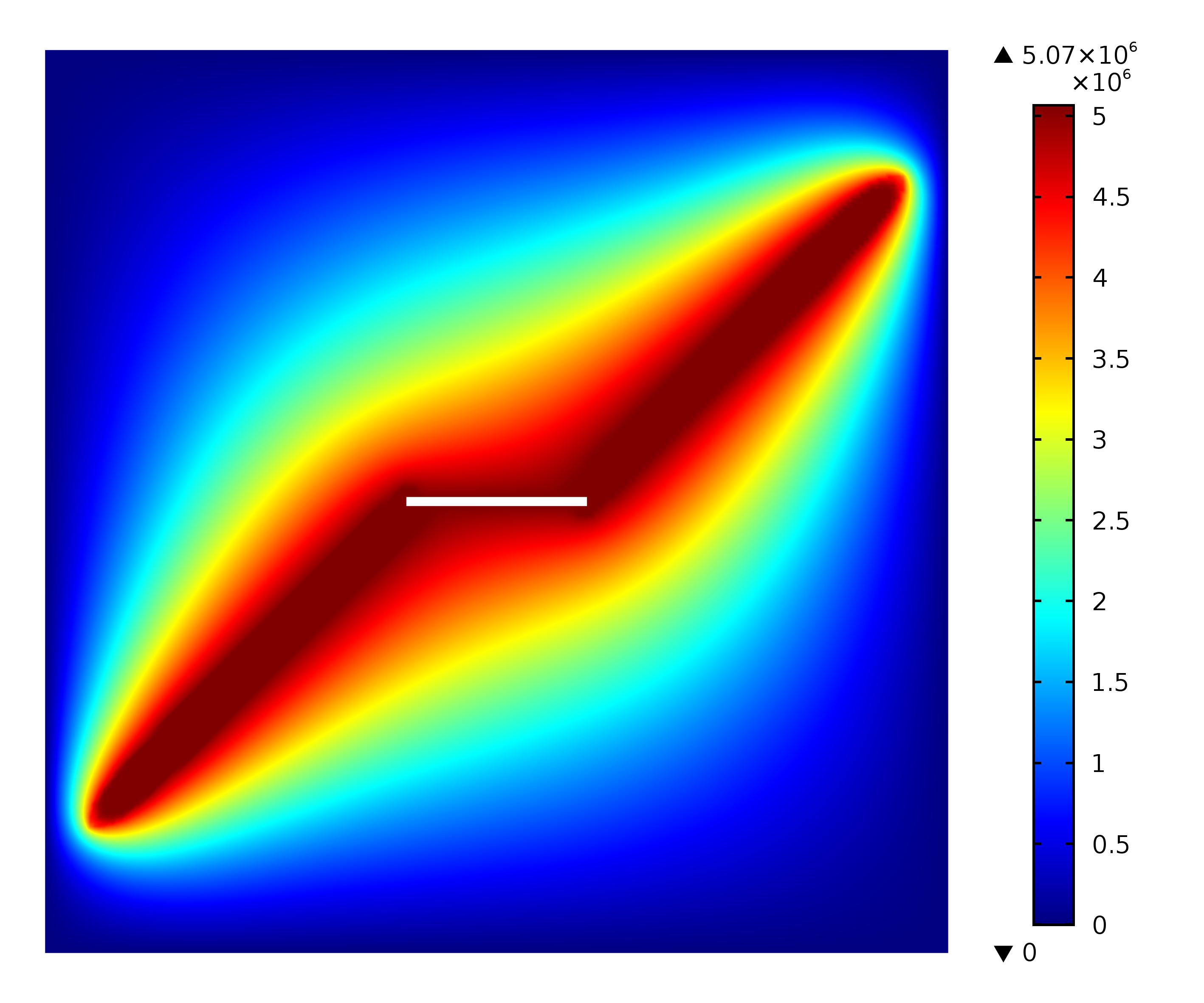}}
	\subfigure[$t=68.6$ s, $E_2/E_1=5$]{\includegraphics[width = 4.2cm]{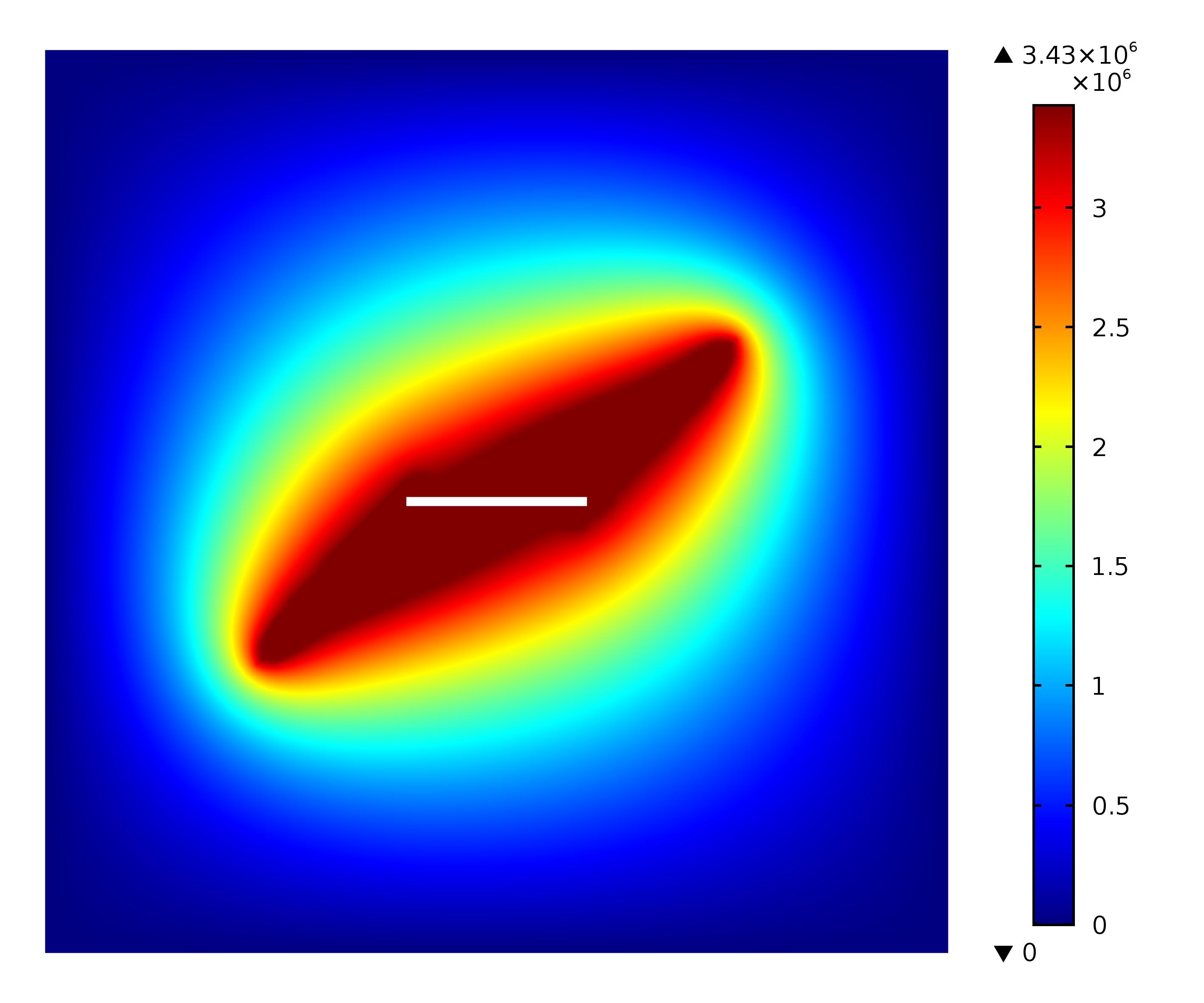}}
	\caption{Fluid pressure field of the transversely isotropic porous medium with an interior notch under different $E_2/E_1$ (unit: Pa)}
	\label{Fluid pressure field of the transversely isotropic porous medium with an interior notch under different E_2E_1}
\end{figure}

\begin{figure}[htbp]
	\centering
	\includegraphics[width = 10cm]{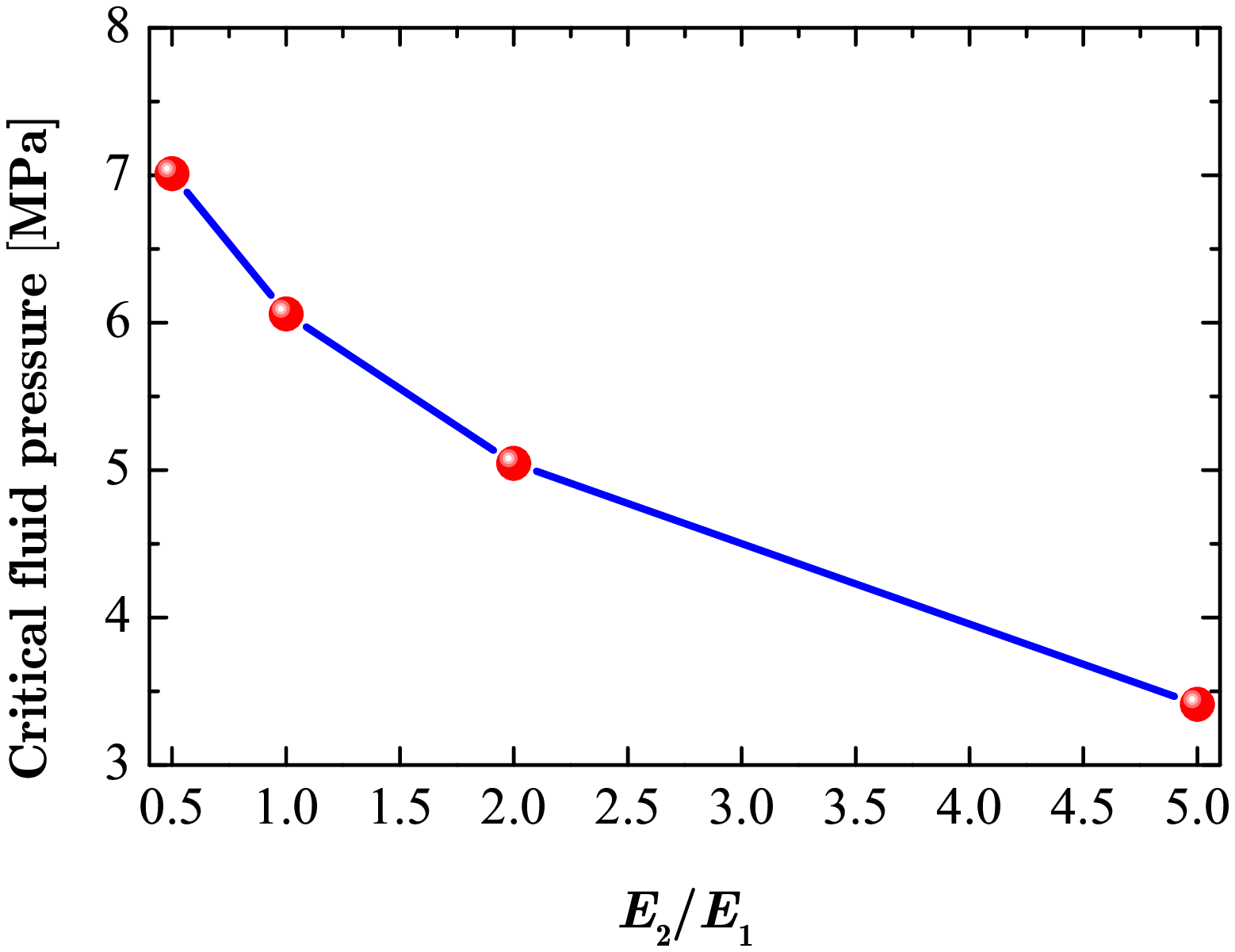}
	\caption{Critical fluid pressure for fracture initiation under different $E_2/E_1$}
	\label{Critical fluid pressure for fracture initiation under different E_2/E_1}
\end{figure}

Figure \ref{Fluid pressure field of the transversely isotropic porous medium with an interior notch under different E_2E_1} shows the fluid pressure field for the transversely isotropic medium with an interior notch under different $E_2/E_1$. As expected, the fluid pressure field is consistent with the phase field as shown in Fig. \ref{Fracture evolution of the transversely isotropic porous medium with an interior notch under different E_2E_1} and it has a maximum value in the fracture domain. However, the pressure field is more diffused around the notch for $E_2/E_1=5$ because in this case the fracture domain is much wider than in the other cases. In addition, Fig. \ref{Critical fluid pressure for fracture initiation under different E_2/E_1} presents the influence of the ratio $E_2/E_1$ on the critical fluid pressure for fracture initiation. As observed, the critical fluid pressure decreases with the increase in $E_2/E_1$ due to the fact that the transversely isotropic medium gradually loses its resistance to deformation if this ratio increases.

\subsubsection{Influence of permeability anisotropy}

\begin{figure}[htbp]
	\centering
	\subfigure[$r_{21}=0.1$]{\includegraphics[width = 5cm]{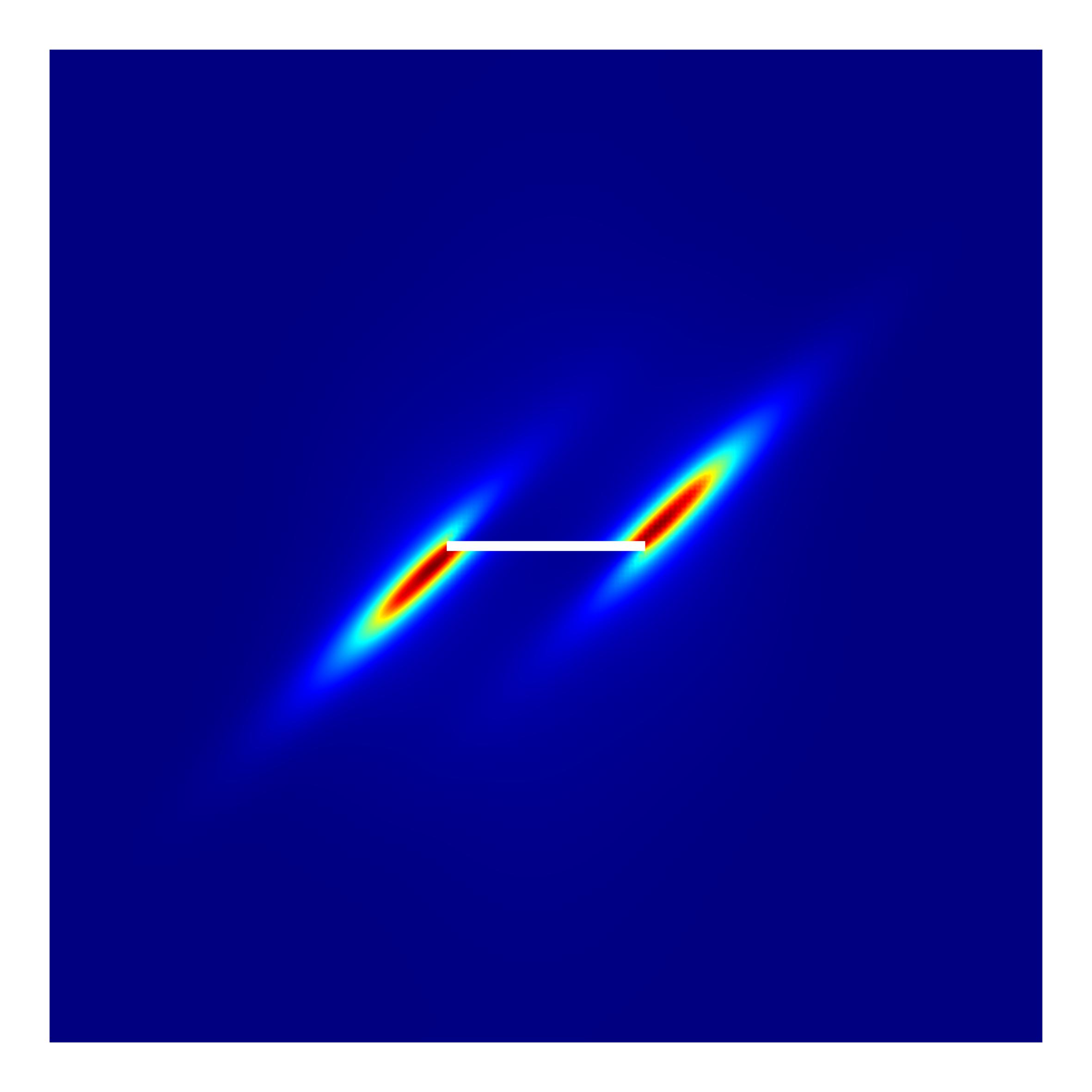}}
	\subfigure[$r_{21}=1$]{\includegraphics[width = 5cm]{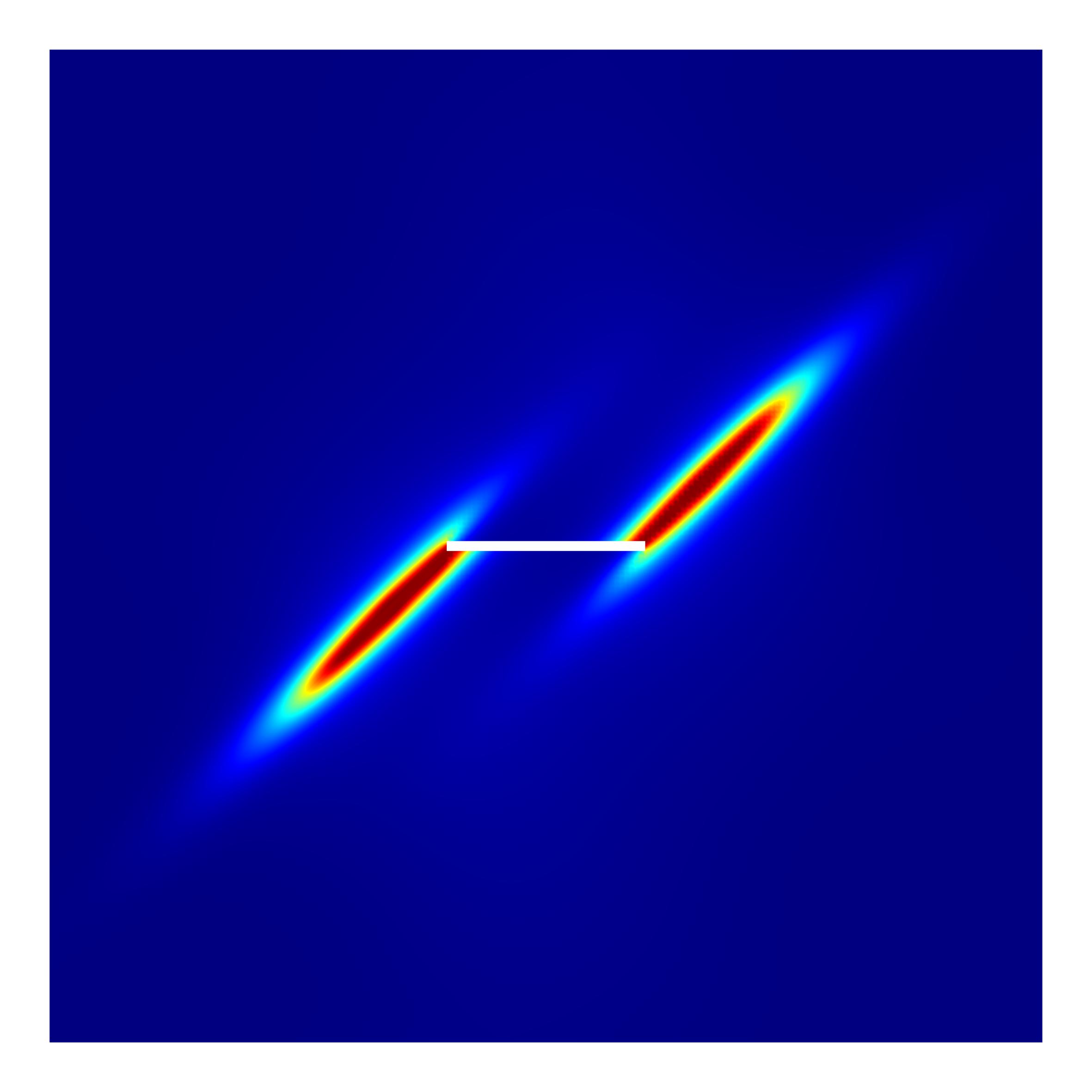}}
	\subfigure[$r_{21}=10$]{\includegraphics[width = 5cm]{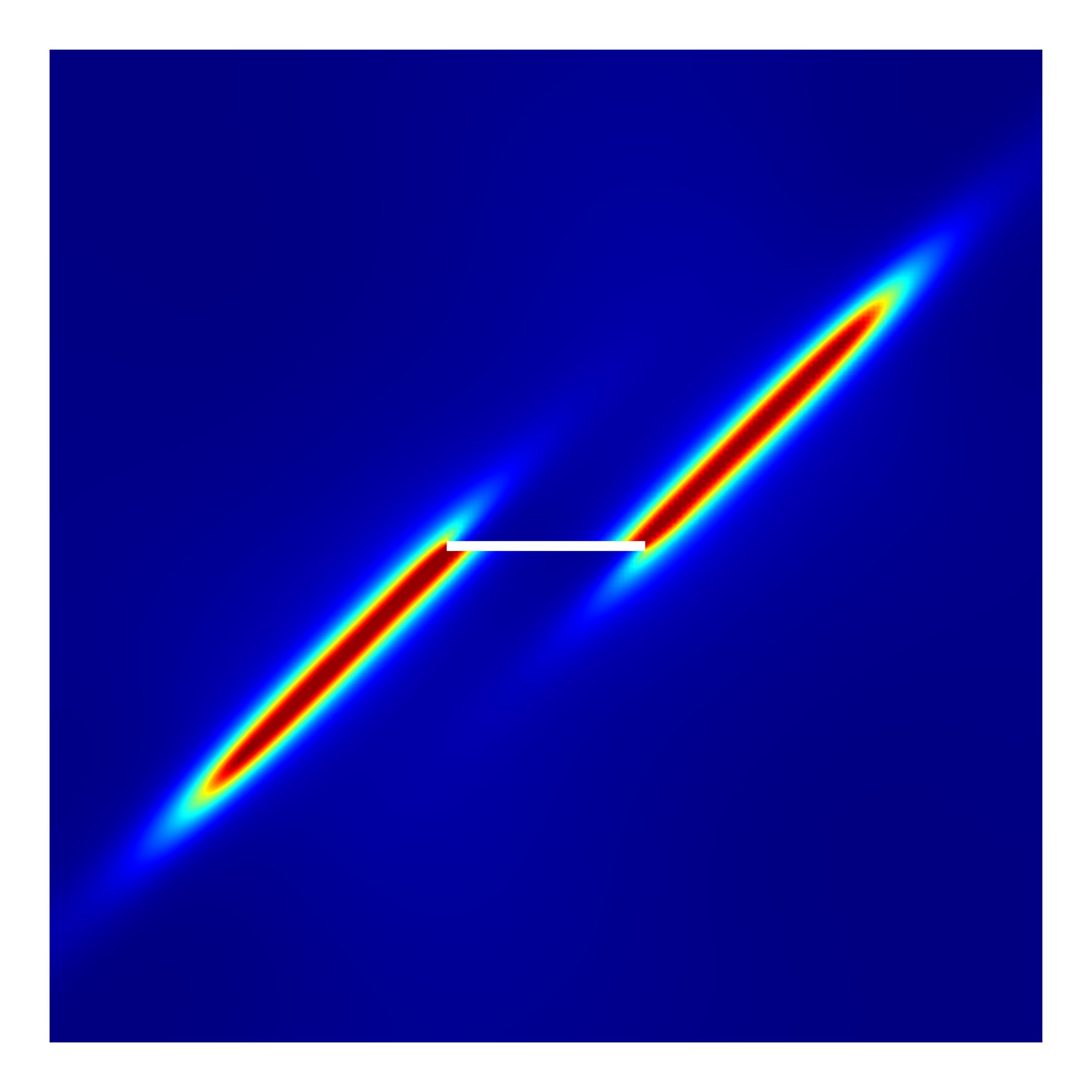}}
	\caption{Fracture evolution of the transversely isotropic porous medium with an interior notch under different $r_{21}$ (unit: Pa)}
	\label{Fracture evolution of the transversely isotropic porous medium with an interior notch under different r_21}
\end{figure}

The influence of the difference between the permeabilities along the rotation axis $\bm e_1$ and in the transverse plane is also tested; we therefore define a ratio $r_{21}=k_{r2}/k_{r1}=k_{f2}/k_{f1}$ accordingly to show this influence. It should be noted only $k_{r2}$ and $k_{f2}$ are changed correspondingly in this investigation while $k_{r1}$, $k_{f1}$ and the other parameters are the same as those in Table \ref{Parameters for an isotropic specimen subjected to internal fluid pressure}. Figure \ref{Fracture evolution of the transversely isotropic porous medium with an interior notch under different r_21} represents the fracture patterns of the transversely isotropic medium with an interior notch under different ratios of $k_{r2}$ to $k_{r1}$ and at a fixed time of $t=121.4$ s. As expected, the fracture pattern is not affected by the permeability anisotropy; however, it is observed that the decrease in the permeability in the transverse plane ($k_{r2}$ and $k_{f2}$) suppresses the fracture propagation along the direction $\bm e_2$. Therefore, the lowest $r_{21}$ achieves the smallest fracture length as shown in Fig. \ref{Fracture evolution of the transversely isotropic porous medium with an interior notch under different r_21}.

\begin{figure}[htbp]
	\centering
	\subfigure[$r_{21}=0.1$]{\includegraphics[width = 5cm]{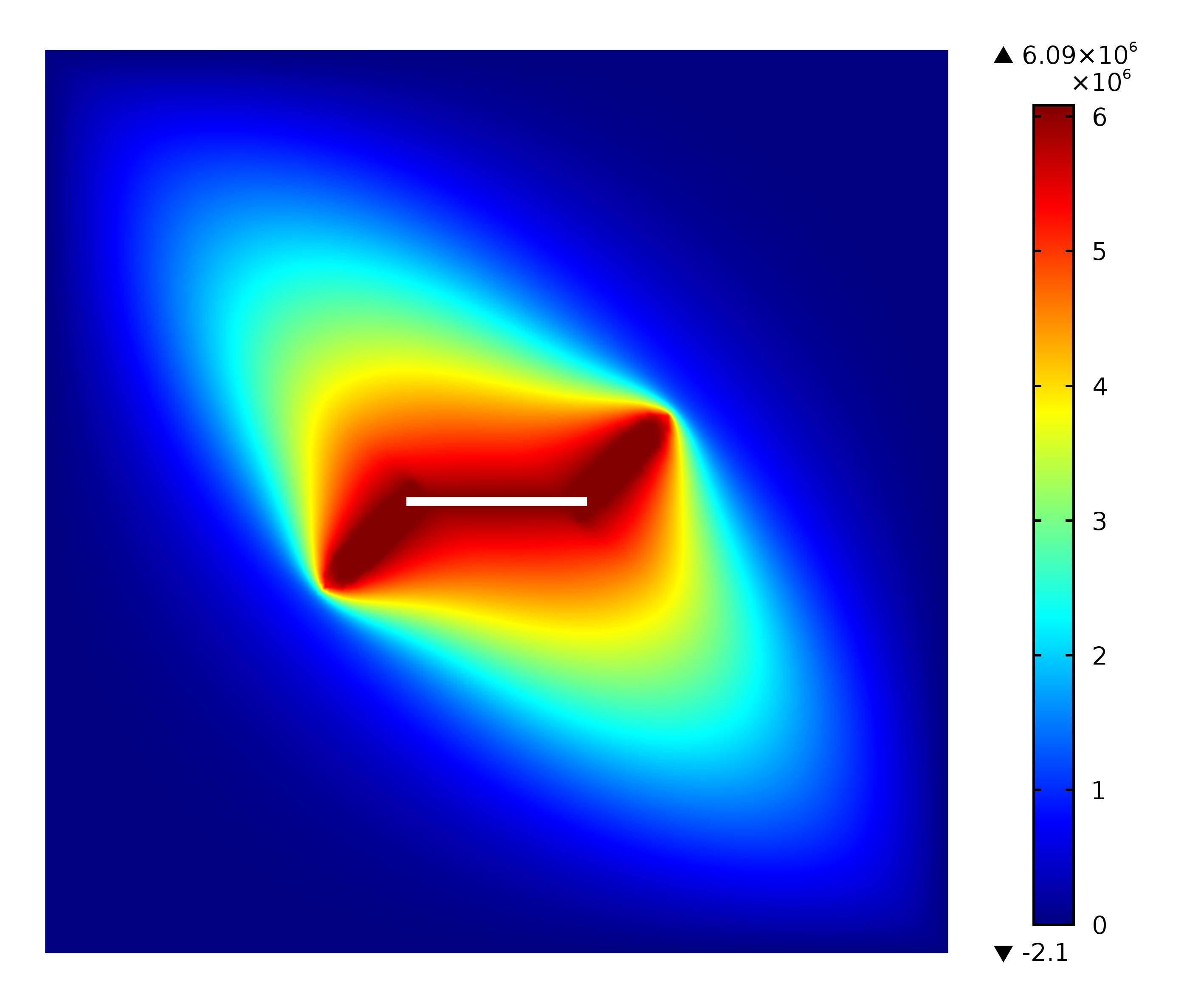}}
	\subfigure[$r_{21}=1$]{\includegraphics[width = 5cm]{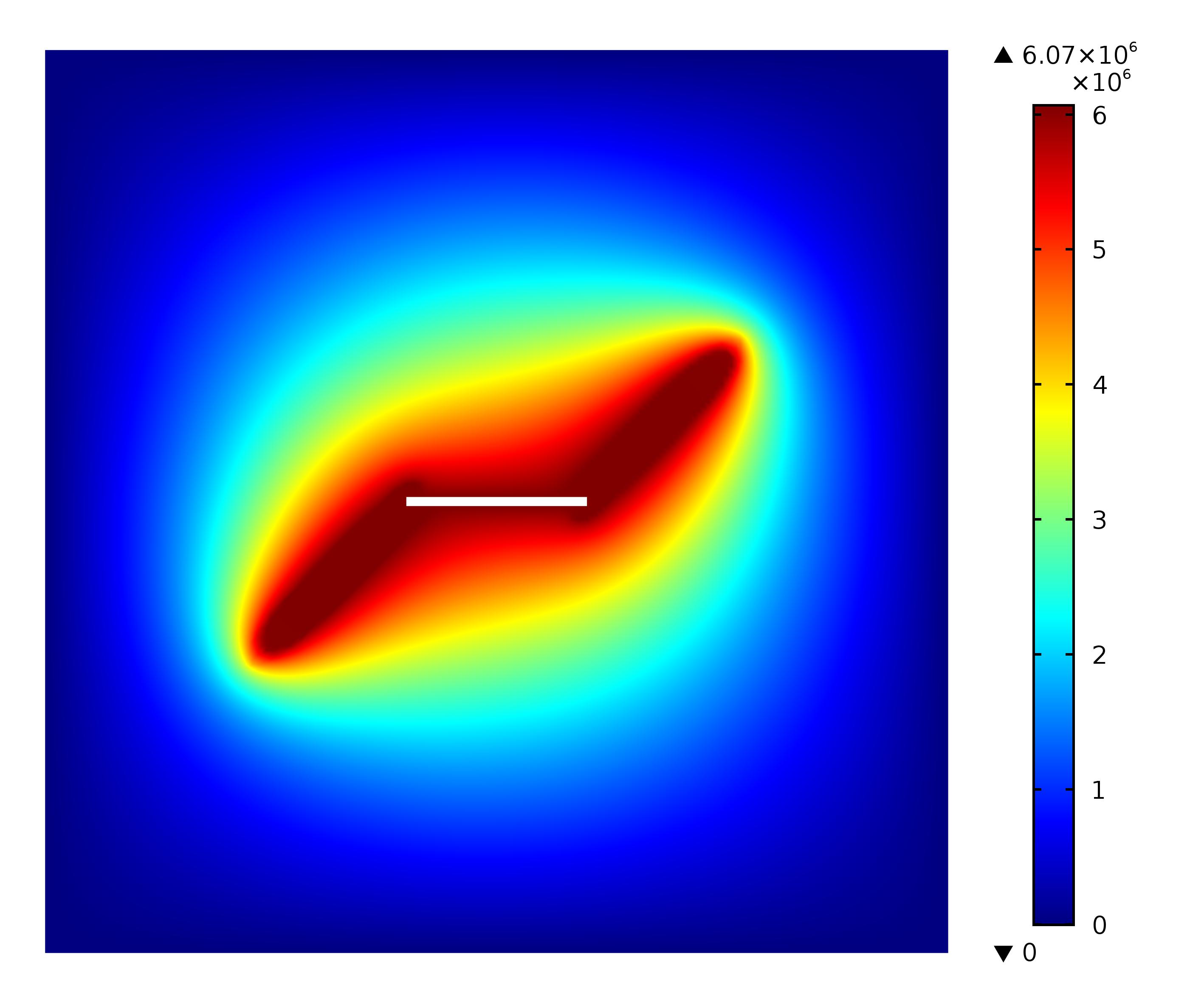}}
	\subfigure[$r_{21}=10$]{\includegraphics[width = 5cm]{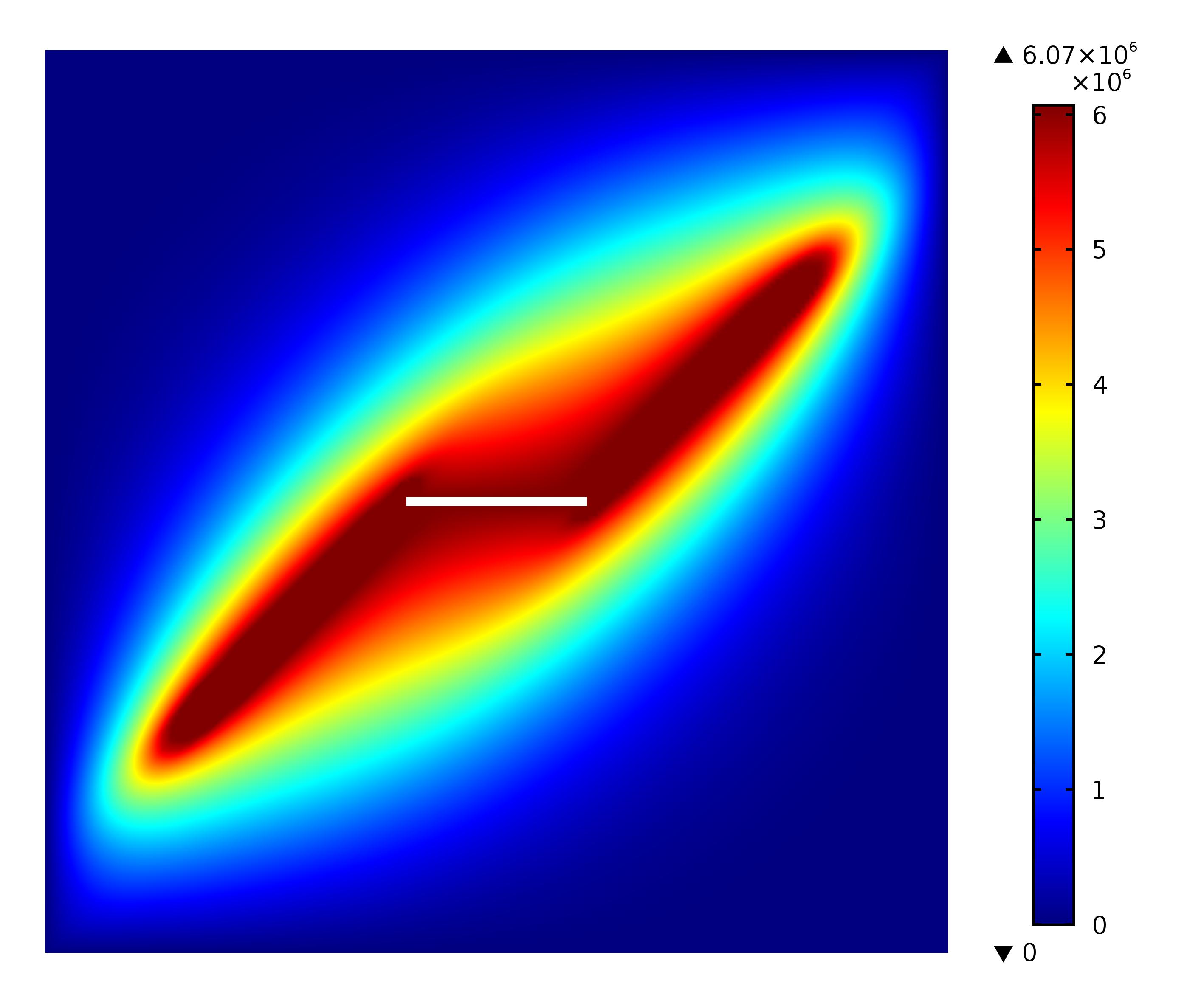}}
	\caption{Fluid pressure field of the transversely isotropic porous medium with an interior notch under different $r_{21}$}
	\label{Fluid pressure field of the transversely isotropic porous medium with an interior notch under different r_21}
\end{figure}

\begin{figure}[htbp]
	\centering
	\includegraphics[width = 10cm]{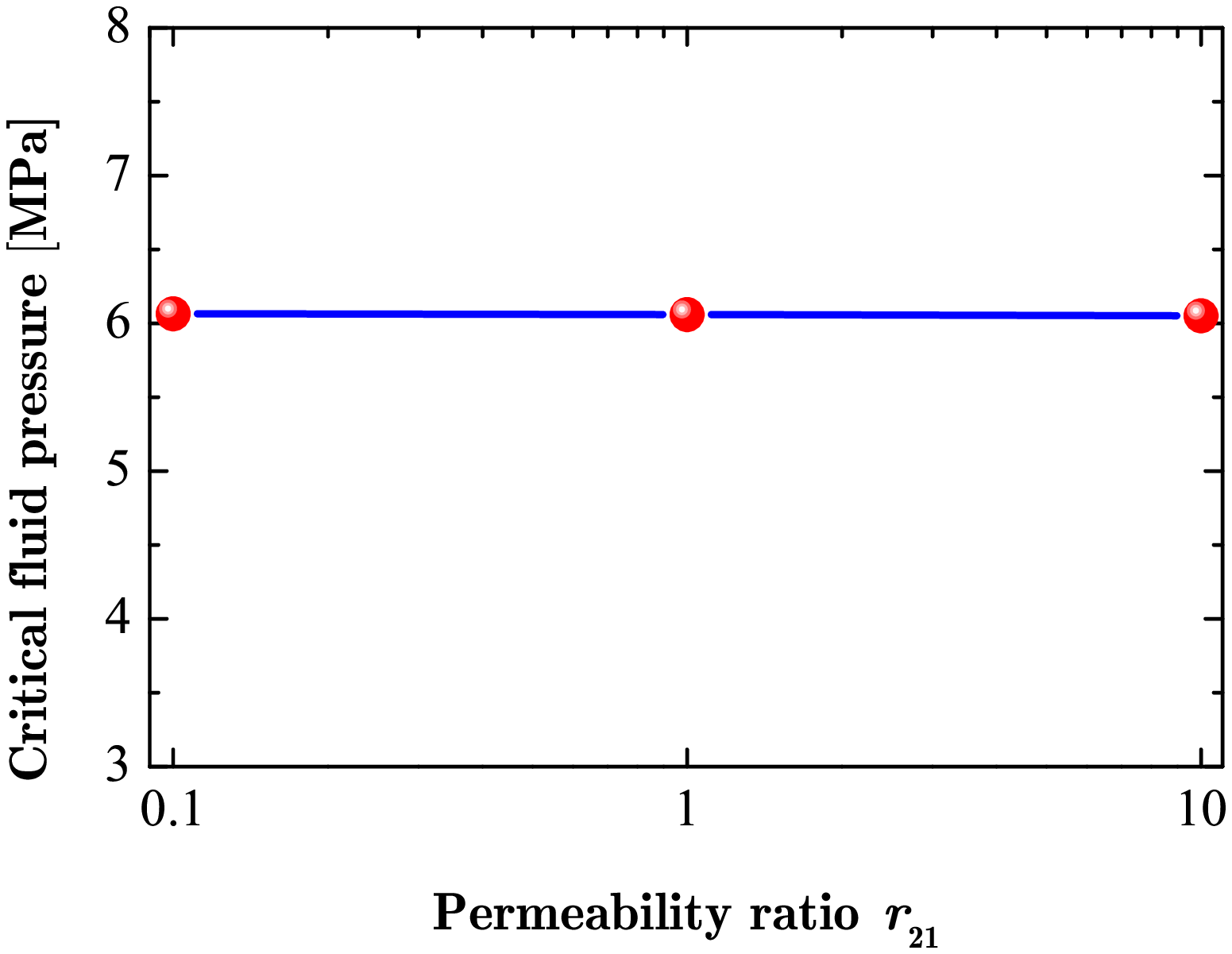}
	\caption{Critical fluid pressure for fracture initiation under different $r_{21}$}
	\label{Critical fluid pressure for fracture initiation under different r_21}
\end{figure}

Figure \ref{Fluid pressure field of the transversely isotropic porous medium with an interior notch under different r_21} shows the fluid pressure field for the transversely isotropic medium with an interior notch under different $r_{21}$. This figure indicates that the fluid pressure propagate along the $\bm e_1$ direction more easily with a lower $r_{21}$ due to a relatively higher permeability. In addition, Fig. \ref{Critical fluid pressure for fracture initiation under different r_21} presents the influence of the ratio $r_{21}$ on the critical fluid pressure for fracture initiation. It is inferred from this figure that the permeability anisotropy has a negligible effect on the critical fluid pressure. 

\subsection{2D medium with two parallel interior notches}

\begin{figure}[htbp]
	\centering
	\includegraphics[width = 8cm]{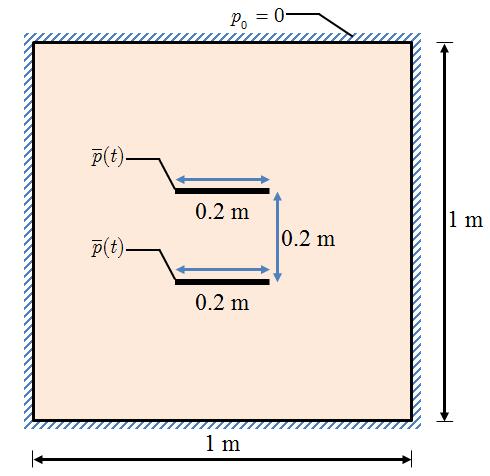}
	\caption{Geometry and boundary conditions for the 2D transversely isotropic porous medium with two parallel notches}
	\label{Geometry and boundary conditions for the 2D transversely isotropic porous medium with two parallel notches}
\end{figure}

In this example, we investigate the hydraulic fracture propagation in a 2D transversely isotropic medium with two parallel interior notches subjected to an increasing fluid pressure of $\bar{p}=5\times10^{4}$ $\mathrm{Pa/s}\cdot t$. The corresponding geometry and boundary condition are illustrated in \ref{Geometry and boundary conditions for the 2D transversely isotropic porous medium with two parallel notches}. In addition, the parameters used in this example are the same as those in Table \ref{Parameters for an isotropic specimen subjected to internal fluid pressure} while the critical energy release rate $G_{c2}$ being fixed as 50 N/m.

\begin{figure}[htbp]
	\centering
	\subfigure[$t=158.2$ s, $\beta=0^\circ$]{\includegraphics[width = 4.2cm]{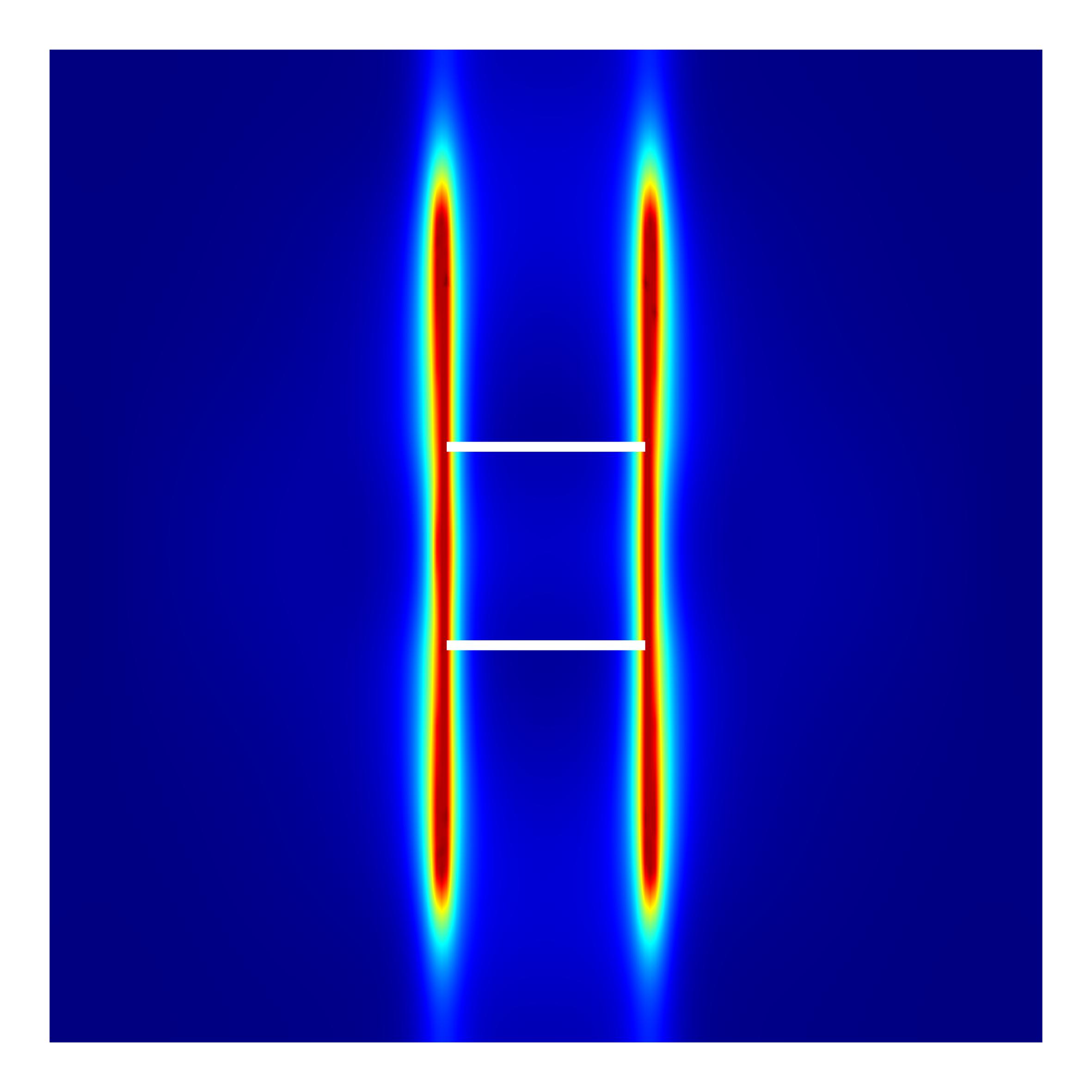}}
	\subfigure[$t=142.4$ s, $\beta=-45^\circ$]{\includegraphics[width = 4.2cm]{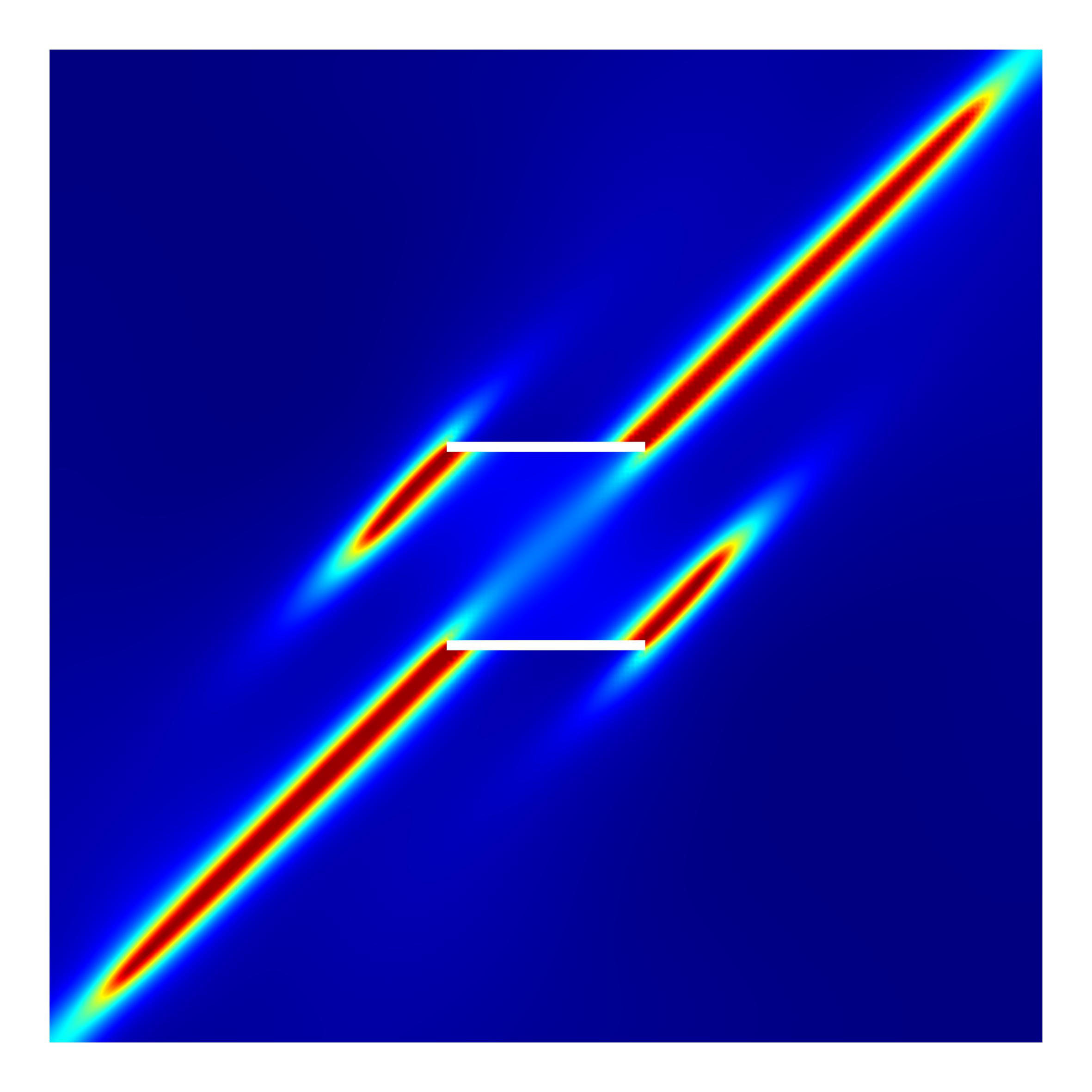}}
	\subfigure[$t=139.7$ s, $\beta=-90^\circ$]{\includegraphics[width = 4.2cm]{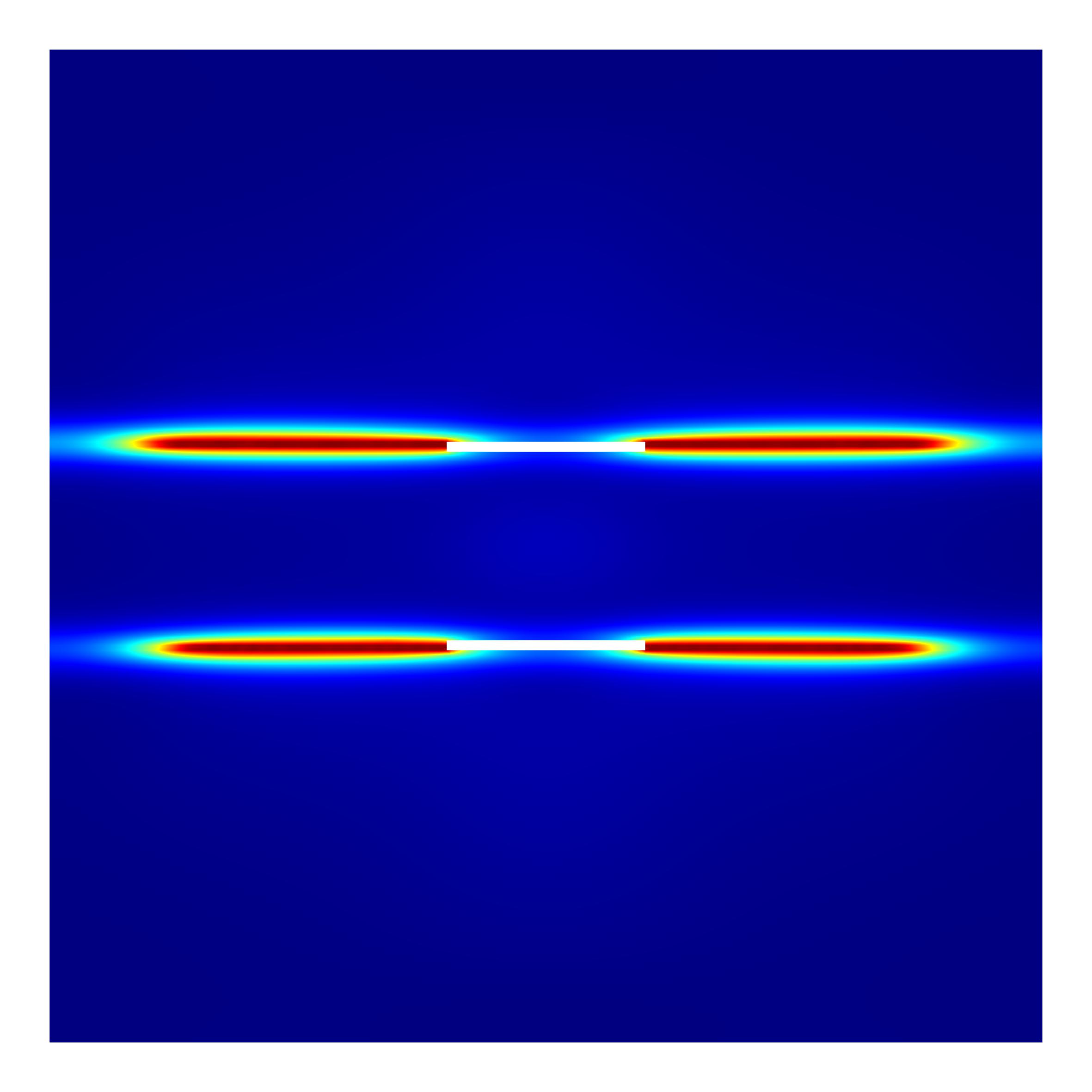}}
	\caption{Fracture evolution of the transversely isotropic porous medium with two parallel interior notches under different $\beta$}
	\label{Fracture evolution of the transversely isotropic porous medium with two parallel interior notches under different beta}
\end{figure}

The length and width of the notches are 0.2 m and 0.01 m and the notch spacing is set as 0.2 m. Regular Q4 elements are used to discretize the domain with an identical element size $h=5$ mm. In addition, the time step $\Delta t =0.02$ s is adopted for the phase field modeling. We test the fracture patterns under three rotation angles $\beta=0^\circ$, $-45^\circ$, and $-90^\circ$, and the simulated results are shown in Fig. \ref {Fracture evolution of the transversely isotropic porous medium with two parallel interior notches under different beta}. The fractures are observed to propagate in the transverse plane and perpendicular to the rotation axis (along $\bm e_2$). However, for $\beta=-45^\circ$, the lengths of fractures initiating from one notch are not the same due to the interaction between the fractures. Note that for $\beta=0^\circ$ and $-90^\circ$, newly initiated parallel fractures are observed and the stress-shadowing effect \citep{lee2016pressure}, which causes the two fractures to curve away from each other, is not observed. The main reason for this is the anisotropy in fracture toughness; the critical energy release rate is set as $G_{c1}=10$ N/m along the rotation axis and $G_{c2}=50$ N/m in the transverse plane. Therefore, the fracture perpendicular to the rotation axis has the highest priority, and the large difference between $G_{c1}$ and $G_{c2}$ impedes the deflection of the fracture in the transverse plane to the direction of the rotation axis.

\subsection{3D medium with a penny-shaped notch}
\subsubsection{Influence of rotation angle}\label{3D Influence of rotation angle}

The first 3D example is a 3D transversely isotropic medium with a penny-shaped notch where an internal fluid pressure $\bar{p}=5\times10^{4}$ $\mathrm{Pa/s}\cdot t$ is applied. By using this example, the capability of our proposed approach for 3D problems are proved. The dimension of the domain is $0.4\times0.4\times0.4$ m$^3$ and its center coincides with the notch center. More specifically, the notch is parallel to the $x-y$ plane and have a radius of 0.06 m and a height of 0.01 m. The same parameters in Table \ref{Parameters for an isotropic specimen subjected to internal fluid pressure} are used in this example except that $G_{c2}=500$ N/m is adopted. In addition, we set the length scale parameter to $l_0 = 5\times10^{-3}$ m. We use 6-node prism elements to discretize the 3D domain and the element size is controlled within $h=5\times10^{-3}$ m for reducing computational cost. On the other hand, the time increment is set as 0.1 s.
 
\begin{figure}[htbp]
	\centering
	$\beta=-15^\circ$\subfigure[$t=295$ s]{\includegraphics[width = 5cm]{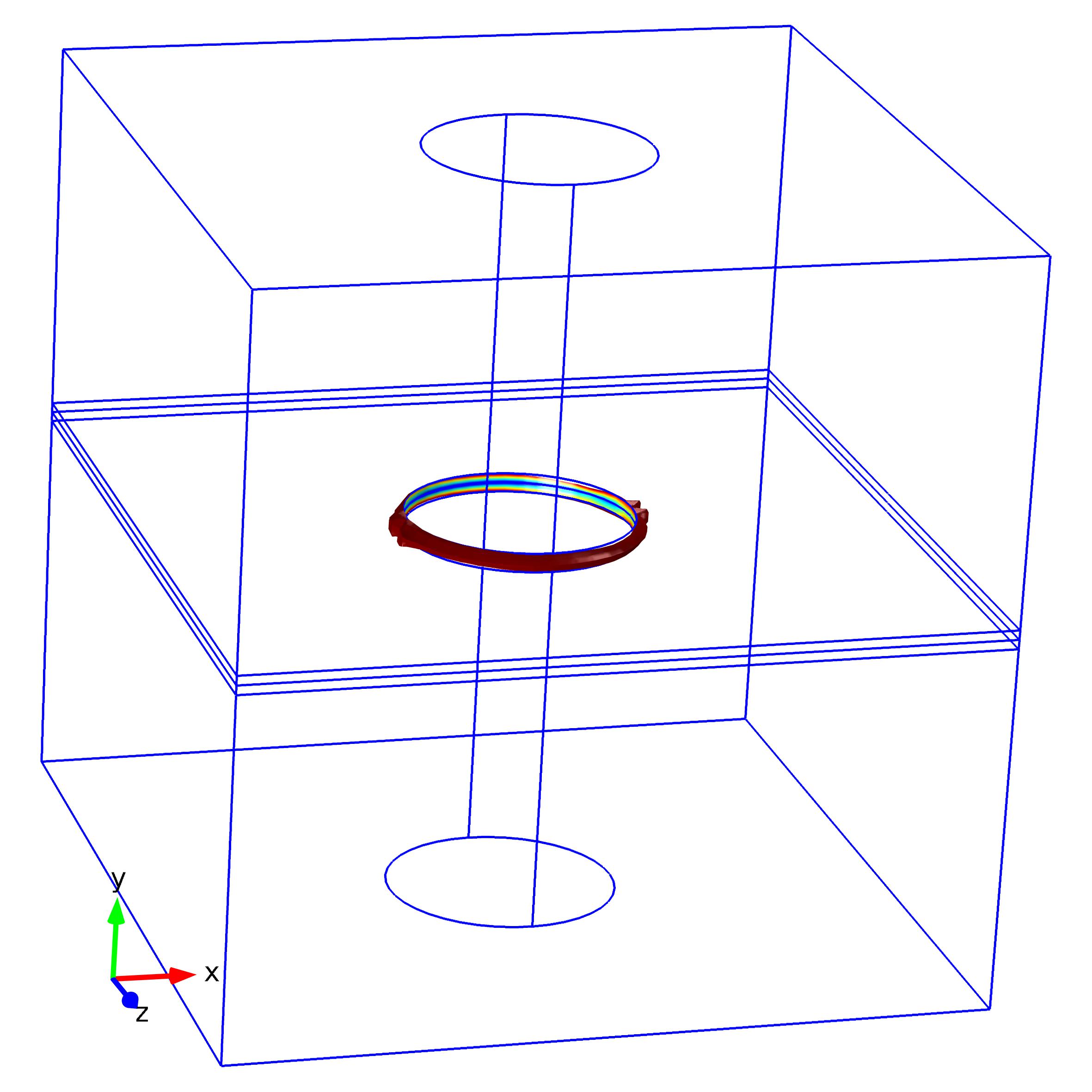}}
	\subfigure[$t=305$ s]{\includegraphics[width = 5cm]{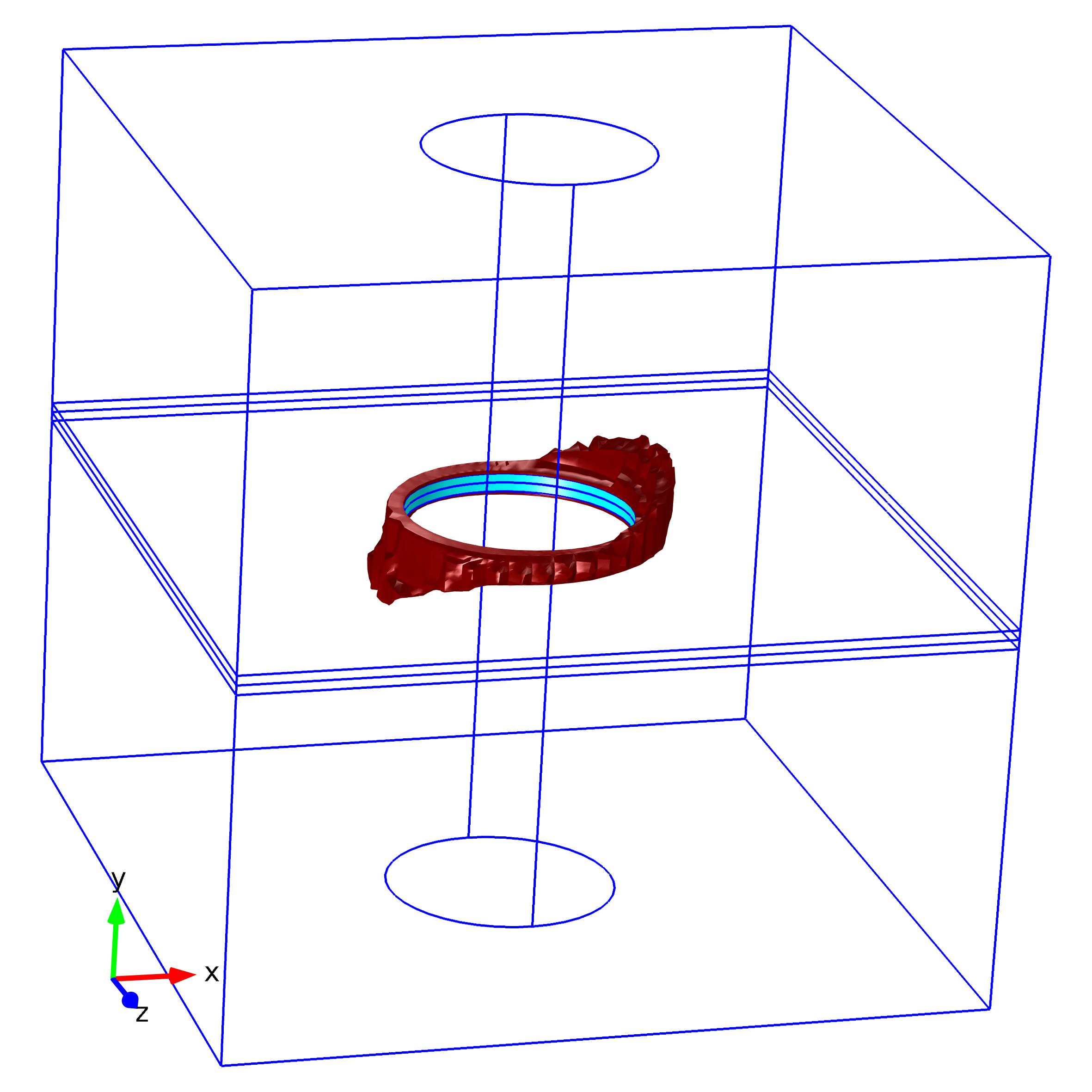}}
	\subfigure[$t=311.5$ s]{\includegraphics[width = 5cm]{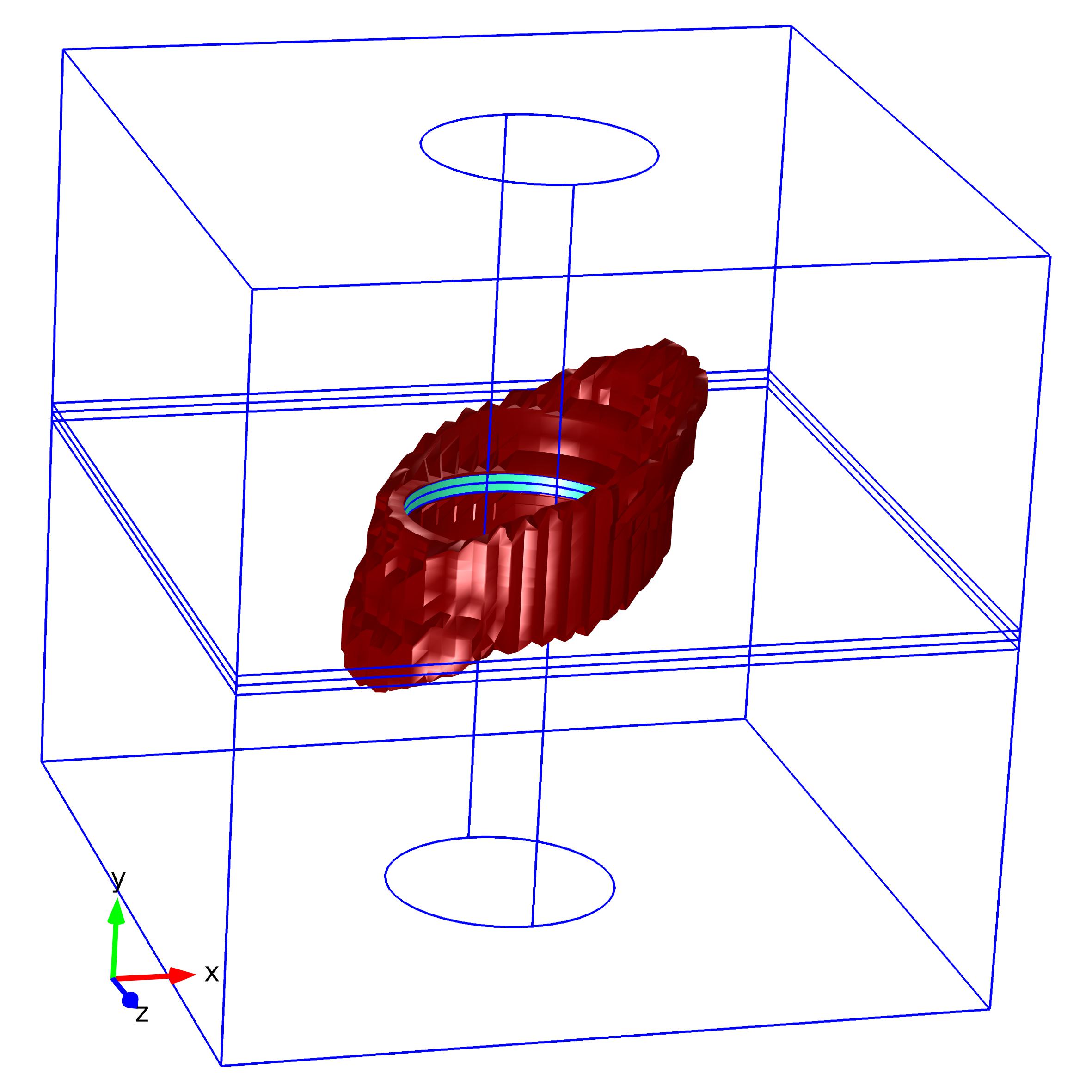}}\\
	$\beta=-45^\circ$\subfigure[$t=300$ s]{\includegraphics[width = 5cm]{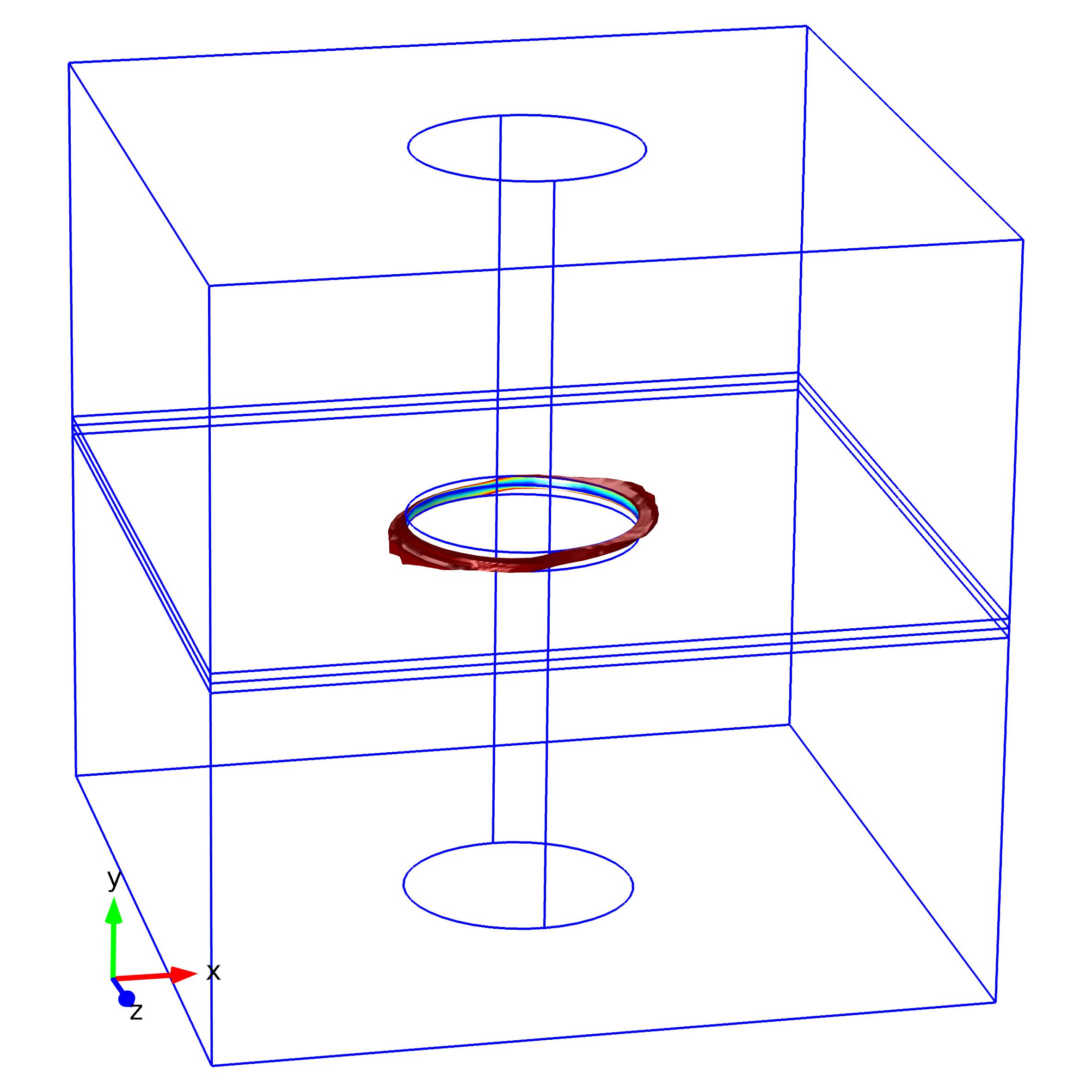}}
	\subfigure[$t=305$ s]{\includegraphics[width = 5cm]{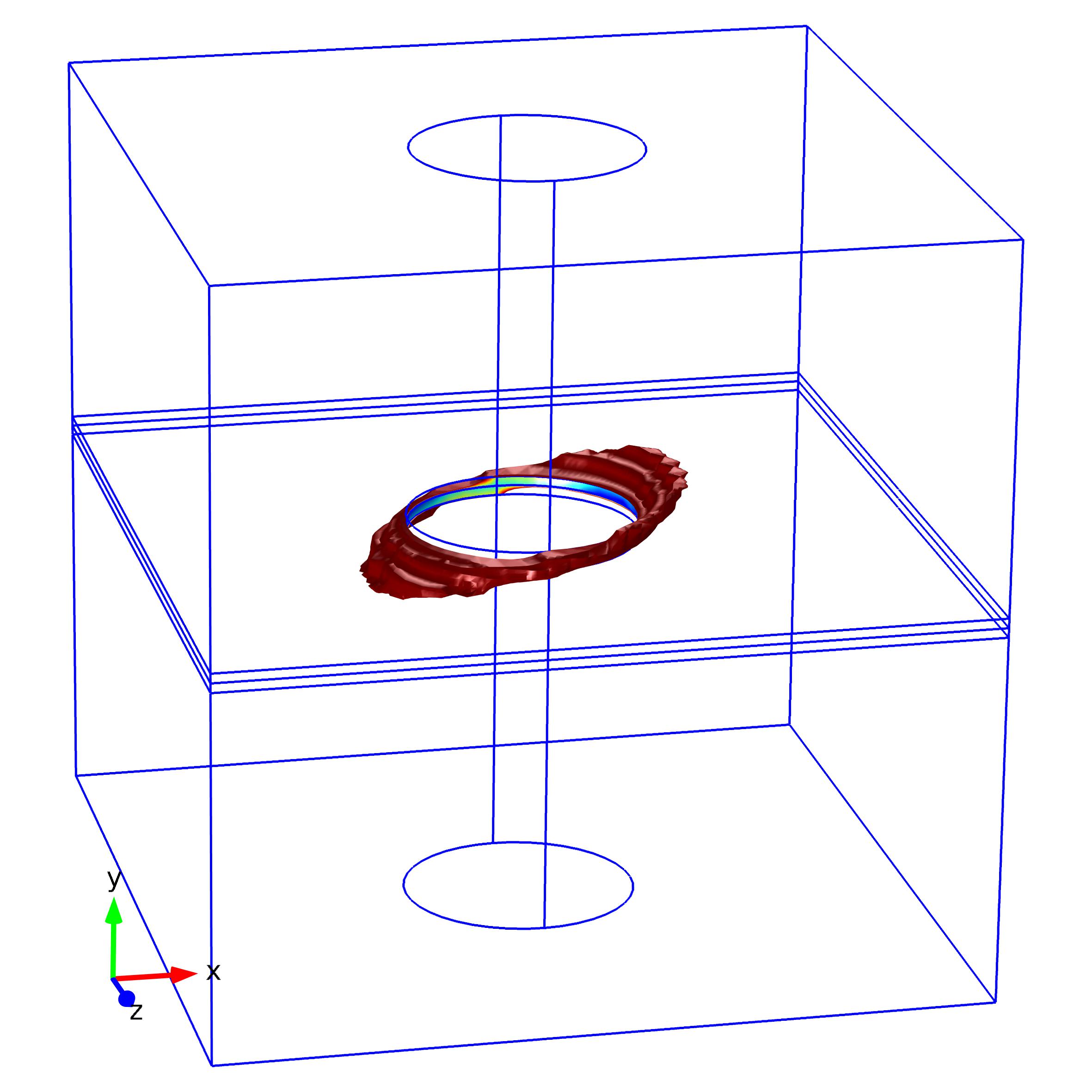}}
	\subfigure[$t=309$ s]{\includegraphics[width = 5cm]{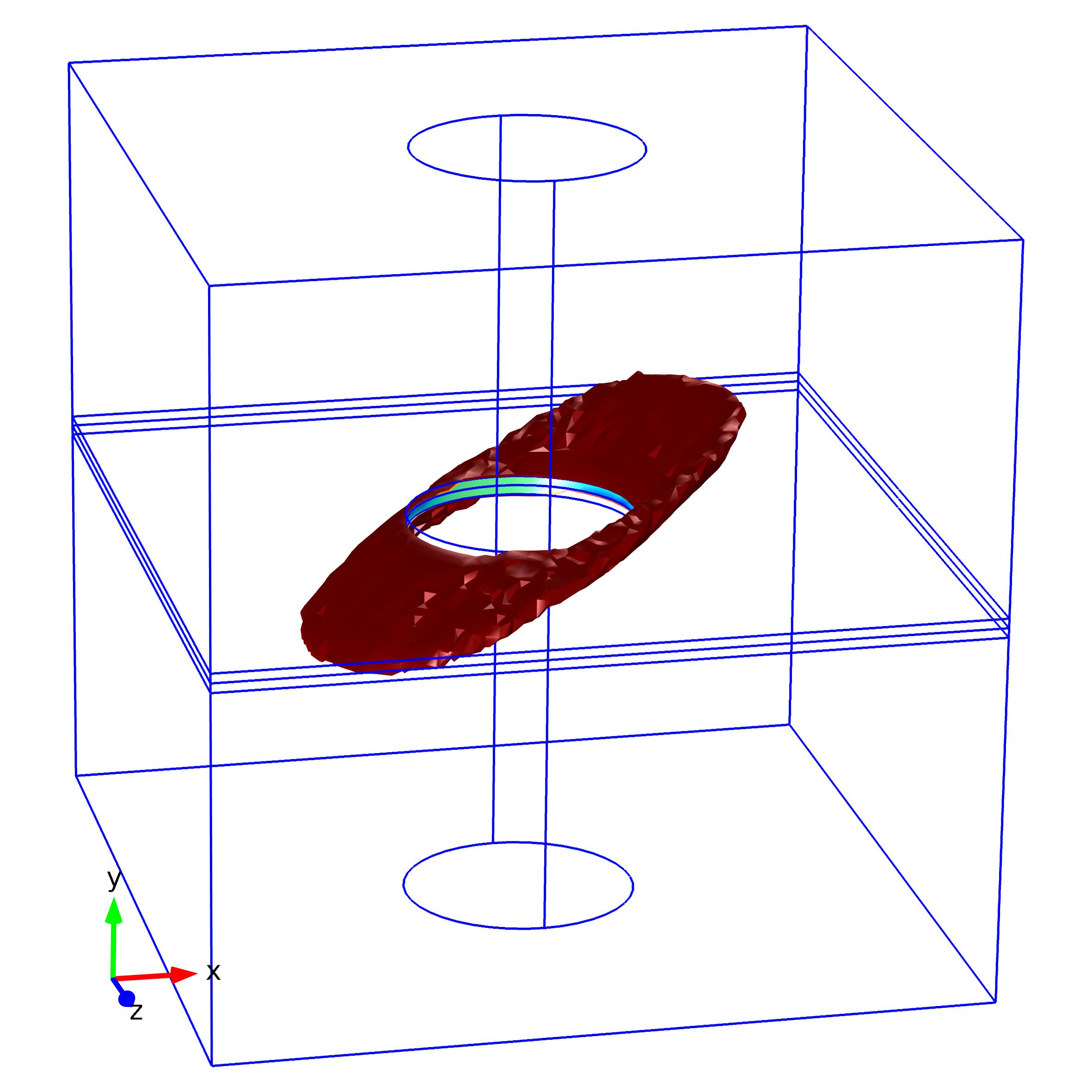}}\\
	$\beta=-75^\circ$\subfigure[$t=240$ s]{\includegraphics[width = 5cm]{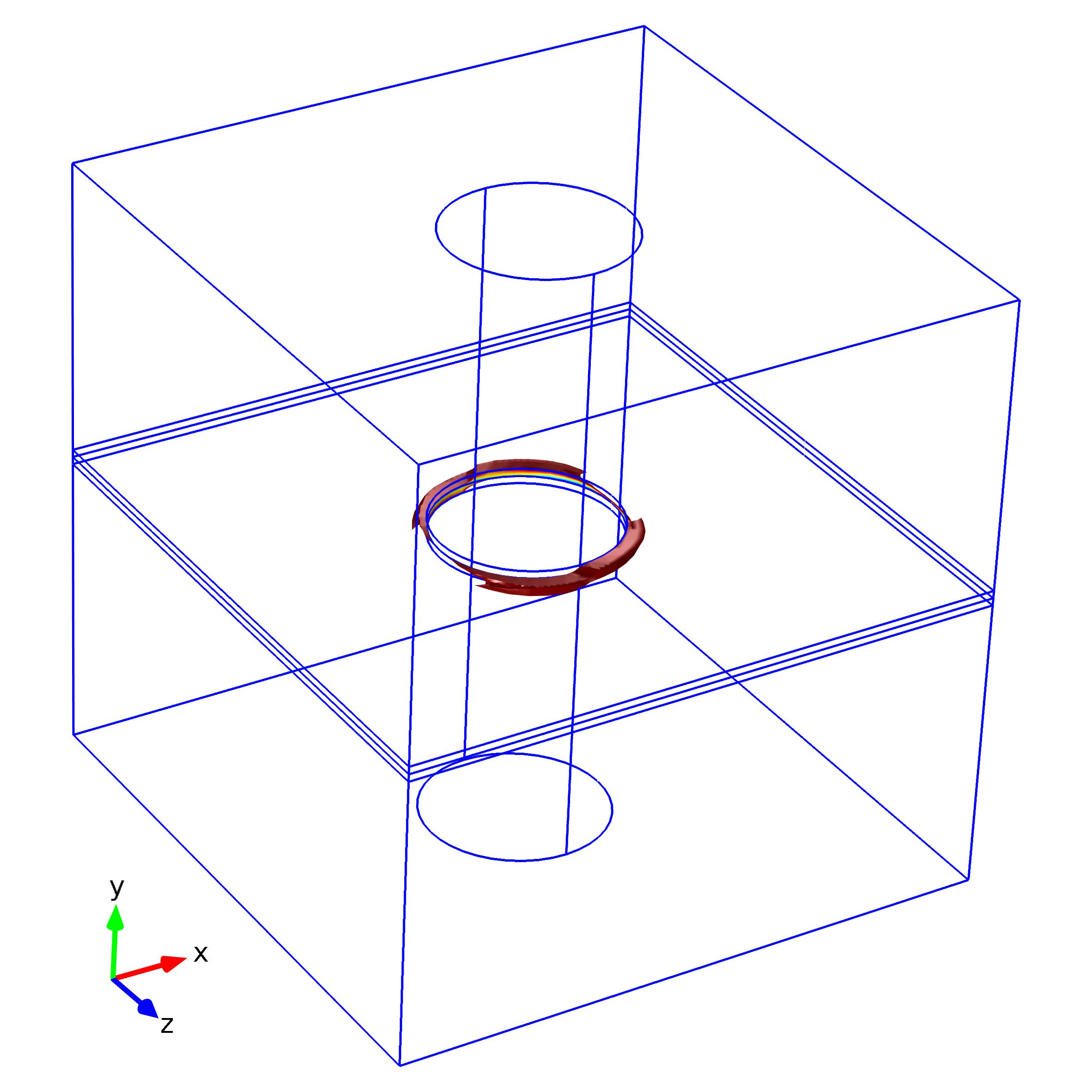}}
	\subfigure[$t=245$ s]{\includegraphics[width = 5cm]{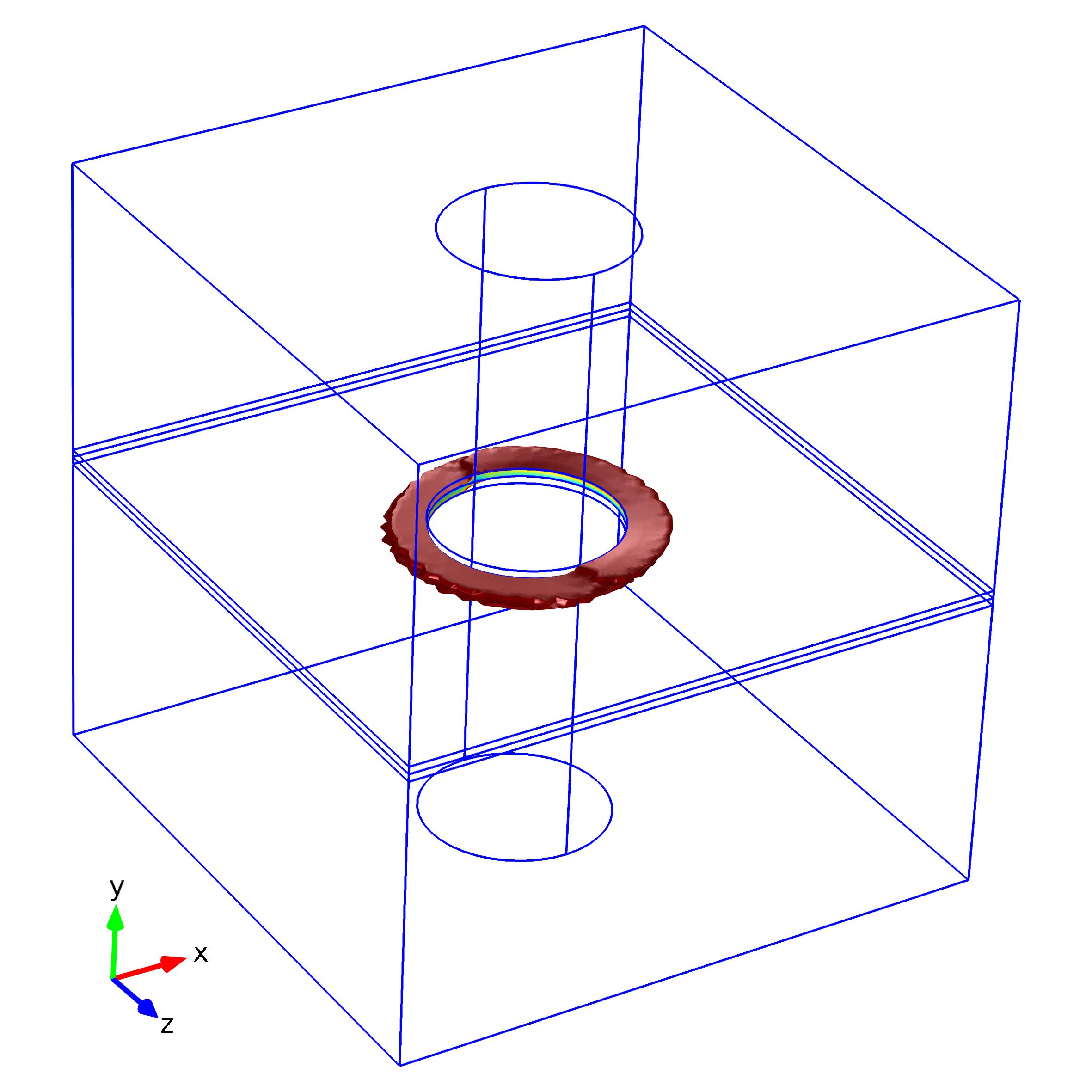}}
	\subfigure[$t=248.1$ s]{\includegraphics[width = 5cm]{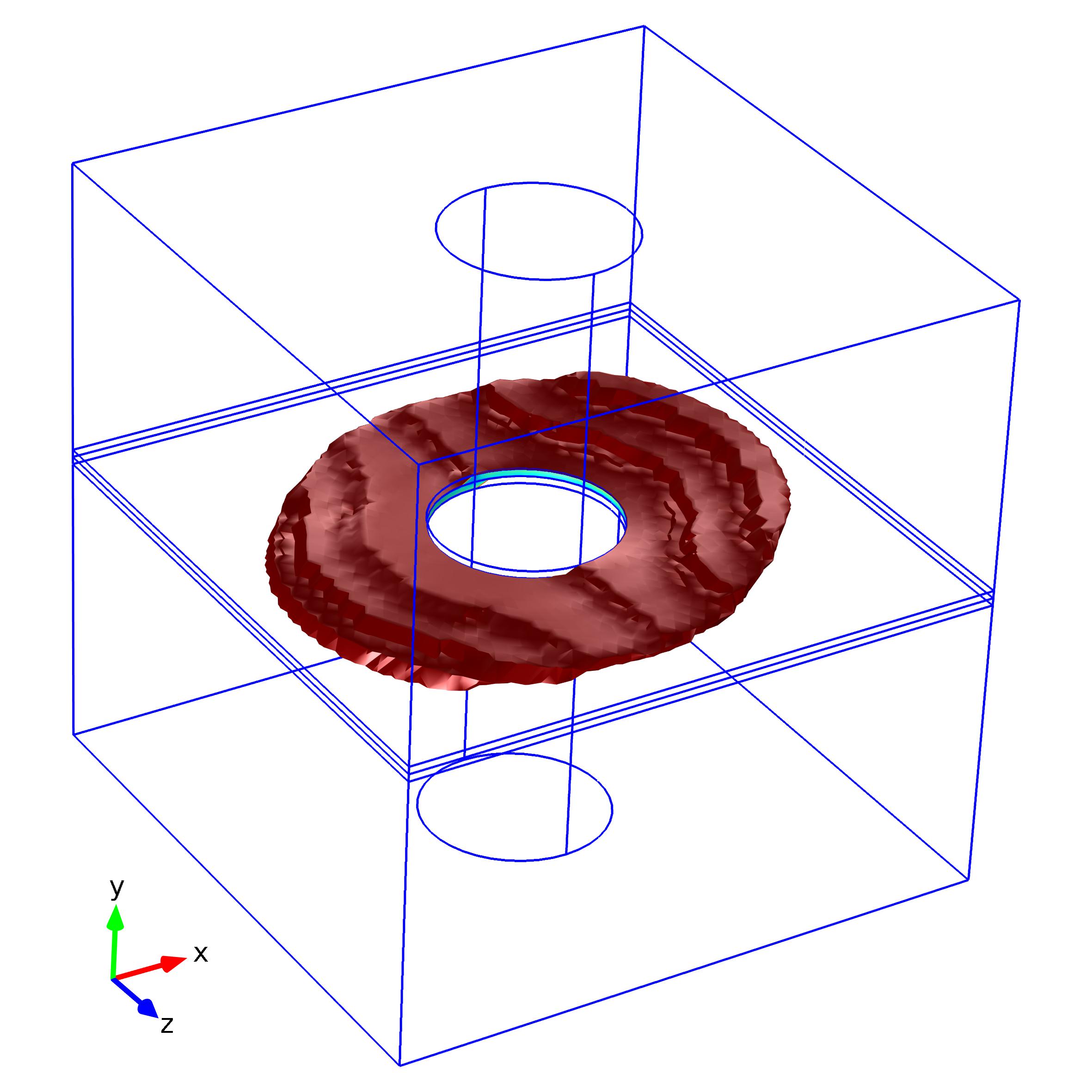}}\\		
	\caption{Fracture evolution of the 3D transversely isotropic porous medium with an interior notch under different $\beta$}
	\label{Fracture evolution of the 3D transversely isotropic porous medium with an interior notch under different beta}
\end{figure}

Hydraulic fracture propagation patterns in the 3D transversely isotropic medium are shown in Fig. \ref{Fracture evolution of the 3D transversely isotropic porous medium with an interior notch under different beta}. It should be noted that we totally test three rotation angles, namely, $\beta=-15^\circ$, $-45^\circ$, and $-75^\circ$. Figures \ref{Fracture evolution of the 3D transversely isotropic porous medium with an interior notch under different beta}a, b, and c indicate that for $\beta=-15^\circ$ the fractures initiate at $t=295$ s and propagate at $t=305$ s and $311.5$ s. In addition, as shown in Figs. \ref{Fracture evolution of the 3D transversely isotropic porous medium with an interior notch under different beta}d, e, and f, the fractures initiate at $t=300$ s for $\beta=-45^\circ$ and then propagate at $t=305$ s and $309$ s. Moreover, Figs. \ref{Fracture evolution of the 3D transversely isotropic porous medium with an interior notch under different beta}g, h, and i represent the fracture evolution for $\beta=-75^\circ$.

It can also be seen from Fig. \ref{Fracture evolution of the 3D transversely isotropic porous medium with an interior notch under different beta} that the hydraulic fractures are perpendicular to the rotation axis of the transversely isotropic medium and the projected area of the fracture domain to the $x-y$ plane increases as the rotation angle between the rotation axis $\bm e_1$ and $x-$axis $\beta$ increases. It should be noted that in the 3D example, only a coarse mesh is adopted without any re-meshing and adaptive techniques. This fully reflects the strong capability of our proposed phase field model for simulating complex hydraulic fracture propagation in transversely isotropic porous media.

\subsubsection{Influence of fracture toughness anisotropy}

The second 3D example is similar to that in Subsection \ref{3D Influence of rotation angle} while the rotation angle $\beta$ is fixed to $-45^\circ$. Therefore, the influence of fracture toughness anisotropy on 3D hydraulic fracture propagation can be investigated. The dimension of the analysis domain is changed to $4\times4\times4$ m$^3$ for a better comparison with \citet{lee2017iterative} and the initial notch has a radius of 0.6 m and a height of 0.1 m. In addition, an internal fluid pressure $\bar{p}=1\times10^{3}$ $\mathrm{Pa/s}\cdot t$ is applied on the notch surfaces. The same parameters as those in the 3D example of \citet{lee2017iterative} are applied except that the element size is controlled within $h=5\times10^{-2}$ m (6-node prism elements are also used to discretize the 3D domain).
	
\begin{figure}[htbp]
	\centering
	\subfigure[$G_{c2}=10$N/m, $t=31.1$ s]{\includegraphics[width = 5cm]{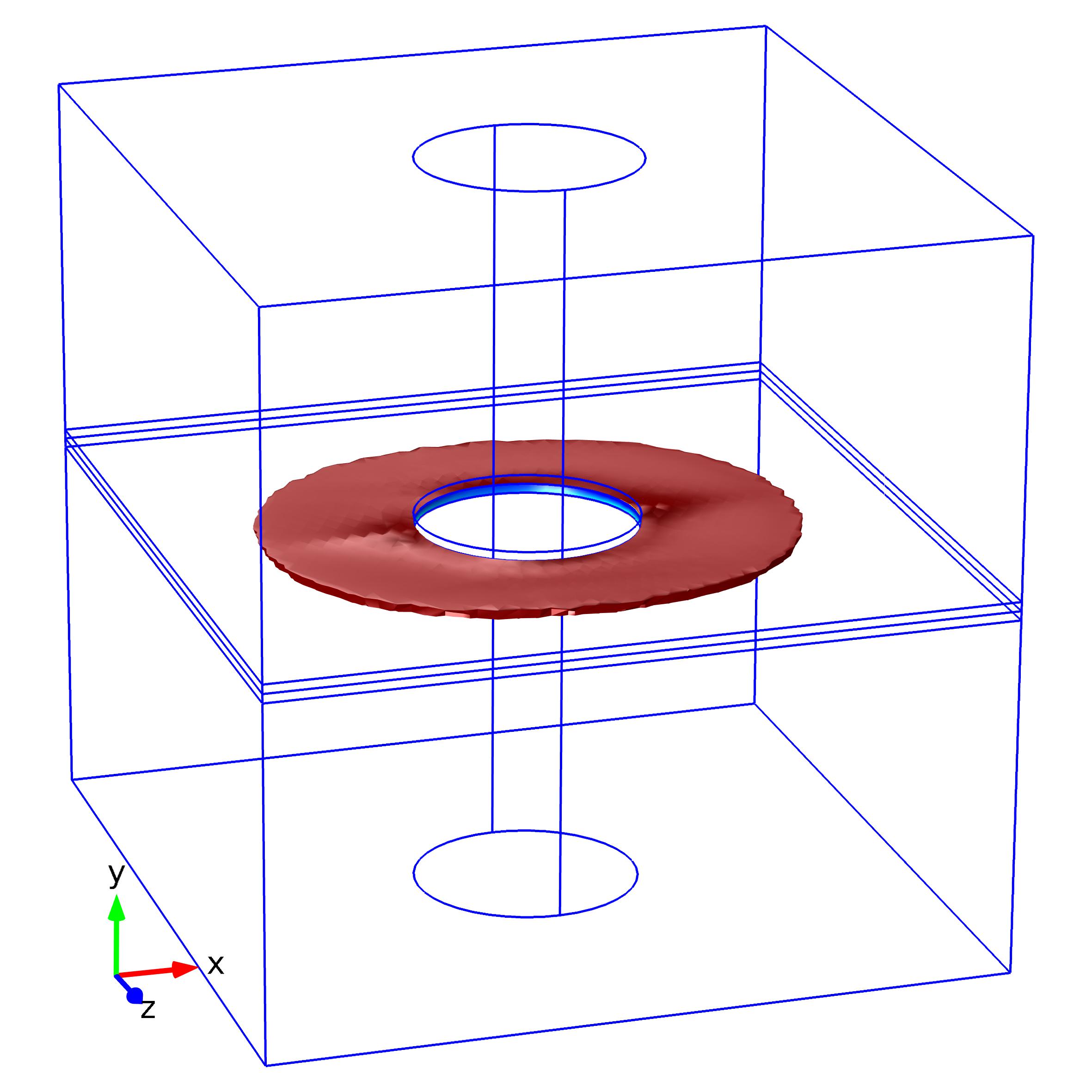}}
	\subfigure[$G_{c2}=30$N/m, $t=33.3$ s]{\includegraphics[width = 5cm]{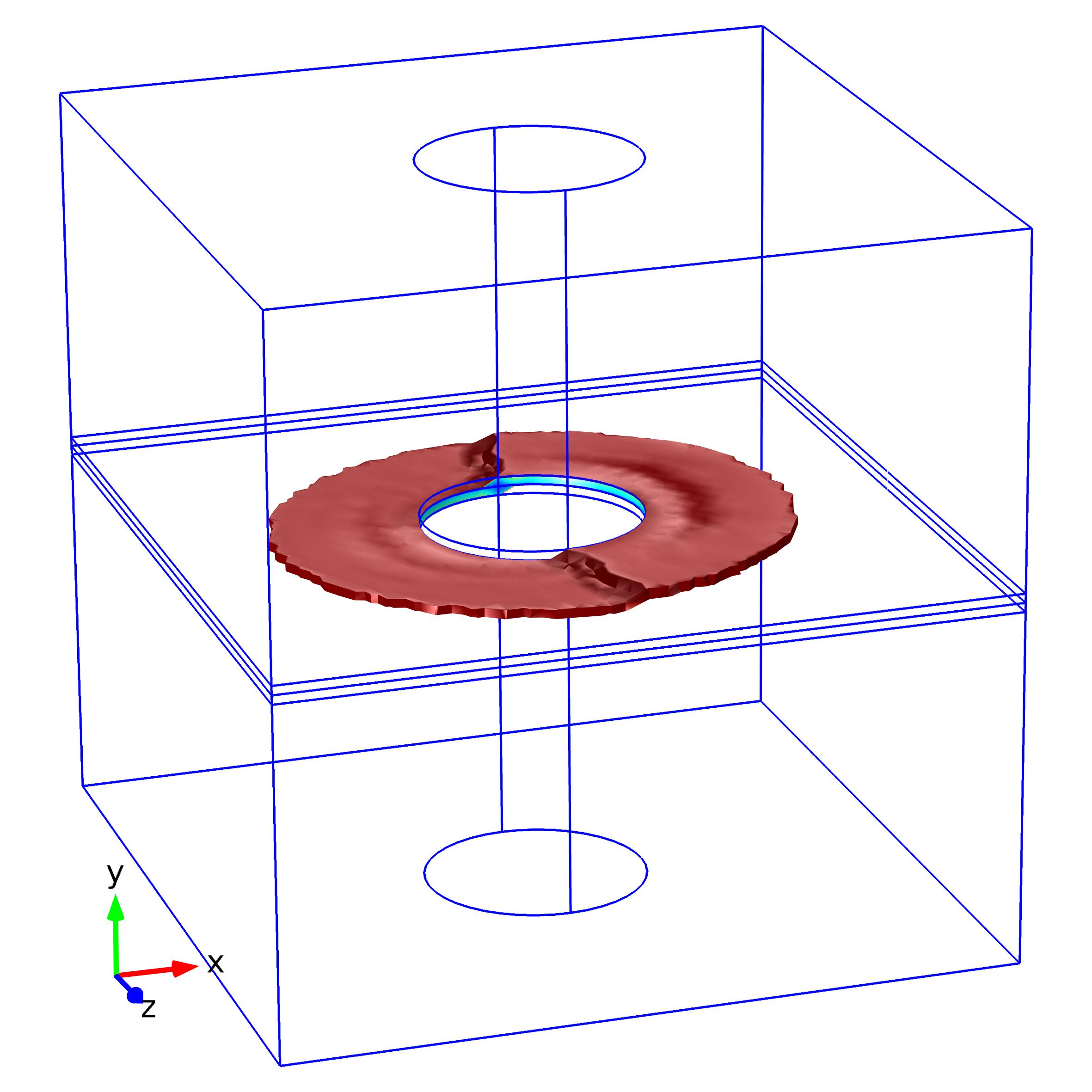}}
	\subfigure[$G_{c2}=100$N/m, $t=37.0$ s]{\includegraphics[width = 5cm]{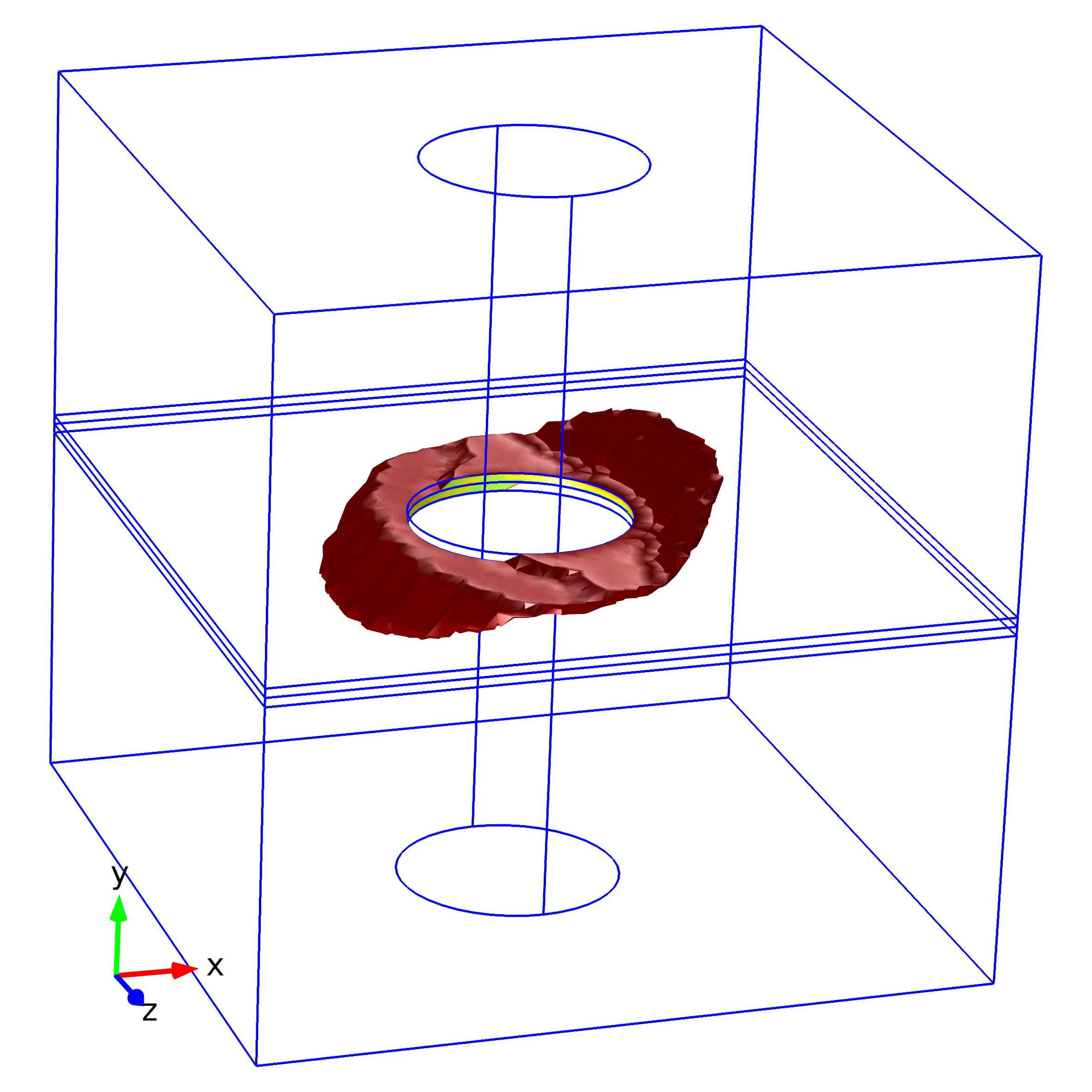}}\\		
	\caption{Fracture evolution of the 3D transversely isotropic porous medium with an interior notch under different $G_{c2}$}
	\label{Fracture evolution of the 3D transversely isotropic porous medium with an interior notch under different Gc2}
\end{figure}

The 3D fracture patterns under $G_{c2} = $ 10, 30, 100 N/m are investigated while $G_{c1}$ = 10 N/m is fixed and the time increment is set as 0.05 s. The corresponding hydraulic fractures are shown in Fig. \ref{Fracture evolution of the 3D transversely isotropic porous medium with an interior notch under different Gc2}, which indicates that the changes of 3D hydraulic fracture path induced by the presence of fracture toughness anisotropy can be captured by the proposed PFM. As depicted in Fig. \ref{Fracture evolution of the 3D transversely isotropic porous medium with an interior notch under different Gc2}a, for $G_{c2} = 10$ N/m, the porous media is isotropic and the hydraulic fracture propagates in the $x-y$ plane. This fracture type is in good agreement with the observations in \citet{lee2017iterative}. However, owing to the increase of $G_{c2}$, the hydraulic fracture propagating in the $x-y$ plane will deflect towards the direction of the rotation angle, as shown in Fig. \ref{Fracture evolution of the 3D transversely isotropic porous medium with an interior notch under different Gc2}b and c.	

\section {Conclusions}\label{Conclusions}

In this paper, a phase field model for hydraulic fracture propagation in transversely isotropic porous media is proposed. The coupling between the fluid flow and displacement fields is established based on the classical Biot poroelasticity theory, while the phase field model is used to reflect the fracture behavior. The transversely isotropic constitutive relationship between stress and strain is applied, and the anisotropy in fracture toughness and permeability is also considered. In addition, an additional pressure-related term and an anisotropic fracture toughness tensor are introduced in the energy functional, which is then used to achieve the governing equations of strong form by using the variational method. On the other hand, the phase field value is used to construct indicator functions that transit the fluid property from the intact domain to the fully fractured one. 

The phase field model is implemented in COMSOL Multiphysics by using the finite element method where a staggered scheme is applied. In addition, two examples are used to initially verify the proposed PFM: a transversely isotropic single-edge-notched square plate subjected to tension and an isotropic porous medium subjected to internal fluid pressure. Finally, three representative numerical examples are presented to further prove the capability of the proposed PFM in 2D and 3D problems: internal fluid-driven fracture propagation in a 2D transversely isotropic medium with an interior notch, a 2D medium with two parallel interior notches, and a 3D medium with a penny-shaped notch.

All the numerical examples indicate that the proposed phase field model can characterize complex hydraulic fracture propagation patterns in the transversely isotropic medium, which verifies the strong capability of the proposed phase field model in 2D and 3D cases. In addition, all the examples also indicate the hydraulic fracture paths are highly related to the rotation angle between the material direction and global coordinate system, reflecting one of the basic features in transversely isotropy. The presented numerical examples only use constant permeability in the fracture because the fracture opening cannot be directly extracted in a PFM. Therefore, future research will incorporate more reasonable permeability models \citep{witherspoon1980validity} into PFMs with reference to more benchmark examples \citep{spence1985self, schrefler2006adaptive}. In addition, in future research, the proposed PFM should be extended to more complex cases such as inelastic, partially saturated, or heterogeneous transversely isotropic porous media. 

\section{Acknowledgement}
The authors gratefully acknowledge financial support provided by the Natural Science Foundation of China (51474157), and RISE-project BESTOFRAC (734370).

\bibliography{references}

%\onecolumn
%\tableofcontents
\end{document}